\documentclass[aps,prd,floatfix,showpacs,showkeys,superscriptadress,amsmath,unsortedaddress,nofootinbib]{revtex4-1}
\pdfoutput=1
\usepackage{amsmath,mathtools}
\usepackage{amssymb}
\usepackage{bbm}
\usepackage{graphicx}
\usepackage{float}
\usepackage{braket}
\usepackage{slashed}
\usepackage{tensor}
\usepackage{caption}
\usepackage{stix}
\usepackage{enumerate}
\usepackage{graphicx}
\usepackage{subfig}
\usepackage{color}
\usepackage{titlesec}
\usepackage{hyperref}
\hypersetup{
    colorlinks=true,
    linkcolor=red,
    filecolor=magenta,      
    urlcolor=blue,
}
\def\be {\begin{equation}}
\def\ee {\end{equation}}
\def\bea {\begin{eqnarray}}
\def\eea {\end{eqnarray}}
\def\bc {\begin{center}}
\def\ec {\end{center}}
\def\nn {\nonumber}
\newcommand \Tr{\operatorname{\text{Tr}}}

\newcommand{\p}{\prime}
\newcommand{\sss}[1]{{\scriptscriptstyle #1}}

\begin{document}

\title{\textbf{General structure of  fermion two-point function  and its  spectral representation 
		in a hot magnetised medium }}
\author{Aritra Das,}
\email[]{aritra.das@saha.ac.in} 
\affiliation{HENPP Division, Saha Institute of Nuclear Physics, HBNI, 
	1/AF Bidhan Nagar, Kolkata 700064, India.}
\author{Aritra Bandyopadhyay,}
\email[]{aritra.bandyopadhyay@saha.ac.in}
\affiliation{Theory Division, Saha Institute of Nuclear Physics, HBNI, 
	1/AF Bidhan Nagar, Kolkata 700064, India} 
\affiliation{Departamento de Fisica, Universidade Federal de Santa Maria, 
	Santa Maria, RS 97105-900, Brazil.} 
\author{Pradip K. Roy,}
\email[]{pradipk.roy@saha.ac.in}
\affiliation{HENPP Division, Saha Institute of Nuclear Physics, HBNI, 
	1/AF Bidhan Nagar, Kolkata 700064, India.} 
\author{Munshi G. Mustafa}
\email[]{munshigolam.mustafa@saha.ac.in}
\affiliation{Theory Division, Saha Institute of Nuclear Physics, HBNI, 
	1/AF Bidhan Nagar, Kolkata 700064, India.}
	\date{\today}
\begin{abstract}
We have systematically constructed the general structure of the fermion self-energy and the effective quark propagator 
in presence of a nontrivial background like hot magnetised medium. This is applicable to both QED and QCD.  The hard thermal loop 
approximation has been used for the heat bath.
We have also examined   transformation properties  of the effective fermion propagator under some of the discrete symmetries of the system.
Using the effective fermion propagator we have analysed the fermion dispersion spectra in a hot magnetised medium  along with the spinor for
each fermion mode  obtained by solving the  modified Dirac equation. The fermion spectra is found to reflect the discrete 
symmetries of the two-point functions. We note that for a chirally symmetric theory the degenerate left and right handed chiral modes 
in vacuum or in a heat bath get separated and become asymmetric in presence of magnetic field without disturbing the chiral invariance. 
The obtained general structure of the two-point functions  is verified by computing the three-point function, which agrees with the existing 
results  in one-loop order.  Finally, we have computed explicitly the spectral representation of the two-point functions which would be very
important to study the spectral properties of the hot magnetised medium corresponding to QED and QCD with background magnetic field.
\end{abstract}

\maketitle
            
	
\section{Introduction} \label{introduction}
In non-central
heavy ion collision (HIC) experiments in LHC at CERN and in RHIC at BNL, it is
believed that a very strong magnetic field is created in the direction perpendicular to the reaction plane due to the 
spectator particles that are not participating in the collisions.  The experiments conducted 
by PHENIX Collaboration~\cite{phenixphoton} showed direct-photon anisotropy 
which has posed a serious challenge to the present theoretical models. It is 
conjectured that this excess elliptic flow may be due to the excess
photons produced by the decay $\rho\rightarrow\,\pi(\eta)\,\gamma$ and the
branching ratio of which increases in presence of the magnetic field near the
critical value where the condensate of $\rho$ is found.
The estimated strength of this magnetic field depends on collision energy and 
impact parameter between the colliding nuclei and is about several times the
pion mass squared, \textit{i.e.}, \( eB \sim 15m^{2}_{\pi}\) at LHC in 
CERN~\cite{VSkokov2009}. Also, a class of neutron star called magnetar 
exhibits ~\cite{VdelaIncera,Bandyopadhyay:1997kh,Chakrabarty:1997ef}
a magnetic field  of $10^{18}-10^{20}$ Gauss at the inner core and 
~$10^{12}-10^{13} $ Gauss at the surface. These observations motivate 
to study the properties of hot magnetised medium
using  both phenomenology and quantum field theory. 

The presence of a strong magnetic field in HIC influences the QCD phase transitions~\cite{andersenphase} and 
particle productions, especially the production of the soft photon~\cite{Basar:2012bp} and
dileptons~\cite{sadooghidilepton,aritraemspectrum,aritradilepton,Tuchin:2013apa,Tuchin:2012mf,Tuchin:2013bda}, 
which act as a probe of the medium.  
Apart from these, there is a large class of other phenomena that take place 
in presence of background magnetic field like chiral magnetic effects  
due to axial anomalies~\cite{dmitrikharzeevCME,FukushimaCME,CMEaxial},
magnetic catalysis~\cite{miranskydimred,igormagneticcatalysis}, 
inverse magnetic catalysis~\cite{inversemagneticcatalysis,Farias:2014eca}, 
superconductivity of the vacuum~\cite{vacuumsuperconductivity}. 
It further influences the thermal chiral and deconfining phase 
transition~\cite{chiralconfinedeconfine}, change of topological charge~\cite{kharzeevtopo}, 
anomalous transports~\cite{anomaloustransport}, refractive indices~\cite{Hattori:2012je,Hattori:2012ny} and 
screening mass~\cite{mesonrefractive}, decay constant~\cite{mesondecay} of 
neutral mesons etc. In addition,  efforts  were also made  to study the bulk properties for  
a fermi gas~\cite{Strickland:2012vu}, low lying hadrons~\cite{Andersen:2012zc} and
 strongly coupled systems~\cite{Mamo:2013efa}, 
collective excitations in  magnetised QED medium~\cite{Sadooghi:2015hha} using Ritus method 
and QCD medium~\cite{elmforsfermion} using Furry's picture,
neutrino  properties~\cite{Bhattacharya:2002qf,Bhattacharya:2002aj} and Field theory of the Faraday effects~\cite{Ganguly:1999ts,DOlivo:2002omk}. 

The magnetic field created in HIC lasts for very short time ( $\sim$ a few fm). The strength of the field decays 
rapidly with time after $\tau\sim 1-2$ $fm/c$. However, the medium effects like electric conductivity can delay 
the decay and by the time deconfined quarks and gluons equilibrate with QGP medium, the magnetic field strength
gets sufficiently weak. At that time the relevant energy scales of the system
can be put in this way: $q_{f}B < m^{2}_{\pi}\ll T^{2}$. In this low field limit the properties of the 
deconfined medium are also affected. So, it becomes important to treat the weak field limit separately. 
Fermion propagator in presence of a uniform background magnetic 
field has been derived first by Schwinger~\cite{schwinger1951}. Using this, one loop fermion self-energy and the vacuum polarisation was calculated in 
double parameter integral in~\cite{tsaifermion} and~\cite{tsaivacuumpol}, respectively. The weak field expansion of this propagator was calculated 
order by order in powers  of $q_{f}B$ in~\cite{chyiweak}. Recently, the pion self-energy and its dispersion property  have been studied 
at zero temperature~\cite{pradiproypionself} in weak field approximation and using full propagator at finite temperature~\cite{Mukherjee:2017dls}.  Also a detailed study of 
the spectral properties of $\rho$ mesons has been performed in presence of magnetic field both  at zero~\cite{arghyarho,Bandyopadhyay:2016cpf} and at non-zero temperature~\cite{arghyarhothermal}.

For hot and dense medium (\textit{e.g.}, QED and QCD plasma), it is well know that a bare perturbation theory 
breaks down due to infrared divergences.  
A reorganisation of the  perturbation theory has been
done  by performing the expansion around a system of massive quasiparticles~\cite{Andersen:1999fw}, where  mass is generated through 
thermal fluctuations.  This requires a resummation of certain class of diagrams, known as hard thermal loop (HTL) resummation~\cite{brateennucl337},
when  the loop momenta are of the order of the temperature. This reorganised  perturbation theory,
 known as HTL perturbation theory (HTLpt),  leads to
 gauge independent results for various physical quantities~\cite{braatendilepton,HTLgluondamping,Haque:2011iz,Haque:2011vt,Haque:2010rb,Andersen:2002ey,Andersen:2003zk,Haque:2012my,Haque:2013qta,Andersen:2009tc,Andersen:2010ct,Andersen:2010wu,Andersen:2011sf,Andersen:2011ug,Haque:2013sja,Haque:2014rua,Mustafa:2004hf}.  Within this one-loop HTLpt, the thermomagnetic 
 correction  to the quark self-energy~\cite{ayalafermionself}, quark-gluon three point~\cite{ayalafermionself} function at zero chemical potential and four point~\cite{Haque:2017nxq} function at finite chemical potential in weak field limit  have been computed. The fermion self-energy 
 has also been extended to the case of non-zero chemical potential and  the pressure of a weakly magnetised QCD 
plasma~\cite{aritraweakpressure} has also been obtained.

In recent years  a huge amount of activity  is underway to explore the properties of a hot medium  
with a background magnetic field using phenomenology  as well as using thermal field theory.  
In a thermal medium  the bulk and dynamical properties~\cite{brateennucl337,weldonfermion,Weldon:1982aq} are 
characterised by the collective excitations in a time like region and the Landau damping in 
a space-like domain. The basic quantity  associated with these medium properties 
is the two point correlation function.  In this work we construct the  general structure of the fermionic 
two point functions  (e.g., self-energy and the effective propagator) in a nontrivial background 
like a hot magnetised medium. 
We then analyse  its property under the transformation of some discrete symmetries of the system,  
the  collective fermionic spectra, QED like three-point functions and the spectral representation of 
the two point function and its consequences in a hot magnetised medium. 
The formulation is applicable equally well 
to both QED and QCD. 
		
The paper is organised as follows; In section \ref{fer_prop_section}, the notation and set up  are briefly
discussed through a  fermion 
propagator in a constant background field  using Schwinger formalism.
Section \ref{gen_2pt}  has number of parts in which we obtain  
the general structure of the self-energy  (subsec.\ref{self_gen_structure}), the effective fermion propagator (subsec.\ref{eff_fer_prop} ),
the transformation properties and discrete symmetries of the effective propagator (subsec.\ref{prop_trans}), the modified Dirac equations
in general and for lowest Landau level (subsec.\ref{dirac_mod}) and 
the dispersion properties of the various collective modes (subsec.\ref{disp_rep}) in time-like region.  In section ~\ref{vert_func} the general structure
of the self-energy and the propagator has been verified from one-loop direct calculation. The spectral
representation of the effective propagator in space-like domain has been obtained in section \ref{spec_rep}. We have presented
some detailed calculations for various sections and subsections in Appendix \ref{append_A}-\ref{spec_htl}. Finally, we conclude in section \ref{remarks}.

		
\section{Charged Fermion Propagator in Background Magnetic Field within Schwinger Formalism} 
\label{fer_prop_section}

In this section we set the notation and briefly outline the fermionic propagator in presence of a background  magnetic field 
following Schwinger formalism~\cite{schwinger1951} . Without any loss of generality,  the background 
magnetic field is  chosen along the $z$ direction, \(\vec{B}=B\hat{z}\),  
and the  vector potential  in a symmetric gauge reads as 
\be
A^\mu =(0,-\frac{yB}{2},\frac{xB}{2}, 0) \, . \label{symm_g}
\ee

 Below we also outline the  notation we shall be using throughout:  
 \bea
&& a\indices{^\mu}=(a\indices{^0},a\indices{^1},a\indices{^2},a\indices{^3})=(a_{0},\vec{a}); ~~ a\cdot b\equiv a\indices{_0}b\indices{_0}-\vec{a}\cdot\vec{b};~~
g\indices{^\mu^\nu}=\textsf{diag}\left(1,-1,-1,-1\right),\nn \\
&& a^\mu = a_\shortparallel^\mu + a_\perp^\mu;~~ a_\shortparallel^\mu = (a^0,0,0,a^3) ;~~  a_\perp^\mu = (0,a^1,a^2,0)  \nn\\
&&g^{\mu\nu} = g_\shortparallel^{\mu\nu} + g_\perp^{\mu\nu};~~ g_\shortparallel^{\mu\nu}= \textsf{diag}(1,0,0,-1);~~ g_\perp^{\mu\nu} = \textsf{diag}(0,-1,-1,0),\nn\\
&&(a\cdot b) = (a\cdot b)_\shortparallel - (a\cdot b)_\perp;~~ (a\cdot b)_\shortparallel = a^0b^0-a^3b^3;~~ (a\cdot b)_\perp = a^1b^1+a^2b^2, \nn \\
&& 	\slashed{a}=\gamma\indices{^\mu}a\indices{_\mu}=\slashed{a}_{\shortparallel}+\slashed{a}_{\perp};~~~
\slashed{a}_{\shortparallel} = \gamma\indices{^0}a\indices{_0}-\gamma\indices{^3}a\indices{^3};~~~ 
\slashed{a}_{\perp} = \gamma\indices{^1}a\indices{^1}+\gamma\indices{^2}a\indices{^2}
\eea
where $\shortparallel$ and $\perp$ are, respectively, the parallel and perpendicular 
components, which would be  separated out due to the presence of the background magnetic field.

Now, the fermionic two-point function is written as
\begin{small}
\begin{align}
& S(x,x^{\prime})=-i\,C(x,x^{\prime})\int_{0}^{\infty}\,ds\,\,\frac{1}{s\,\sin(q_{f}\,B\,s)} \exp\left(-im_f^2\,s+i\,q_{f}\,B\,s\,\Sigma_{3}\right)\, \nonumber\\
& \hspace{1cm}\exp\left[-\frac{i}{4\,s}\left((x-x^{\prime})^2_{\shortparallel}-\frac{q_{f}\,B\,s}{\tan(q_{f}\,B\,s)}(x-x^{\prime})^2_{\perp}\right)\right] \nonumber \\ 
& \times \left[m_f+\frac{1}{2\,s}\left((\slashed{x}_{\shortparallel}-\slashed{x}^{\prime}_{\shortparallel})-\frac{q_{f}\,B\,s}{\sin(q_{f}\,B\,s)}\exp\left(-i\,q_{f}\,B\,s\,
\Sigma_{3}\right)\,\left(\slashed{x}_{\perp}-\slashed{x}^{\prime}_{\perp}\right)\right)\right], \label{gxxp}
\end{align}
\end{small}
where  the parameter $s$ is called Schwinger proper time  variable \cite{schwinger1951}. 
We note that $m_f$ and $q_f$ are mass and \textit{absolute charge} of the fermion of flavour $f$, respectively.
The phase factor, $C(x,x^{\prime})$, is independent of $s$  but is 
responsible for breaking of both gauge and  translational invariance.   Remaining part, denoted  as $\mathcal{S}(x-x^{\prime})$,  is  translationally invariant.  
However,  as shown below, $C(x,x^{\prime})$ drops out for a gauge invariant calculation.  Now $C(x,x^{\prime})$ reads as
\begin{align}
C(x,x^{\prime}) &= C\,\exp\left[i\,q_{f}\,\int_{x^{\prime}}^{x}\,d\xi^{\mu}\left(A_{\mu}(\xi)+\frac{1}{2}F_{\mu\nu}(\xi-x^{\prime})^{\nu}\right)\right] , \label{sol_c}
\end{align}
where $C$ is just a number.  The integral in the exponential is independent of the path taken 
between \(x\) and \(x^{\prime}\) and  choosing it as a straight line one  can write 
\begin{align}
C(x,x^{\prime}) = C\,\Phi(x,x^{\prime})=C\,\exp\left[i\,q_{f}\,\int_{x^{\prime}}^{x}\,d\xi^{\mu}A_{\mu}(\xi)\right].
\end{align}
Using the gauge transformation 
$A^{\mu}(\xi) \rightarrow A^{\mu}(\xi)+\partial^\mu\Lambda(\xi)$,
and choosing symmetric gauge as given in (\ref{symm_g}), the phase factor \(\Phi(x,x^{\prime})\) becomes $1$, if we take \cite{ayalafermionself} 
\begin{align}
 \Lambda(\xi)=\frac{B}{2}\left(x^{\prime}_{2}\xi_{1}-x^{\prime}_{1}\xi_{2}\right).
\end{align}
From equation \eqref{gxxp},  the momentum space propagator can be obtained  as
\begin{small}  
\begin{align}
S(K) &= \int d^{4}x\,e^{iK\cdot x}\,\mathcal{S}(x-x^{\prime}) \nonumber \\
&= -i\int_{0}^{\infty}\,ds\,\exp\left[i\,s\,\left(K^{2}_{\shortparallel}-\frac{\tan(q_{f}B\,s)}{q_{f}B\,s}K^{2}_{\perp}-m_f^{2}\right)\right] \nonumber \\  
&\,\,\, \, \, \times 
\left[\left(1+\gamma_{1}\gamma_{2}\tan(q_{f}B\,s)\right)\left(\slashed{K}_{\shortparallel}+m_f\right)-\sec^2(q_fBs){\slashed {K}_{\perp}}\right] \nn \\
&= \,\exp\left(-{{K_\perp}^{2}}/{|q_{f}B|}\right)\sum_{l=0}^{\infty}(-1)^n\frac{D_{n}(q_{f}B,K)}{K^{2}_{\shortparallel}-m_f^{2}-2l|q_{f}B|} , \label{mirprop}
\end{align}
where  $k_\perp^2 = 2l|q_fB|$, is quantised with Landau level $l=0,1, \cdots$,  and 
\begin{align}
D_{l}(q_{f}B,K) &= \left(\slashed{K}_{\shortparallel}+m_f\right)\left[(1-i\gamma_{1}\gamma_{2})L_{l}\left(2\frac{{K_\perp}^{2}}{|q_{f}B|}\right)
- (1+i\gamma_{1}\gamma_{2})L_{l-1}\left(2\frac{{K_\perp}^{2}}{|q_{f}B|}\right)\right] \nonumber \\
& - 4\slashed{K}_{\perp}L^{1}_{l-1}\left(2\frac{{K_\perp}^{2}}{|q_{f}B|}\right), 
\label{lagu}
\end{align}
\end{small}
where \( L_{l}(x) \) is Laguerre polynomial, \( L^{j}_{l}(x) \) is associated Laguerre polynomial 
with \( L^{j}_{-1}(x)=0 \) and both $j$ is a non-negative integer. 	

Below we discuss the structure of the propagator in \eqref{mirprop} in presence of background magnetic field.  Since fermion propagator is
$4\times 4$ matrix,  a new matrix $ (\slashed{K}_{\shortparallel}+m_f) i\gamma_1\gamma_2$  appears in addition to that of the vacuum structure 
$(\alpha'{\slashed{K}}$, $\alpha'(K^2)$  is a Lorentz invariant structure function) for a chirally symmetric  theory.  One can now write 
the new matrix for a chirally  symmetric theory in terms of background electromagnetic field tensor $F^{\rho\lambda}$ as
\bea
i\gamma_1\gamma_2  \slashed{K}_{\shortparallel} \, B &=& -\gamma_5K^\mu {\tilde F}_{\mu\nu} \gamma^\nu,  \label{rel_t0}
\eea
where the background dual field tensor reads as
\be
{\tilde F}_{\mu\nu}= \frac{1}{2} \epsilon_{\mu\nu\rho\lambda} F^{\rho\lambda}. \label{dual} 
\ee
 The structure of a chirally symmetric  free fermion propagator
in presence of only magnetic field can be viewed as  ($\alpha'{\slashed{K}}+\delta' \gamma_5K^\mu {\tilde F}_{\mu\nu} \gamma^\nu $),
where $\delta'$ is a new structure constant that appears due to the presence of background magnetic field. When a fermion propagates only 
in a hot  medium, then  the vacuum part will  be modified  only due to the thermal background~\cite{weldonfermion} 
as $(\alpha'{\slashed{K}}+\beta'\slashed{u})$,  where $u$ is  the four velocity of the heat bath.  When a fermion moves in a nontrivial background 
like  hot magnetised medium then 
one can write \eqref{rel_t0} as
 \begin{align}
 i\gamma_{1}\gamma_{2}\slashed{K}_{\shortparallel}  &= -\gamma_{5}\left[\left(K.n\right)\slashed{u}-\left(K.u\right)\slashed{n}\right]  
 , \label{acom} 
 \end{align}
 where 
 \be
 n_\mu = \frac{1}{2B} \epsilon_{\mu\nu\rho\lambda}\, u^\nu F^{\rho\lambda} =  \frac{1}{B}u^\nu {\tilde F}_{\mu\nu} \, . \label{nmu}
 \ee
and  the four velocity in the rest frame of the heat bath and the direction of the magnetic field $B$, respectively, given as 
\begin{subequations}
 \begin{align}
 u^{\mu} &= (1,0,0,0), \label{fv4} \\
 n^{\mu} &= (0,0,0,1) . \label{mgdir}
 \end{align}
\end{subequations}
One can notice that  in a hot magnetised medium both $u$ and $n$ are correlated as given in \eqref{nmu} 
and the contribution due to magnetic field in \eqref{rel_t0} in presence of  heat bath becomes  a thermo-magnetic contribution.  
We also further note that in absence of heat bath, \eqref{acom} reduces to \eqref{rel_t0}, which is not obvious by inspection
 but we would see later.
\section{General Structure of Fermion Two-point Function  in a Hot Magnetised Medium} 
\label{gen_2pt}
In previous section the modification of a free propagator has been discussed briefly 
in presence of a background magnetic field. 
In this section we would like to obtain  the most general structure of a fermion self-energy,  the effective fermion propagator
and some of its properties  in a nontrivial background like hot magnetised medium.   We would also discuss the modified  Dirac equation 
and the fermion dispersion  spectrum in a hot magnetised medium. For thermal bath we would  use HTL approximation and 
any other approximation required for the purpose will be stated therein.

\subsection{General Structure of the Fermion Self-Energy}  
\label{self_gen_structure}
The fermionic self-energy is a matrix as well as a Lorentz scalar. However, in presence of nontrivial background, \textit{e.g.,}  heat bath 
and magnetic field, the boost and rotational 
symmetries of the system are broken.  The general structure of fermion self-energy for hot magnetised medium  can be written by the following arguments.
The self-energy  $\Sigma(P)$ is a $4\times 4$ matrix which  depends, in present case, on the four momentum of the fermion $P$, 
the  velocity of the heat bath $u$ and  the direction of the magnetic field $n$. 
Now, any $4\times 4$ matrix can be expanded in terms of 16 basis 
matrices: $\{\mathbbm{1},\gamma_{5},\gamma_{\mu},\gamma_{\mu}\gamma_{5},\sigma_{\mu\nu}\}$, which are the unit matrix, the four $\gamma$-matrices, 
the six $\sigma_{\mu\nu}$ matrices, the four $\gamma_{5}\gamma_{\mu}$ matrices and finally 
$\gamma_{5}$.  So, the general structure can be written as
\begin{align}
\Sigma(P) &= -\alpha \mathbbm{1} - \beta \gamma_{5} - a \slashed{P} - b\slashed{u} - c \slashed{n} - a'\gamma_{5}\slashed{P} - b^{\prime}\gamma_{5}\slashed{u} 
- c^{\prime}\gamma_{5}\,\slashed{n} \nonumber \\
& -h\, \sigma_{\mu\nu}P^{\mu}P^{\nu}- h^\prime \sigma_{\mu\nu}u^{\mu}u^{\nu}- \kappa \ \sigma_{\mu\nu}n^{\mu}n^{\nu}
 - d\sigma_{\mu\nu}P^{\mu}u^{\nu}-d^{\prime}\sigma_{\mu\nu}n^{\mu}P^{\nu}-\kappa^\prime\sigma_{\mu\nu}u^{\mu}n^{\nu} 
\, , \label{genstructselfenergy0}
\end{align}
where various coefficients are known as structure functions.
We note that the combinations involving $\sigma_{\mu\nu}$ do not 
appear due to antisymmetric nature of it in any  loop order of self-energy. Also in a chirally invariant 
theory, the terms \( \alpha \mathbbm{1} \)  and \(\gamma_{5}\beta \)  will not appear as they would break the chiral symmetry.  
The term \(  \gamma_5\slashed{P} \) would appear in the self-energy if  fermions interact with an axial vector\footnote{
The presence of an axial gauge coupling leads to chiral or axial anomaly  and a chirally invariant theory does not 
allow this. Other way,  the preservation of both chiral and axial symmetries is impossible,  a choice must be made 
 which one should be preserved.   For a chirally invariant theory this term drops out. Also the presence of $\gamma_5$ 
 in a Lagrangian violates parity invariance.}.
By dropping those  in \eqref{genstructselfenergy0} for chirally symmetric theory, one
can now write
\begin{align}
\Sigma(P) &= - a \,\slashed{P} - b\,\slashed{u} - c \,\slashed{n}  - b^{\prime}\gamma_{5}\,\slashed{u} 
- c^{\prime}\gamma_{5}\,\slashed{n} . \label{genstructselfenergy1}
\end{align}
Now we  point out that some important information is encoded  into the fermion propagator 
in  \eqref{mirprop}  through \eqref{acom} for a hot magnetised medium.  This suggests that $c\slashed{n}$ should not appear
in the fermion self-energy~\footnote{We have checked that even if one keeps $c \,\slashed{n}$, the coefficient $c$ becomes zero 
in one-loop order in the weak field approximation.}  and the most general form of the fermion self-energy for a hot magnetised medium becomes 
\begin{align}
\Sigma(P) &= - a \,\slashed{P} - b\slashed{u} - b^{\prime}\gamma_{5}\,\slashed{u} 
- c^{\prime}\gamma_{5}\,\slashed{n}. \label{genstructselfenergy} 
\end{align}
When a fermion propagates in a  vacuum, then $b=b'=c'=0$ and $\Sigma(P)=-a\slashed{P}$. But when it propagates in a background of pure magnetic
field without any heat bath, then $a\ne 0$, $b=0$ and the structure  functions, $b'$ and $c'$, will depend only on the  
background magnetic field as we will see later. When a fermion propagates in a heat bath, then $a\ne 0$, $b\ne0$ but  both 
$b'$ and $c'$ vanish because there would not be any thermo-magnetic corrections as can also be seen later.

We now write down the  \emph{right chiral} projection operator, $\displaystyle \mathcal{P}_{+}$ and the \emph{left chiral} projection 
operator $\displaystyle \mathcal{P}_{-}$ , respectively,  defined as:
\begin{subequations}
\begin{align}
\mathcal{P}_{+} &= \frac{1}{2}\left(\mathbbm{1}+\gamma_{5}\right) \label{RChPO} , \\
\mathcal{P}_{-} &= \frac{1}{2}\left(\mathbbm{1}-\gamma_{5}\right) \label{LChPO} ,
\end{align}
\end{subequations} 
which  satisfy the usual properties of projection operator:
\begin{equation}
\mathcal{P}^{2}_{\pm} = \mathcal{P}_{\pm}, \quad \mathcal{P}_{+}\,\mathcal{P}_{-}=\mathcal{P}_{-}\,\mathcal{P}_{+} 
= 0, \quad \mathcal{P}_{+}+\mathcal{P}_{-} = \mathbbm{1}, \quad \mathcal{P}_{+}-\mathcal{P}_{-} = \gamma_{5}. \label{PropProj}
\end{equation}
Using the chirality projection operators,  the general structure of the self-energy in \eqref{genstructselfenergy} 
can be casted in the following form		
\begin{align}
\Sigma(P) = -\mathcal{P}_{+}\,{\slashed{C}}\,\mathcal{P}_{-} -\mathcal{P}_{-}\,{\slashed{D}}\,\mathcal{P}_{+},  \label{fergenstruct}
\end{align}
where $\slashed{C}$ and $\slashed{D}$ are defined as
\begin{subequations}
\begin{align}
\slashed{C} &= a\,\slashed{P}+(b+b^{\prime})\,\slashed{u}+c^{\prime}\,\slashed{n} \label{slashc} , \\
\slashed{D} &= a\,\slashed{P}+(b-b^{\prime})\,\slashed{u}-c^{\prime}\,\slashed{n}. \label{slashd}
\end{align}
\end{subequations}  
From \eqref{genstructselfenergy}  one obtains  the general form of the various  structure functions as
\begin{subequations}
\begin{align}
a &= \frac{1}{4}\,\,\frac{\Tr\left(\Sigma\slashed{P}\right)-(P.u)\,\Tr\left(\Sigma\slashed{u}\right)}{(P.u)^{2}-P^{2}} , \label{sta}\\
b &= \frac{1}{4}\,\,\frac{-(P.u)\,\Tr\left(\Sigma\slashed{P}\right)+P^{2}\,\Tr\left(\Sigma\slashed{u}\right)}{(P.u)^{2}-P^{2}} , 
\label{stb} \\
b^{\prime} &= - \frac{1}{4}\,\Tr\left(\slashed{u}\Sigma\gamma_{5}\right) , \label{stbp} \\
c^{\prime} &=  \frac{1}{4}\,\Tr\left(\slashed{n}\Sigma\gamma_{5}\right) , \label{stcp}
\end{align}
\end{subequations}
which are also Lorentz scalars . Beside $T$ and $B$, they would also depend on three Lorentz scalars defined by
\begin{subequations}
\begin{align}
\omega &\equiv P^{\mu}u_{\mu}, \label{ome} \\
p\indices{^3}&\equiv -P^{\mu}n_{\mu} =p_z \, , \label{p3} \\
p_{\perp} &\equiv \left[(P^{\mu}u_{\mu})^{2}-(P^{\mu}n_{\mu})^{2}-(P^{\mu}P_{\mu})\right] ^{1/2}. \label{pperp}
\end{align}
\end{subequations}
Since \(P^{2}=\omega^{2}-p^{2}_{\perp}-{p\indices{_z}}^{2}\), we may interpret $\omega$, $p_{\perp}$, $p\indices{_z}$ as Lorentz invariant 
energy, transverse momentum, longitudinal momentum respectively. 
All these structure functions for 1-loop order  in a weak 
field and HTL approximations have been computed in 
Appendix \ref{append_A} and quoted here~\footnote{In weak field approximation the domain of applicability  
becomes $m_{th}^2 (\sim g^2T^2)  < q_fB < T^2$ 
instead of $m^2 < q_fB< T^2$ as discussed in Appendix~\ref{append_A}.}  as
 \begin{small}
 \begin{subequations}
 \begin{align}
a(p_0,|\vec p|) &
\, = \, -\frac{m^{2}_{th}}{|\vec{p}|^{2}}Q_{1}\left(\frac{p_{0}}{|\vec{p}|}\right), \label{at}\\ 
b(p_0,|\vec p|)&
\, = \, \frac{m^{2}_{th}}{|\vec{p}|}\left[\frac{p_{0}}{|\vec{p}|}Q_{1}\left(\frac{p_{0}}{|\vec{p}|}\right)-Q_{0}\left(\frac{p_{0}}{|\vec{p}|}\right)\right], \label{bt} \\
b^{\prime}(p_0,|\vec p|) &
\, = \, 4C_{F}g^{2}M^{2}(T,m_f,q_{f}B)\frac{p_{z}}{|\vec{p}|^{2}}Q_{1}\left(\frac{p_{0}}{|\vec{p}|}\right), \label{bprime}\\
c^{\prime} (p_0,|\vec p|)&
\, =\, 4C_{F}g^{2}M^{2}(T,m_f,q_{f}B)\frac{1}{|\vec{p}|}Q_{0}\left(\frac{p_{0}}{|\vec{p}|}\right). \label{cprime}
\end{align}
\end{subequations}
\end{small}
We note that the respective vacuum contributions in $a$, $b'$ and $c'$ have been dropped by the choice of the renormalisation prescription, 
and the general structure of the self-energy,  as found in appendix \ref{append_A},  agrees with that in  \eqref{genstructselfenergy}. 
 \subsection{Effective Fermion Propagator}
  \label{eff_fer_prop}	
\begin{figure}[h!]
\centering
\includegraphics[scale=0.9]{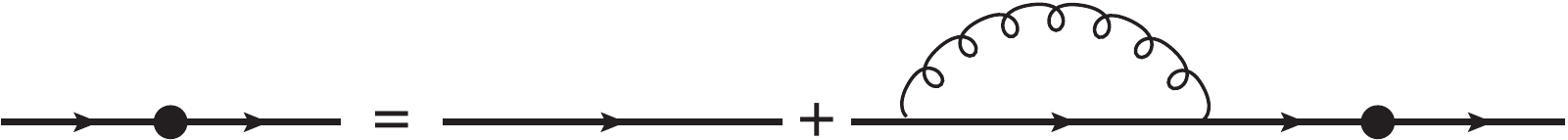}
\caption{\small Diagramatic representation of the Dyson-Schwinger equation for one-loop effective fermion propagator.}
\label{fig:dyson_schwinger}
\end{figure}

The effective fermion propagator is given by Dyson-Schwinger equation (see Fig. \ref{fig:dyson_schwinger}) which reads as 
 \begin{align}
 \,S^{*}(P)=\frac{1}{\slashed{P}-\Sigma(P)}\, , \label{eff_prop0}
 \end{align}
 and the inverse fermion propagator  reads as
 \be
 {S^*}^{-1}(P)  =\slashed{P}-\Sigma(P)\, . \label{inv_prop}
 \ee
 
 Using \eqref{fergenstruct} the inverse fermion propagator can be written as
\begin{align}
{S^*}^{-1}(P) 
&= \mathcal{P}_{+}\left[(1+a(p_{0},|\vec{p}|))\slashed{P}+\left(b(p_{0},|\vec{p}|)
+b^{\prime}(p_{0},p_{\perp},p_{z})\right)\slashed{u}+c^{\prime}(p_{0},|\vec{p}|)\slashed{n}\right] \mathcal{P}_{-} \nonumber \\
 &\,\, +\mathcal{P}_{-}\left[(1+a(p_{0},|\vec{p}|))\slashed{P}+\left(b(p_{0},|\vec{p}|)-b^{\prime}(p_{0},p_{\perp},p_{z})\right)\slashed{u}
 -c^{\prime}(p_{0},|\vec{p}|)\slashed{n}\right]\mathcal{P}_{+} \nn \\
 &= \mathcal{P}_{+}\,\slashed{L}\,\mathcal{P}_{-}+\mathcal{P}_{-}\,\slashed{R}\,\mathcal{P}_{+} \, , \label{eff_in_prop0}
\end{align}
where $\slashed{L}$ and $\slashed{R}$  can be obtained from  two four vectors given by
\begin{subequations}
\begin{align}
L\indices{^\mu}(p_{0},p_{\perp},p_{z}) &= \mathcal{A}(p_{0},|\vec{p}|)\,P^{\mu}
+\mathcal{B}_{+}(p_{0},p_{\perp},p_{z})\,u^{\mu}+c^{\prime}(p_{0},|\vec{p}|)\, n^\mu , \label{l_mu} \\
R\indices{^\mu}(p_{0},p_{\perp},p_{z}) &= \mathcal{A}(p_{0},|\vec{p}|)\,P^{\mu}
+\mathcal{B}_{-}(p_{0},p_{\perp},p_{z})\,u^{\mu}-c^{\prime}(p_{0},|\vec{p}|)\, n^\mu \,  ,
\label{r_mu}
\end{align}
\end{subequations}
with                                           
\begin{subequations}
\begin{align}
\mathcal{A}(p_{0},|\vec{p}|)&=1+a(p_{0},|\vec{p}|), \label{cal_a} \\
\mathcal{B}_{\pm}(p_{0},p_{\perp},p_{z})&=b(p_{0},|\vec{p}|) \pm b^{\prime}(p_{0},p_{\perp},p_{z}) \ . \label{cal_bpm}
\end{align}
\end{subequations}
Using \eqref{eff_in_prop0}  in \eqref{eff_prop0}, the propagator can now be written as
\begin{align}
S^{*}(P) &= \mathcal{P}_{-}\frac{\slashed{L}}{L^{2}}\mathcal{P}_{+} + \mathcal{P}_{+}\frac{\slashed{R}}{R^{2}}\mathcal{P}_{-} \, , \label{eff_prop1}
\end{align}
where we have used the properties of the projection operators
$\, \mathcal{P}_{\pm}\gamma\indices{^\mu}=\gamma\indices{^\mu}\mathcal{P}_{\mp}, \, \mathcal{P}^{2}_{\pm}=\mathcal{P}_{\pm},  \,  \mbox{and} \, 
\mathcal{P}_{+}\mathcal{P}_{-}=\mathcal{P}_{-}\mathcal{P}_{+}=0$.  It can be  checked that  $S^*(P){S^*}^{-1}(P)= \mathcal{P}_{+}
+\mathcal{P}_{-} =  \mathbbm{1}$ .  
 Also we have
\begin{subequations}
\begin{align}
L^2 =L^{\mu}L_{\mu} 
&= \left(\mathcal{A}p_{0}+\mathcal{B}_{+}\right)^{2}-\left[\left(\mathcal{A}p_{z}
+c^{\prime}\right)^{2}+\mathcal{A}^{2}p^{2}_{\perp}\right] =L_0^2-|\vec{L}|^2 \, ,  \label{l2}\\
R^{2} =R^\mu R_\mu &= \left(\mathcal{A}p_{0}+\mathcal{B}_{-}\right)^{2}-\left[\left(\mathcal{A}p_{z}
-c^{\prime}\right)^{2}+\mathcal{A}^{2}p^{2}_{\perp}\right] = R_0^2-|\vec{R}|^2 \, ,\label{r2}
\end{align}
\end{subequations}
where we have used $u^{2}=1, \, n^{2}=-1, \,u\cdot n=0, \,  P\cdot u=p_{0},\, \, \mbox{and}\, \,   P\cdot n=-p_z$. Note that we have suppressed 
the functional dependencies of $L,\, R, \, \mathcal{A}$, $\mathcal{B}_{\pm}$ and $c^{\prime}$  and would bring them back 
whenever necessary. 

For the lowest Landau Level (LLL),  $l=0 \, \Rightarrow p_\perp=0$, and these relations reduce to
\begin{subequations}
\begin{align}
L^2_{LLL} &= \left(\mathcal{A}p_{0}+\mathcal{B}_{+}\right)^{2}-\left(\mathcal{A}p_{z} +c^{\prime}\right)^{2}=L_0^2-L_z^2 \, ,  \label{defLsquare_l0}\\
R^{2}_{LLL}  &= \left(\mathcal{A}p_{0}+\mathcal{B}_{-}\right)^{2}-\left(\mathcal{A}p_{z}-c^{\prime}\right)^{2} = R_0^2-R_z^2 \, .\label{defRsquare_r0}
\end{align}
\end{subequations}
The poles  of the effective  propagator,  $ L^2=0$  and  $ R^2 =0$,   give rise to quasi-particle dispersion relations in
a hot magnetised medium. There will be four collective modes with positive energies:  two from $L^2=0$ and two from $R^2=0$.  
Nevertheless,  we will discuss dispersion properties later.
\subsection{Transformation Properties of Structure Functions  and Propagator}
\label{prop_trans}
First, we outline some transformation properties of the various structure functions as obtained in 
\eqref{at}, \eqref{bt}, \eqref{bprime} and  \eqref{cprime}.
\begin{enumerate}
\item  Under the transformation  ${\vec p} \rightarrow -{\vec p}= (p_\perp,-p_z)$:
\begin{subequations}
\begin{align}
a(p_{0},|-\vec{p}|) &= a(p_{0},|\vec{p}|),  \label{amvp} \\
b(p_{0},|-\vec{p}|) &= b(p_{0},|\vec{p}|), \label{bmvp} \\
b^{\prime}(p_{0},p_{\perp},-p_{z}) &= -b^{\prime}(p_{0},p_{\perp},p_{z}), \label{bpmvp} \\
c^{\prime}(p_{0},|-\vec{p}|) &= c^{\prime}(p_{0},|\vec{p}|)  . \label{cpmvp}
\end{align}
\end{subequations}

\item For $p_0 \rightarrow -p_0$: 
\begin{subequations}
\begin{align}
a(-p_{0},|\vec{p}|) &= a(p_{0},|\vec{p}|), \label{amp0} \\
b(-p_{0},|\vec{p}|) &= -b(p_{0},|\vec{p}|), \label{bmp0} \\
b^{\prime}(-p_{0},p_{\perp},p_{z}) &= b^{\prime}(p_{0},p_{\perp},p_{z}), \label{bpmp0} \\
c^{\prime}(-p_{0},|\vec{p}|) &= -c^{\prime}(p_{0},|\vec{p}|) . \label{cpmp0} 
\end{align} 
\end{subequations}

\item For $P \rightarrow -P =(-p_0,-{\vec p})$: 
\begin{subequations}
\begin{align}
a(-p_{0},|-\vec{p}|) &= a(p_{0},|\vec{p}|), \label{amp} \\
b(-p_{0},|-\vec{p}|) &= -b(p_{0},|\vec{p}|),  \label{bmp} \\
b^{\prime}(-p_{0},p_{\perp},-p_{z}) &= -b^{\prime}(p_{0},p_{\perp},p_{z}), \label{bpmp} \\
c^{\prime}(-p_{0},|-\vec{p}|) &= -c^{\prime}(p_{0},|\vec{p}|) .  \label{cpmp} 
\end{align}
\end{subequations}
We have used the fact that $Q_{0}(-x) = -Q_{0}(x)$ and $Q_{1}(-x) = Q_{1}(x)$.
\end{enumerate}

Now based on the above we also note down the transformation properties of those quantities appearing in the 
propagator: .
\begin{enumerate}  
\item For $\mathcal A$:   
\begin{subequations}
 \begin{align}
\mathcal{A}(p_{0},p_{\perp},p_{z}) &\xrightarrow[]{\text{$\vec{p}\rightarrow -\vec{p}$}} \mathcal{A}(p_{0},p_{\perp},p_{z}),  \label{calamvp} \\
\mathcal{A}(p_{0},p_{\perp},p_{z}) &\xrightarrow[]{\text{$p_{0}\rightarrow -p_{0}$}} \mathcal{A}(p_{0},p_{\perp},p_{z}) ,  \label{calamvp} \\
\mathcal{A}(p_{0},p_{\perp},p_{z}) &\xrightarrow[\text{$\vec{p}\rightarrow -\vec{p}$}]{\text{$p_{0}\rightarrow -p_{0}$}} \mathcal{A}(p_{0},p_{\perp},p_{z}).
 \label{calcpmp} 
\end{align}
\end{subequations} 

\item   For ${\mathcal B}_\pm$: 
\begin{subequations}
\begin{align}
\mathcal{B}_{\pm}(p_{0},p_{\perp},p_{z}) &\xrightarrow[]{\text{$\vec{p}\rightarrow -\vec{p}$}} \mathcal{B}_{\mp}(p_{0},p_{\perp},p_{z}), \label{calbmvp} \\
\mathcal{B}_{\pm}(p_{0},p_{\perp},p_{z}) &\xrightarrow[]{\text{$p_{0}\rightarrow -p_{0}$}} -\mathcal{B}_{\mp}(p_{0},p_{\perp},p_{z}), \label{calbmp0} \\
\mathcal{B}_{\pm}(p_{0}, p_{\perp},p_{z}) &\xrightarrow[\text{$\vec{p}\rightarrow -\vec{p}$}]{\text{$p_{0}\rightarrow -p_{0}$}} -\mathcal{B}_{\pm}(p_{0},p_{\perp},p_{z}).
\label{calbmp} 
\end{align}
\end{subequations}
\end{enumerate}

Using the above  transformation properties,  it can be shown that 
$\slashed{L}, \, \slashed{R}, \, L^2$ and $R^2$, respectively given in  \eqref{l_mu},  \eqref{r_mu}, \eqref{l2} and \eqref{r2}   transform as
\begin{subequations}
\begin{align} 
 \slashed{L}(p_{0},p_{\perp},p_{z}) &\xrightarrow[]{\text{$\vec{p}\rightarrow -\vec{p}$}} \mathcal{A}(p_{0},|\vec{p}|)(p_{0}\gamma\indices{^0}+\vec{p}\cdot\vec{\gamma})
 +\mathcal{B}_{-}(p_{0},p_{\perp},p_{z})\slashed{u}+c^{\prime}(p_{0},|\vec{p}|)\slashed{n} \, , \label{l_mp}\\
\slashed{R}(p_{0},p_{\perp},p_{z}) &\xrightarrow[]{\text{$\vec{p}\rightarrow -\vec{p}$}} \mathcal{A}(p_{0},|\vec{p}|)(p_{0}\gamma\indices{^0}+\vec{p}\cdot\vec{\gamma})
+\mathcal{B}_{+}(p_{0},p_{\perp},p_{z})\slashed{u}-c^{\prime}(p_{0},|\vec{p}|)\slashed{n}\, ,  \label{r_mp} \\
L^{2}(p_{0},p_{\perp},p_{z}) &\xrightarrow[]{\text{$\vec{p}\rightarrow -\vec{p}$}}R^{2}(p_{0},p_{\perp},p_{z})\,  , \label{l2_mp} \\
R^{2}(p_{0},p_{\perp},p_{z}) &\xrightarrow[]{\text{$\vec{p}\rightarrow -\vec{p}$}}L^{2}(p_{0},p_{\perp},p_{z})\,  , \label{r2_mp} 
\end{align}
\end{subequations}
and 
\begin{subequations}
\begin{align} 
\slashed{L}(p_{0},p_{\perp},p_{z}) &\xrightarrow[\text{$\vec{p}\rightarrow -\vec{p}$}]{\text{$p_{0}\rightarrow -p_{0}$}}-\slashed{L}(p_{0},p_{\perp},p_{z}) , \label{l_mp_p_mp} \\
\slashed{R}(p_{0},p_{\perp}, p_{z}) &\xrightarrow[\text{$\vec{p}\rightarrow -\vec{p}$}]{\text{$p_{0}\rightarrow -p_{0}$}} -\slashed{R}(p_{0},p_{\perp},p_{z}) , \label{r_mp0_mp} \\
L^{2}(p_{0},p_{\perp}, p_{z})& \xrightarrow[\text{$\vec{p}\rightarrow -\vec{p}$}]{\text{$p_{0}\rightarrow -p_{0}$}} L^{2}(p_{0},p_{\perp},p_{z}), \label{l2_mp0_mp} \\
R^{2}(p_{0},p_{\perp},p_{z}) &\xrightarrow[\text{$\vec{p}\rightarrow -\vec{p}$}]{\text{$p_{0}\rightarrow -p_{0}$}}  R^{2}(p_{0},p_{\perp},p_{z}) . \label{r2_mp0_mp}
\end{align}
\end{subequations}
Now we are in a position to check the transformation properties of the effective propagator under some of the discrete symmetries:

\subsubsection{Chirality}
\label{chirality_trans}
Under chirality the fermion propagator transform as~\cite{Weldon:1999th}
\begin{equation}
S(p_{0},\vec{p}) \longrightarrow  -\gamma_{5}\,S(p_{0},\vec{p})\,\gamma_{5} . \label{chi_def}
\end{equation}

The effective propagator, $S^*(p_{0},p_{\perp},p_{z})$,  in  \eqref{eff_prop1} transforms  under chirality as 
\begin{align}
-\gamma_{5}\,S^*(p_{0},p_{\perp},p_{z})\,\gamma_{5} &= -\gamma_{5}\mathcal{P}_{-}\frac{\slashed{L}(p_{0},p_{\perp},p_{z})}{L^{2}(p_{0},p_{\perp},p_{z})}\mathcal{P}_{+}\gamma_{5}-\gamma_{5}\mathcal{P}_{+}\frac{\slashed{R}(p_{0},p_{\perp},p_{z})}{R^{2}(p_{0},p_{\perp},p_{z})}\mathcal{P}_{-}\gamma_{5} \nonumber \\
&= \mathcal{P}_{+}\frac{\slashed{L}(p_{0},p_{\perp},p_{z})}{L^{2}(p_{0},p_{\perp},p_{z})}\mathcal{P}_{+} + \mathcal{P}_{-}\frac{\slashed{R}(p_{0},p_{\perp},p_{z})}{R^{2}(p_{0},p_{\perp},p_{z})}\mathcal{P}_{-} \nn  \\
&= S^*(p_{0},p_{\perp},p_{z}), \label{chi_inv}
\end{align}
which satisfies \eqref{chi_def} and indicates  that it   is chirally invariant.

\subsubsection{Reflection} 
\label{ref_sym}
Under reflection the fermion propagator transforms~\cite{Weldon:1999th}
as
\begin{equation}
S(p_{0},\vec{p}) \longrightarrow  S(p_{0},- \vec{p}) . \label{ref_def}
\end{equation}

The effective propagator, $S^*(p_{0},p_{\perp},p_{z})$,  in  \eqref{eff_prop1} transforms  under reflection as 
\begin{align}
S^*(p_{0},p_{\perp},-p_{z})&= \mathcal{P}_{-}\frac{\slashed{L}(p_{0},p_{\perp},-p_{z})}{L^{2}(p_{0},p_{\perp},-p_{z})}\mathcal{P}_{+}
 + \mathcal{P}_{+}\frac{\slashed{R}(p_{0},p_{\perp},-p_{z})}{R^{2}(p_{0},p_{\perp},-p_{z})}\mathcal{P}_{-} \nonumber \\
&= \mathcal{P}_{-}\frac{\mathcal{A}(p_{0},|\vec{p}|)(p_{0}\gamma\indices{^0}+\vec{p}\cdot\vec{\gamma})
+\mathcal{B}_{-}(p_{0},p_{\perp},p_{z})\slashed{u}+c^{\prime}(p_{0},|\vec{p}|)\slashed{n}}{R^{2}(p_{0},p_{\perp},p_{z})}\mathcal{P}_{+} \nonumber \\
&\hspace{0.35cm} +\mathcal{P}_{+}\frac{\mathcal{A}(p_{0},|\vec{p}|)(p_{0}\gamma\indices{^0}+\vec{p}\cdot\vec{\gamma})
+\mathcal{B}_{+}(p_{0},p_{\perp},p_{z})\slashed{u}-c^{\prime}(p_{0},|\vec{p}|)\slashed{n}}{L^{2}(p_{0},p_{\perp},p_{z})}\mathcal{P}_{-} \nonumber \\
& \ne S^*(p_{0},p_{\perp},p_{z}) . \label{ref_gen_ninv}
\end{align}

However, now considering  the rest frame of the heat bath,  \(u\indices{^\mu} = (1,0,0,0)\), 
and the background magnetic field along $z$-direction, \(n\indices{^\mu} = (0,0,0,1)\), 
one can write \eqref{ref_gen_ninv}   as
\begin{align}
S^*(p_{0},p_{\perp},-p_{z})&= \mathcal{P}_{-}\frac{\mathcal{A}(p_{0},|\vec{p}|)(p_{0}\gamma\indices{^0}+\vec{p}\cdot\vec{\gamma})
+\mathcal{B}_{-}(p_{0},p_{\perp},p_{z})\gamma_0-c^{\prime}(p_{0},|\vec{p}|)\gamma^3}{R^{2}(p_{0},p_{\perp},p_{z})}\mathcal{P}_{+} \nonumber \\
&\hspace{0.35cm} +\mathcal{P}_{+}\frac{\mathcal{A}(p_{0},|\vec{p}|)(p_{0}\gamma\indices{^0}+\vec{p}\cdot\vec{\gamma})
+\mathcal{B}_{+}(p_{0},p_{\perp},p_{z})\gamma_0+c^{\prime}(p_{0},|\vec{p}|)\gamma^3}{L^{2}(p_{0},p_{\perp},p_{z})}\mathcal{P}_{-} \nonumber \\
& \ne S^*(p_{0},p_{\perp},p_{z}) . \label{ref_rest_ninv}
\end{align}
As seen in both cases the reflection symmetry is violated as we will see later while discussing the dispersion property of a fermion.

\subsubsection{Parity}
\label{parity_trans}
Under parity  a fermion propagator  transforms~\cite{Weldon:1999th} as 
\begin{equation}
S(p_{0},\vec{p}) \longrightarrow \gamma\indices{_0}\,S(p_{0},-\vec{p})\,\gamma\indices{_0} . \label{parity_def}
\end{equation}
The effective propagator, $S^*(p_{0},p_{\perp},p_{z})$,  in  \eqref{eff_prop1} under parity  transforms as 
\begin{align}
\gamma\indices{_0}\,S^*(p_{0},p_{\perp},-p_{z})\,\gamma\indices{_0} &= \gamma\indices{_0}\mathcal{P}_{-}
\frac{\slashed{L}(p_{0},p_{\perp},-p_{z})}{L^{2}(p_{0},p_{\perp},-p_{z})}\mathcal{P}_{+}\gamma\indices{_0} 
+ \gamma\indices{_0}\mathcal{P}_{+}\frac{\slashed{R}(p_{0},p_{\perp},-p_{z})}{R^{2}(p_{0},p_{\perp},-p_{z})}\mathcal{P}_{-}\gamma\indices{_0} \nonumber \\
&= \mathcal{P}_{+}\gamma\indices{_0}\frac{\slashed{L}(p_{0},p_{\perp},-p_{z})}{R^{2}(p_{0},p_{\perp},p_{z})}\gamma\indices{_0}\mathcal{P}_{-} + \mathcal{P}_{-}\gamma\indices{_0}\frac{\slashed{R}(p_{0},p_{\perp},-p_{z})}{L^{2}(p_{0},p_{\perp},p_{z})}\gamma\indices{_0}\mathcal{P}_{+} \nonumber \\
&\ne  S^*(p_{0},p_{\perp},p_{z}) \, , \label{parity_ninv}
\end{align}
which  does not obey \eqref{parity_def}, indicating that the effective propagator in general frame of reference  is not parity invariant due to the background medium.

However, now considering  the rest frame of the heat bath,  \(u\indices{^\mu} = (1,0,0,0)\), 
and the background magnetic field along $z$-direction, \(n\indices{^\mu} = (0,0,0,1)\), 
one can write \eqref{parity_ninv} by using \eqref{l_mp},  \eqref{r_mp} and 
\( \gamma\indices{_0}\,\gamma\indices{^i} = -\gamma\indices{^i}\,\gamma\indices{_0}\) as
\begin{align}
\gamma\indices{_0}\,S^*(p_{0},p_{\perp},-p_{z})\,\gamma\indices{_0} 
&= \mathcal{P}_{+}\frac{\slashed{R}(p_{0},p_{\perp},p_{z})}{R^{2}(p_{0},p_{\perp},p_{z})}\mathcal{P}_{-}+\mathcal{P}_{-}\frac{\slashed{L}(p_{0},p_{\perp},p_{z})}{L^{2}(p_{0},p_{\perp},p_{z})}\mathcal{P}_{+} \nonumber \\
&= S^*(p_{0},p_{\perp},p_{z}),
\end{align}
which indicates that the propagator is parity invariant in the rest frame of the magnetised heat bath. We note that other discrete symmetries 
can also be checked  but leave them on the readers. 
\subsection{Modified Dirac Equation}       
\label{dirac_mod}
\subsubsection{For General Case }
\label{hll_modes}
The  effective propagator that satisfy the modified Dirac equation with spinor $U$ is given by 
\begin{align}
\left(\mathcal{P}_{+}\,\slashed{L}\,\mathcal{P}_{-}+\mathcal{P}_{-}\,\slashed{R}\,\mathcal{P}_{+}\right)U &= 0 . \label{moddireqn}
\end{align} 
Using the chiral basis 
\begin{align}
\gamma_{0} = \begin{pmatrix}
0 && \mathbbm{1} \\
\mathbbm{1} && 0
\end{pmatrix},\hspace{1cm}\vec{\gamma} = \begin{pmatrix}
0 && \vec{\sigma} \\
-\vec{\sigma} && 0
\end{pmatrix},\hspace{1cm}\gamma_{5} = \begin{pmatrix}
-\mathbbm{1} && 0 \\
0 && \mathbbm{1}
\end{pmatrix}, \hspace{1cm} U = \begin{pmatrix}
\psi_{L} \\
\psi_{R}
\end{pmatrix}\, , \label{chi_basis}
\end{align}
one can write  \eqref{moddireqn}  as
\begin{align}
\begin{pmatrix}
0 && \sigma \cdot R \\
\bar{\sigma}\cdot L && 0
\end{pmatrix}\begin{pmatrix}
\psi_{L}\\
\psi_{R}
\end{pmatrix} = 0\,,
\end{align}
where $\psi_{R}$ and  $\psi_{L}$ are two component Dirac spinors with \(\sigma\equiv (1,\vec{\sigma})\) 
and \( {\bar \sigma} \equiv (1,-\vec{\sigma})\), respectively.  One can obtain nontrivial  solutions  with the condition 
\begin{align}
\mbox{det}\begin{pmatrix}
0 && \sigma \cdot R \nonumber\\
\bar{\sigma}\cdot L && 0
\end{pmatrix} &= 0 \nonumber \\
\mbox{det}[L\cdot \bar{\sigma}]\,\mbox{det}[R\cdot {\sigma}] &= 0 \nonumber \\
L^{2}R^{2} &= 0  \, . \label{det0}
\end{align}
We note that  for a given $p_0\ (=\omega)$, either $L^{2}=0$, or $R^{2}=0$, but not both of them are simultaneously zero.
This implies that i) when \(L^{2}=0\),   \(\psi_{R}=0\) ; ii) when \(R^{2}=0\),  \(\psi_{L}=0\). These dispersion conditions are same as 
obtained from the poles of the effective propagator in \eqref{eff_prop1}  as obtained in subsec.~\ref{eff_fer_prop}.
\begin{enumerate}
\item For  \(R^{2}=0\) but \(L^{2}\neq 0\), the right chiral equation is given by 
		\begin{align}
		\left ( R\cdot \sigma\right )\,\psi_{R}=0 . \label{chiral_rt}
		\end{align}
		Again \(R^{2}=0\) \,  $\Rightarrow$ \, \( R_{0}=\pm |\vec{R}| = \pm \sqrt{R^{2}_{x}+R^{2}_{y}+R^{2}_{z}} \) and the 
		corresponding dispersive modes are denoted by $R^{(\pm)}$. So the solutions of \eqref{chiral_rt} are 
		\begin{subequations}		
		\begin{align}
		{\mbox {(i)}} \, \, R_{0}&= |\vec{R}|; \hspace{0.7 cm} {\mbox{mode}} \, \,  R^{(+)};  \hspace{0.7cm}
		U_{R^{(+)}} = \sqrt{\frac{|\vec{R}|+R_{z}}{2|\vec{R}|}}\begin{pmatrix}
		0\\
		0\\
		1 \\
		\frac{R_{x}+iR_{y}}{|\vec{R}|+R_{z}} 
	 	\end{pmatrix}\, 
		= \begin{pmatrix}
		0 \\
		\, \, \, \, \,  \psi_R^{(+)} 
		\end{pmatrix} \, ,
		 \label{r0+}\\ 
		{\mbox {(ii)}} \, \, R_{0}&= -|\vec{R}|; \hspace{0.4 cm} {\mbox{mode}} \, \,  R^{(-)};  \hspace{0.4cm}
		U_{R^{(-)}} = -\sqrt{\frac{|\vec{R}|+R_{z}}{2|\vec{R}|}}\begin{pmatrix}
		0\\
		0\\
		\frac{R_{x}-iR_{y}}{|\vec{R}|+R_{z}} \, . \\
		-1
		\end{pmatrix} \, 
		= \begin{pmatrix}
		0 \\
		\,\, \, \, \, \psi_R^{(-)} 
		\end{pmatrix} \,
		. \label{r0-}
		\end{align}
		\end{subequations}		
 \item  For \(L^{2}=0\) but \(R^{2}\neq 0\),  the left chiral equation is given by 
		\begin{align}
		(L \cdot \bar{\sigma}) \,\psi_{L}=0 , \label{chiral_lt}
		\end{align}
               where \(L^{2}=0\) implies two conditions; 
\( L_{0}=\pm |\vec{L}| = \pm \sqrt{L^{2}_{x}+L^{2}_{y}+L^{2}_{z}}\) and the 
		corresponding dispersive modes are denoted by $L^{(\pm)}$.  The two solutions of \eqref{chiral_lt} are  obtained as 
		\begin{subequations}
		\begin{align}
		{\mbox {(i)}} \,\, L_{0}= |\vec{L}|; \hspace{0.7 cm} {\mbox{mode}} \, \,  L^{(+)};  \hspace{0.7cm}
		 U_{L^{(+)}} = -\sqrt{\frac{|\vec{L}|+L_{z}}{2|\vec{L}|}}\begin{pmatrix}
		\frac{L_{x}-iL_{y}}{|\vec{L}|+L_{z}} \\
		-1 \\
		0\\
		0
		\end{pmatrix} \, 
		= \begin{pmatrix}
		\,\, \, \, \,  \psi_L^{(+)} \\
		0
		\end{pmatrix} \, ,
		  \label{l0+}\\
		  {\mbox {(i)}} \,\, L_{0}= -|\vec{L}|;\hspace{0.7 cm} {\mbox{mode}} \, \,  L^{(-)};  \hspace{0.7cm}
		  		U_{L^{(-)}} = \sqrt{\frac{|\vec{L}|+L_{z}}{2|\vec{L}|}}\begin{pmatrix}
		1 \\
		\frac{L_{x}+iL_{y}}{|\vec{L}|+L_{z}} \\
		0\\
		0
		\end{pmatrix}\, 
		= \begin{pmatrix}
		\,\, \, \, \, \psi_L^{(-)} \\
		0
		\end{pmatrix} \, .  \label{l0-}
		\end{align}
        	\end{subequations}
\end{enumerate}
We note here that $\psi^{(\pm)}_L$ and $\psi^{(\pm)}_R$ are only chiral eigenstates but neither  the spin nor the helicity  eigenstates. 	
\subsubsection{For lowest Landau level (LLL)}
\label{lll_modes}	
\begin{enumerate}	
\item For  $R_{LLL}^2=0$  in \eqref{defRsquare_r0}  indicates that  $R_0=\pm R_z, \, R_x=R_y=0$.  The two solutions obtained, respectively, 
in  \eqref{ll_sp3} and \eqref{ll_sp4} in Appendix~\ref{lll_app}  are given as 
		\begin{subequations}		
		\begin{align}
		{\mbox {(i)}} \, \, R_{0}&= R_z; \hspace{0.7 cm} {\mbox{mode}} \, \,  R^{(+)};  \hspace{0.7cm}
		U_{R^{(+)}} = \begin{pmatrix}
		0\\
		0\\
		1 \\
		0
	 	\end{pmatrix} 
		=\begin{pmatrix}
		0\\
		\chi_+
		\end{pmatrix} \ . 
		\, \label{llr0+}\\
		{\mbox {(ii)}} \, \, R_{0}&= -R_z; \hspace{0.7 cm} {\mbox{mode}} \, \,  R^{(-)};  \hspace{0.7cm}
		U_{R^{(-)}} = \begin{pmatrix}
		0\\
		0\\
		0 \\
		1
		\end{pmatrix} \, 
		=\begin{pmatrix}
		0\\
		\chi_-
		\end{pmatrix} \ , \label{llr0+}
		\end{align}
		\end{subequations}
where \(\displaystyle \chi_{+} = \begin{pmatrix} 1 \\ 0 \end{pmatrix}\) and \(\displaystyle \chi_{-} = \begin{pmatrix} 0 \\ 1 \end{pmatrix}\).

\item  For LLL,   $L_{LLL}^2=0$  in \eqref{defLsquare_l0}  indicates that  $L_0=\pm L_z, \, L_x=L_y=0$.  The two solutions obtained, respectively, 
in  \eqref{ll_sp5} and \eqref{ll_sp6} in Appendix~\ref{lll_app}  are given as 
\begin{subequations}
		\begin{align}
		{\mbox {(i)}} \, \, L_{0}&= L_z; \hspace{0.7 cm} {\mbox{mode}} \, \,  L^{(+)};  \hspace{0.7cm}
		U_{L^{(+)}} = \begin{pmatrix}
		0 \\
		1 \\
		0\\
		0
		\end{pmatrix} 
		=\begin{pmatrix}
		\chi_-\\
		0
		\end{pmatrix} 
		\, ,  \label{lll0+}\\
		{\mbox {(i)}} \, \, L_{0}&= -L_z; \hspace{0.7 cm} {\mbox{mode}} \, \,  L^{(-)};  \hspace{0.7cm}
		U_{L^{(-)}} = \begin{pmatrix}
		1 \\
		0 \\
		0\\
		0
		\end{pmatrix}\, 
		=\begin{pmatrix}
		\chi_+\\
		0
		\end{pmatrix} \,
		.  \label{lll0-}
		\end{align}
        	\end{subequations}
\end{enumerate}
The spin operator along the $z$ direction is given by 
\be
\Sigma^{3} = \mathcal{\sigma}\indices{^1^2}=\frac{i}{2}\left [\gamma\indices{^1},\gamma\indices{^2}\right ]=i\,\gamma\indices{^1}\gamma\indices{^2}
=\begin{pmatrix} \sigma\indices{^3} && 0 \\ 0 && \sigma\indices{^3}\end{pmatrix},
\ee
where $\sigma$ with single index denotes Pauli spin matrices whereas that with  double indices denote 
generator of Lorentz group in spinor representation.
Now,
\begin{align}
\Sigma\indices{^3}\,\,{U}_{R^{(\pm)}} &= \begin{pmatrix} \sigma\indices{^3} && 0 \\ 0 && \sigma\indices{^3} \end{pmatrix}\,
\begin{pmatrix} 0 \\ \chi_{\pm} \end{pmatrix}=\begin{pmatrix} 0 \\ \sigma\indices{^3}\,\chi_{\pm}\end{pmatrix} = 
\pm\,\begin{pmatrix} 0 \\ \chi_{\pm} \end{pmatrix} = \pm\, {U}_{R^{(\pm)}} , \label{lll_spin_sol} \\
\Sigma\indices{^3}\,\,{U}_{L^{(\pm)}} &= \begin{pmatrix} \sigma\indices{^3} && 0 \\ 0 && \sigma\indices{^3} \end{pmatrix}\,
\begin{pmatrix} \chi_{\mp} \\ 0 \end{pmatrix} = \begin{pmatrix} \sigma\indices{^3}\,\chi_{\mp} \\ 0 \end{pmatrix} = 
\mp \begin{pmatrix} \chi_{\mp} \\ 0 \end{pmatrix} = \mp\,{U}_{L^{(\pm)}} . \label{llr_spin_sol}
\end{align}
So, the modes $L^{(-)}$ and $R^{(+)}$ have spins along the direction of magnetic field whereas  $L^{(+)}$ and $R^{(-)}$ have spins opposite to the direction of magnetic field.
Now we discuss the helicity eigenstates of the various modes in LLL.
The helicity operator is defined as 
\begin{align}
\mathcal{H}_{\vec{p}} = \mathbf{\hat{p}}\cdot\vec{\Sigma} \, .
\end{align}
When a particle moves along $+z$ direction, $\mathbf{\hat{p}}=\mathbf{\hat{z}}$ and when it 
moves along $-z$ direction, $\mathbf{\hat{p}}=-\mathbf{\hat{z}}$. \\
Thus 
\begin{align}
\mathcal{H}_{\vec{p}} = \begin{cases}\,\,\,\,  \Sigma\indices{^3},\qquad\text{for}\qquad p_{z}>0 ,
\\ -\Sigma\indices{^3},\qquad\text{for}\qquad p_{z}<0 .\end{cases}
\end{align}
Thus,
\begin{align}
\mathcal{H}_{\vec{p}}\,\,{\displaystyle U}_{R^{(\pm)}}= \begin{cases} \pm {\displaystyle U}_{R^{(\pm)}}, \qquad\text{for}\qquad p_{z}>0 ,\\
 \mp {\displaystyle U}_{R^{(\pm)}}, \qquad\text{for}\qquad p_{z}<0 . \end{cases}
\end{align}
and
\begin{align}
\mathcal{H}_{\vec{p}}\,\,{\displaystyle U}_{L^{(\pm)}}= \begin{cases} \mp {\displaystyle U}_{L^{(\pm)}}, \qquad\text{for}\qquad p_{z}>0 \, , \\ 
\pm {\displaystyle U}_{L^{(\pm)}}, \qquad\text{for}\qquad p_{z}<0 \, . \end{cases}
\end{align}
\subsection{Dispersion}
\label{disp_rep}
\begin{figure}[H]
\centering
\includegraphics[width=0.43\textwidth]{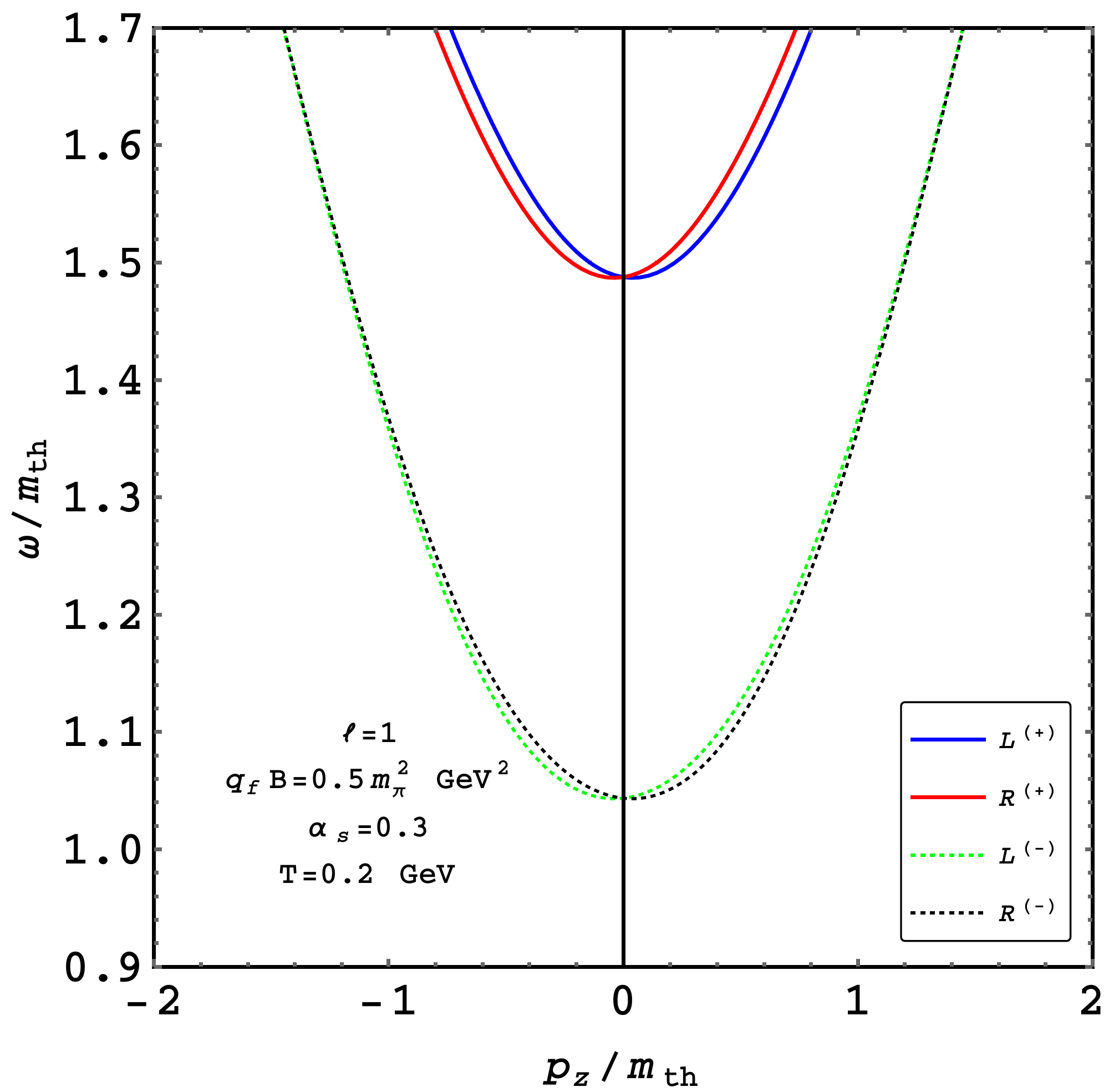}
\includegraphics[width=0.43\textwidth]{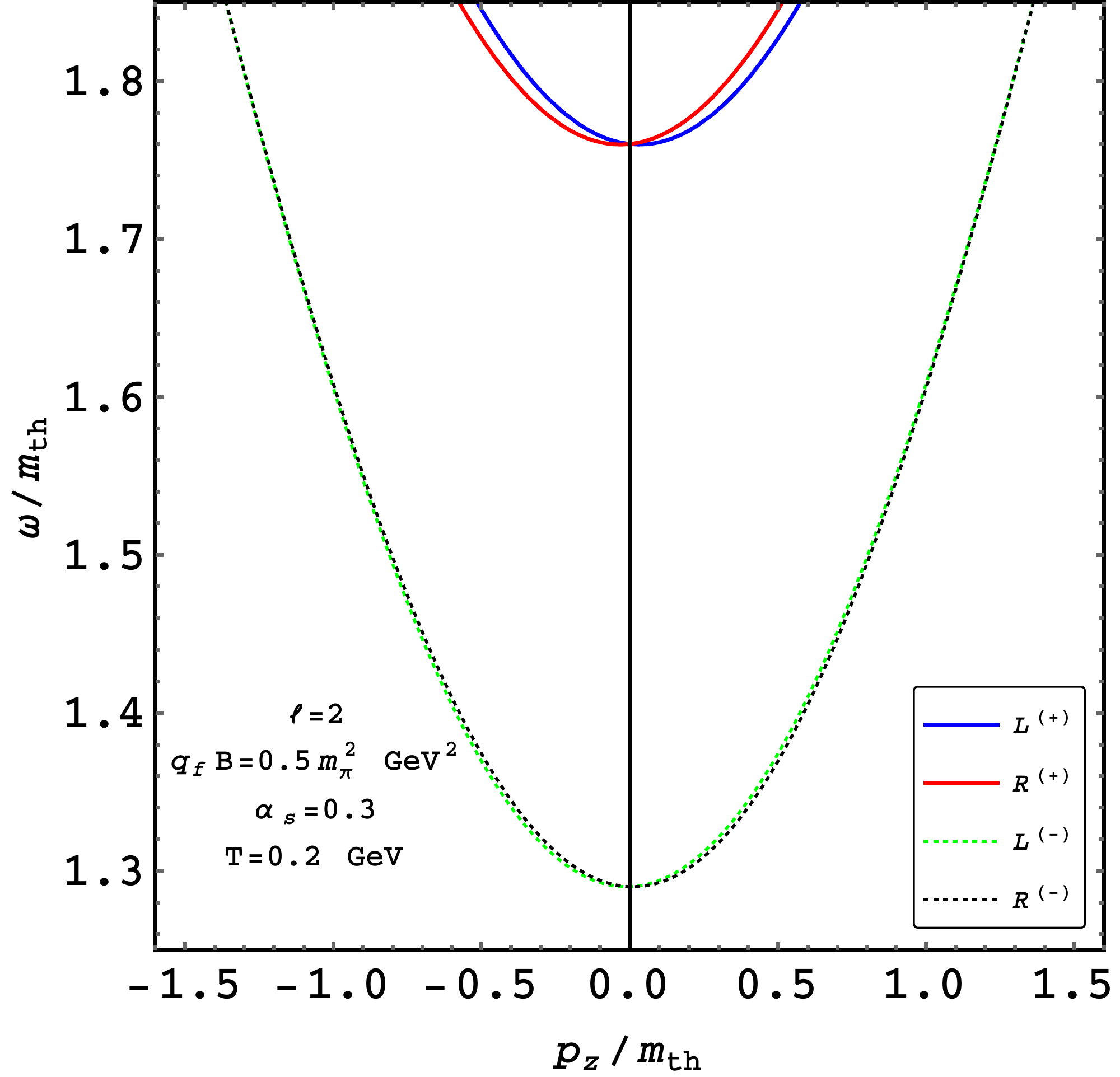}
\caption{\small Dispersion plots for  higher Landau level, $l\ne 0$. The energy  $\omega$
is scaled with the thermal mass $m_{th}$ for convenience}
\label{fig:HLLfig}
\end{figure}
In presence of magnetic field, the component of momentum transverse to the magnetic field is Landau quantised and takes discrete 	values given by \( \displaystyle p^{2}_{\perp} = 2 l |q_{f} B|\), where \( l \) is a given Landau levels.   In presence of pure background magnetic 
field and no heat bath ($T=0$), the Dirac equation gives rise a dispersion relation
\begin{align}
E^{2}=p^{2}_{z} + m_{f}^{2} + (2\,\nu + 1)\,q_{f}\,|Q|B-q_{f}\,Q\,B\,\sigma  \, , \label{disp_B}
\end{align}
where  $\nu = 0,1,2,\cdots$, $Q=\pm 1$, $\sigma=+1$ for spin up  and $\sigma = -1$ for spin down.
The solutions are classified by energy eigenvalues 
\begin{align}
E_{l}^{2} = p^{2}_{z} + m_{f}^{2} + 2\,l \,q_{f}\,B \ . \label{disp_pure_m}
\end{align} 
where one can define 
\begin{equation}
2\,l = (2\,\nu + 1)\,|Q| - Q\, \sigma \, .
\end{equation}	
Now we discuss the dispersion properties of a fermions in a hot magnetised medium.  For general case (for higher LLs, $l\ne 0$) the dispersion 
curves  obtained by solving, $L^2=0$ and $R^2=0$ given in \eqref{l2} and \eqref{r2},  numerically. 
We note that the roots of $L_0=\pm |\vec L | \, \Rightarrow \, L_0 \mp |\vec L|=0 $ are represented by $L^{(\pm)}$  with energy 
$\omega_{L^{(\pm)}}$ whereas  those  for 
$R_0=\pm |\vec R | \, \Rightarrow \, R_0 \mp |\vec R |=0 $ by $R^{(\pm)}$ with energy 
$\omega_{R^{(\pm)}}$ . The corresponding eigenstates are obtained in \eqref{l0+}, \eqref{l0-},
\eqref{r0+} and \eqref{r0-} in subsection~\ref{hll_modes}.  We have chosen $T=0.2$ GeV, $\alpha_s=0.3$ and $q_fB=0.5 m_\pi^2$, where 
$m_\pi$ is the pion mass. In Fig.~\ref{fig:HLLfig} the  dispersion curves for higher Landau levels are shown  where  all four  modes can 
propagate for a given choice of $Q$. This is because the corresponding states for these modes are neither spin nor helicity eigenstates 
as shown in subsec.~\ref{hll_modes}. We also note that  there will be negative energy modes which are not displayed here but would be discussed  
in the analysis of the spectral representation of the effective propagator  section~\ref{spec_rep}.
 \begin{figure}
\includegraphics[width=0.43\textwidth]{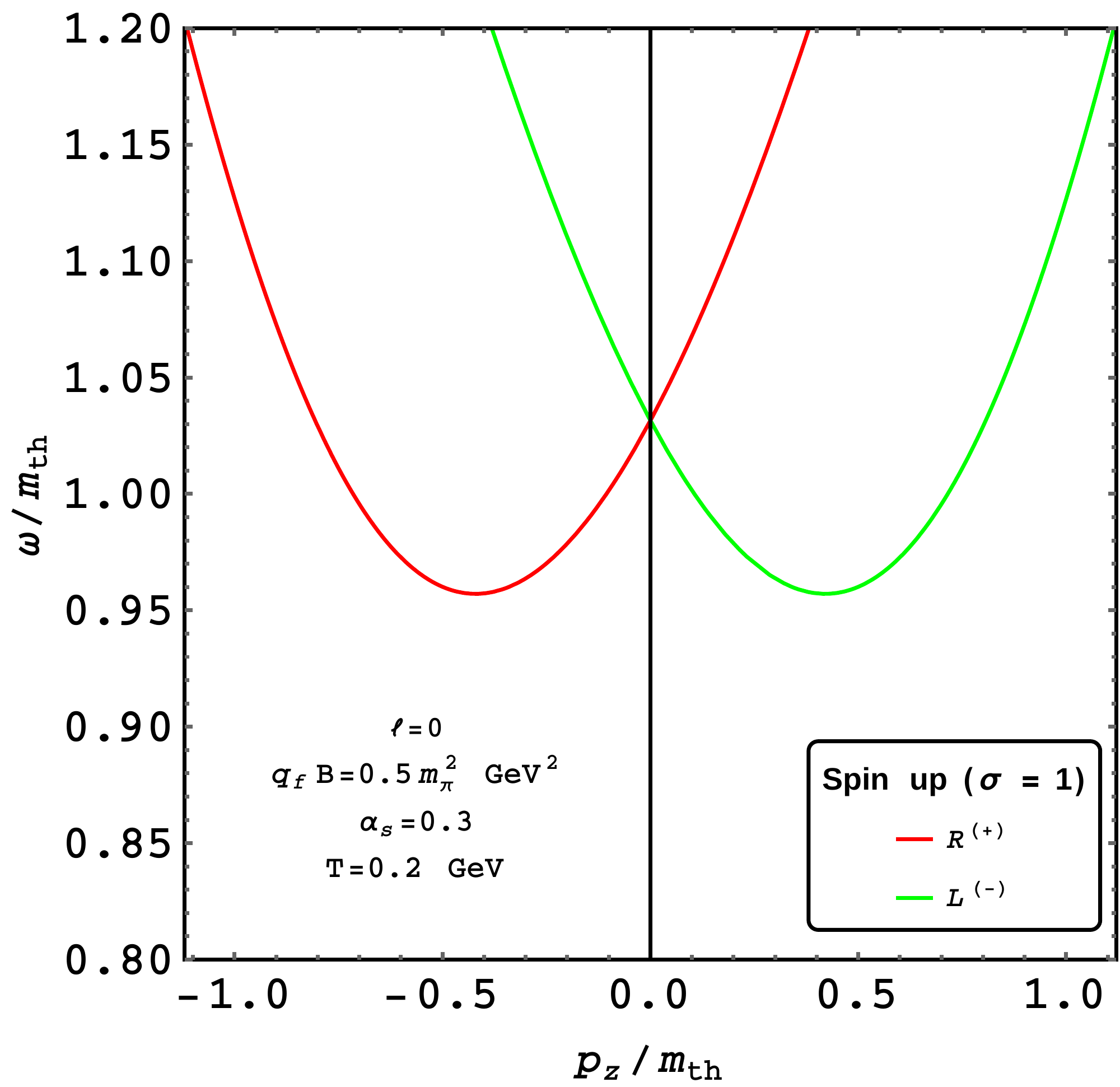}
\includegraphics[width=0.43\textwidth]{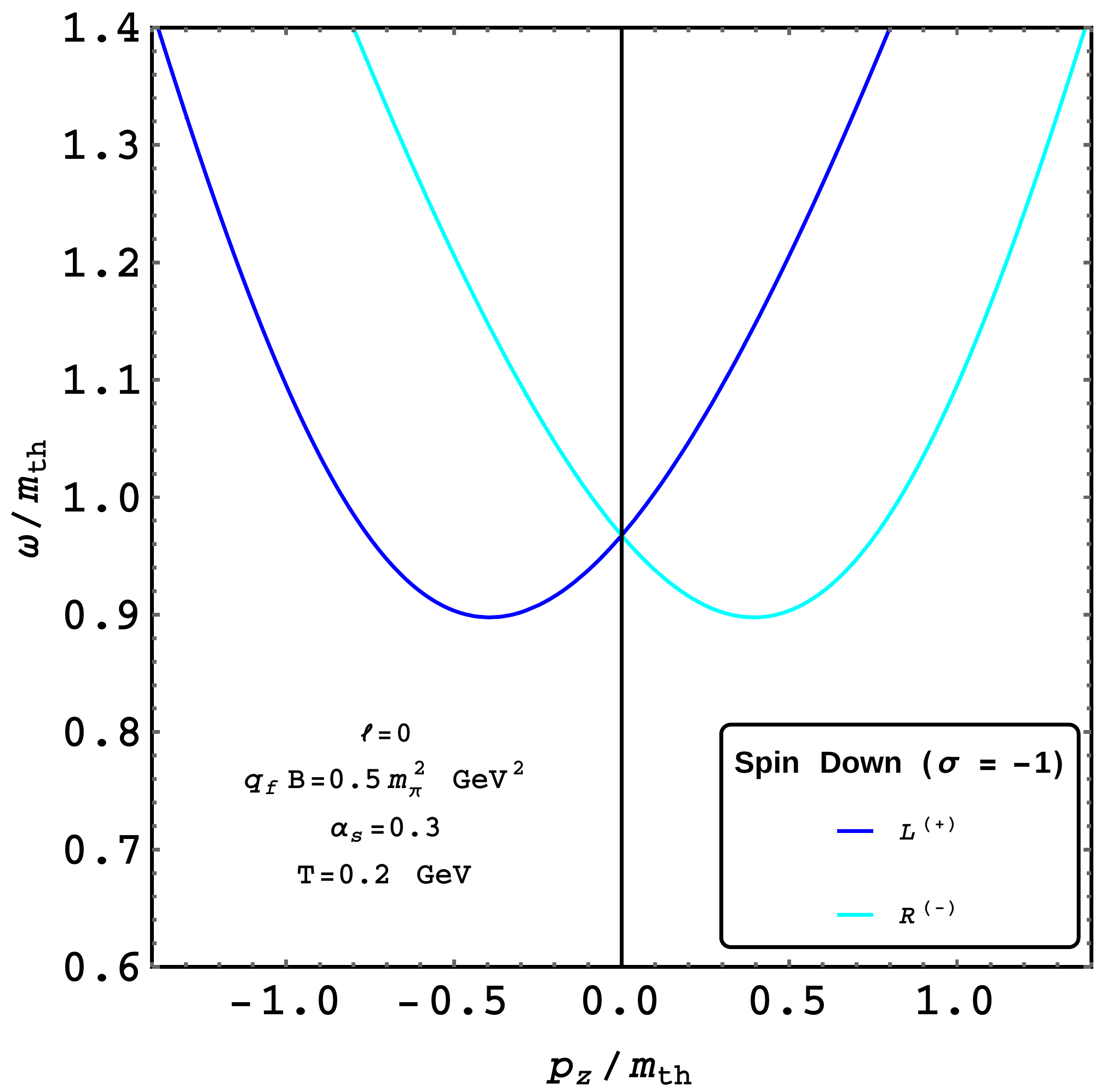}
\caption{\small Dispersion plots  for  LLL, \, $l=0$. The energy  $\omega$ is scaled with the thermal mass $m_{th}$ for convenience. For details see the text.}
\label{fig:lll_disp}
\end{figure}

At LLL  $l=0 \, \rightarrow  \, p_{\perp}=0$ and the roots of $ R_0=\pm R_z $  give rise to two right handed modes $R^{(\pm)}$  with energy 
$\omega_{R^{(\pm)}}$ whereas those for  $L_0=\pm L_z $ produce~\footnote{We make a general note here for left handed modes at LLL.  
 At small $p_z$,  $L_z$ itself is negative for LLL and becomes positive after a moderate value of $p_z$.  This makes  the left handed modes 
 $L^{(+)}$ and $L^{(-)}$ to flip in LLL than those in higher Landau levels. For details see Appendix~\ref{eff_lll_mass}.} two  left handed 
 modes  $L^{(\pm)}$ with energy 
$\omega_{L^{(\pm)}}$  . 
In Appendix~\ref{eff_lll_mass}  the analytic solutions for the dispersion relations in LLL  are presented which show four different modes and 
the corresponding eigenstates are obtained in subsec.~\ref{lll_modes}. 
 Now at LLL we discuss  two possibilities below:
\begin{enumerate}
\item[(i)] for positively charged fermion $Q=1$, $\sigma=1$ implies $\nu = 0$ and $\sigma=-1$ implies $\nu=-1$.  Now we note that  $\nu$ can never be negative. 
This implies that  the modes with  $Q=1$ and  $\sigma=-1$ (spin down) cannot propagate in LLL. Now, the right handed mode $R^{(+)}$ and the 
left handed mode  $L^{(-)}$ have  
spin up as shown 
in subsec.~\ref{lll_modes}, will propagate in LLL for $p_z>0$. The $R^{(+)}$ mode has helicity to chirality ratio $+1$ is a quasiparticle 
whereas the mode $L^{(-)}$ left handed has that of $-1$ known as plasmino (hole).  However,  for $p_z<0$, the right handed mode flips to plasmino (hole) 
as its chirality to helicity ratio becomes -1 whereas the left handed mode becomes particle as its chirality to helicity ratio becomes $+1$.
The dispersion behaviour of the two modes are shown in the left panel of Fig.~\ref{fig:lll_disp} which begins at mass 
$\left. m_{LLL}^{*-}\right |_{p_z=0}$  as given in \eqref{mp}.

\item[(ii)] for negatively charged fermion  $Q=-1$, $\sigma=1$ implies $\nu = -1$ and $\sigma=-1$ implies $\nu=0$.  Thus, the
modes  with  $Q=-1$ and  $\sigma=+1$ (spin up) cannot propagate in LLL.  However, the modes  $L^{(+)}$ and $R^{(-)}$ have  spin down as found 
in subsec.~\ref{lll_modes} will propagate in LLL.  Their dispersion
are shown in the right  panel of Fig.~\ref{fig:lll_disp} which begin at mass $m_{LLL}^{*+} $ as given in \eqref{mp}. 
For $p_z>0$ the mode $L^{(+)}$ has helicity to chirality  ratio $+1$ whereas $R^{(-)}$ has  that of $-1$ and vice-versa for $p_z<0$.
\end{enumerate}
\begin{figure}[h]
\begin{center}
\includegraphics[width=0.5\textwidth]{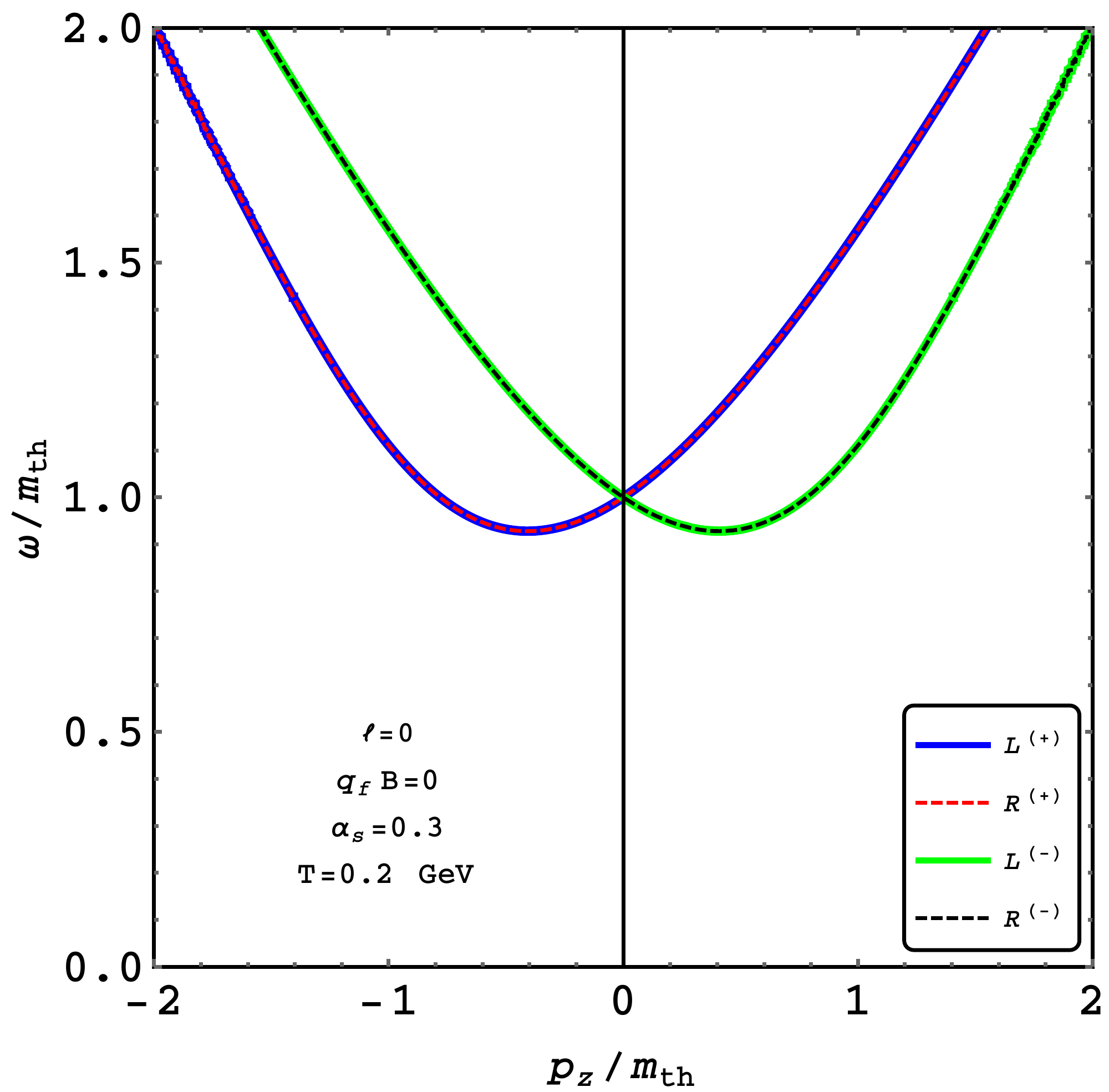}
\caption{\small The dispersion plots corresponding to HTL  propagator in absence of magnetic field, \textit{i.e.}, $B=0$.}
\label{fig:htl_disp}
\end{center}
\end{figure}
In  the absence of the background magnetic field ($B=0$),  the two modes, the  left handed  $L^{(+)}$ and the right handed $R^{(+)}$ fermions,
 merge together whereas the other two modes,  the  left handed  $L^{(-)}$ and the right handed $R^{(-)}$ fermions, also merge together.
 This  leads  to degenerate  (chirally symmetric)  modes for which the dispersion plots start at $m_{th}$ and  one gets back the  
 usual HTL result~\cite{braatendilepton} with quasiparticle and plasmino modes in presence of heat bath as shown in Fig.~\ref{fig:htl_disp}. 

As evident from the dispersion plots (Figs.~\ref{fig:HLLfig} and \ref{fig:lll_disp})
both left and right handed modes are also degenerate at $p_z=0$  in presence of magnetic field but at non-zero $|p_z|$ both left
and right handed modes get separated from each others, causing a chiral asymmetry without disturbing the chiral invariance 
(subsec.~\ref{chirality_trans}) in the system.
Also in subsec.~\ref{ref_sym} it was shown that the fermion propagator does not obey the reflection symmetry in presence of 
medium,  which is now clearly evident from all  dispersion plots as displayed above.  
\section{Three Point Function}
\label{vert_func}
The $(N+1)$-point functions are related to the $N$-point functions through Ward-Takahashi (WT)  identity.
The 3-point function is related to the 2-point function as
\bea
Q_\mu \Gamma^\mu(P,K;Q) &=& S^{-1}(P) -S^{-1} (K) = \slashed{P} -\slashed{K}  - \Sigma(P) + \Sigma(K) \nn \\
&=& \underbrace{(\slashed{P}-\slashed{K})}_{\mbox{Free}}  -\underbrace{\left (\Sigma^{B=0}(P,T) -\Sigma^{B=0}(K,T)\right )}_{\mbox{Thermal or HTL correction}} 
- \,  \underbrace{\left (\Sigma^{B\ne 0}(P,T) -\Sigma^{B\ne 0}(K,T)\right )}_{\mbox{Thermo-magnetic correction}}\nn \\
&=& \slashed{Q} + a(p_0,|\vec p|) \slashed{P} + b(p_0,|\vec p|) \slashed{u} 
 - \, a(k_0,|\vec k|) \slashed{K} -  b(k_0,|\vec k |) \slashed{u}  + b'(p_0, p_\perp,p_z) \gamma_5 \slashed{u}  \nn \\ 
&& +  c'(p_0, p_\perp,p_z) \gamma_5 \slashed{n}
 - b'(k_0, k_\perp,k_z) \gamma_5 \slashed{u} - c'(k_0, k_\perp,k_z) \gamma_5 \slashed{n}
\, , \label{wi_3pt}
\eea
where $Q=P-K$.  We note that  recently the general form of the thermo-magnetic corrections for 3-point~\cite{ayalafermionself,Haque:2017nxq} and 4-point~\cite{Haque:2017nxq} functions have
been given in terms of the involved angular integrals, which satisfy WT identies. Nevertheless,  to validate the general structure of the self-energy in
\eqref{genstructselfenergy} vis-a-vis the inverse propagator in \eqref{inv_prop},  we  obtain  below the temporal component of the
 3-point function at $\vec q=0; \, \vec p= \vec k $ and  $p=k$ . 

Using \eqref{at}, \eqref{bt}, \eqref{bprime} and \eqref{cprime}, we can obtain
\bea
\Gamma^0(P,K;Q)\big |_{{\vec q} =0} &=& \gamma_0 \,
\underbrace{- \,  \frac{m^2_{th}}{pq_0} \, \delta Q_0 
 \, \gamma^0  + \frac{m^2_{th}}{pq_0} \,  \delta Q_1 \,  (\hat p \cdot \vec \gamma)  }_{\mbox{Thermal or HTL correction}} \nn \\
 &&  \underbrace{ - \, \frac{{M'}^2}{pq_0} \, \left [  \delta Q_0 \, \gamma_5  \, 
+  \, \frac{p_z}{p} \,  \delta Q_1 \,  \left (i\gamma^1\gamma^2 \right ) \right ] \gamma^3  }_{\mbox{Thermo-magnetic correction}}\  \nn \\
&=& \gamma^0 \, +\delta \Gamma^0_{\tiny \mbox{HTL}} (P,K;Q) \, + \, \delta \Gamma^0_{\tiny \mbox{TM}} (P,K;Q) \, 
 \, , \label{wi_3pt_g0}
\eea
with 
\bea
\gamma_5\gamma^0&=& -i \gamma^1\gamma^2\gamma^3 \, , \nn \\
{M'}^2 &=& 4 \, C_F \, g^2 \, M^2(T,m,q_fB) \, , \nn \\
\delta Q_j &=& Q_j\left ( \frac{p_0}{p} \right ) - Q_j\left ( \frac{k_0}{p} \right )  \, . 
\eea
where $Q_j$ are the  Legendre functions of the second kind given in \eqref{Q0} and \eqref{Q1}. Important to note that
the thermo-magnetic (TM)  correction $\delta \Gamma^0_{\tiny \mbox{TM}} $ matches exactly with that  from direct calculation in \eqref{tm_g_0}
in Appendix~\ref{vert_direct}.
The result also agrees with the HTL  3-point function~\cite{ayalafermionself,Haque:2017nxq}  in absence of background magnetic field  by setting $B=0\, \Rightarrow M'=0$ as
\bea
\Gamma^0_{\tiny \mbox{HTL}} (P,K;Q)\big |_{{\vec q} =0} &=& \left [ 1\,- \,  \frac{m^2_{th}}{pq_0} \, \delta Q_0 
 \right ]\, \gamma^0  \, + \frac{m^2_{th}}{pq_0} \,  \delta Q_1 \,  (\hat p \cdot \vec \gamma) \,  \nn \\
&=& \gamma^0 +\delta \Gamma^0_{\tiny \mbox{HTL}} (P,K;Q) \, ,
 \,  \label{wi_3pt_htl}
\eea
where all components, \textit{i.e.}, $(0,1,2,3)$, are relevant for pure thermal background.
 
Now in absence of heat bath,  setting  $T=0\,  \Rightarrow$  $m_{th}=0$ and ${M'}^2=4 \, C_F \, g^2 \, M^2(T=0,m,q_f,B)$,  the temporal 3-point function 
in \eqref{wi_3pt_g0}  reduces to
\bea
\Gamma^0_B(P,K;Q)\big |_{{\vec q} =0} &=& 
\gamma^0 \, \underbrace { -   \frac{{M'}^2}{pq_0} \left [ \delta Q_0 \, \gamma_5 + \, \frac{p_z}{p} \,  \delta Q_1 \, (i\gamma^1\gamma^2) \right ] \,
 \gamma^3 }_{\mbox{Pure magnetic correction}} \, \\
&=&  \gamma^0 +\delta \Gamma^0_{\tiny \mbox{M}} (P,K;Q) \, . \label{wi_3pt_mag_g0} 
\eea
We now note that this is the three-point function with pure background magnetic field but no heat bath.  The gauge boson is oriented along the field direction 
and there is no polarisation in the transverse direction. Thus, only the longitudinal components (\textit{i.e,}  (0,3)-components) of the 3-point
function  would 
be relevant for pure background magnetic field in contrast to that of  \eqref{wi_3pt_htl}  for pure  thermal background.
\section{Spectral Representation of the Effective Propagator}
\label{spec_rep}
In this section we obtain the spectral representation of the effective propagator in a hot magnetised medium. This quantity
is of immense interest for  studying  the various spectral properties, real and virtual photon production, damping rates  and 
various transport coefficients etc.  of the hot magnetised medium, in particular,  for hot magnetised QCD medium.
 \subsection{General Case}
 \label{spec_rep_gen}
The effective propagator as obtained in \eqref{eff_prop1} is given by
\bea
S^*=\mathcal{P}_-\frac{\slashed{L}}{L^2}\mathcal{P}_+ + \mathcal{P}_+\frac{\slashed{R}}{R^2}\mathcal{P}_- \, ,
\eea
where  $\slashed{L}$ and $\slashed{R}$  can be written  in the rest frame of the heat bath and the magnetic field in the $z$-direction
following  \eqref{l_mu} and \eqref{r_mu}, respectively,  as 
\bea
\slashed{L} &=& \left[(1+a(p_0,p))p_0+b(p_0,p)+b'(p_0,p_\perp,p_z)\right]\gamma^0 - \left[(1+a(p_0,p))p_z+c'(p_0,p_\perp,p_z)\right]\gamma^3\nn\\
&& - (1+a(p_0,p))(\gamma \cdot p)_\perp \nn \\
&=& \left[(1+a(p_0,p))p_0+b(p_0,p)+b'(p_0,p_\perp,p_z)\right]\gamma^0 - \left[p(1+a(p_0,p))\right](\gamma \cdot \hat{p})- c'((p_0,p_\perp,p_z)\gamma^3\nn \\
&=& g_L^1(p_0,p_\perp,p_z)\gamma^0 - g_L^2(p_0,p_\perp,p_z)(\gamma \cdot \hat{p}) - g_L^3(p_0,p_\perp,p_z)\gamma^3, \label{l_rest} \\
\slashed{R} &=& \left[(1+a(p_0,p))p_0+b(p_0,p)-b'(p_0,p_\perp,p_z)\right]\gamma^0 - \left[(1+a(p_0,p))p_z-c'(p_0,p_\perp,p_z)\right]\gamma^3 \, \nn\\
&& (-1+a(p_0,p))(\gamma \cdot p)_\perp \nn \\
&=& \left[(1+a(p_0,p))p_0+b(p_0,p)-b'(p_0,p_\perp,p_z)\right]\gamma^0 - \left[p(1+a(p_0,p))\right](\gamma \cdot \hat{p})+ c'(p_0,p_\perp,p_z)\gamma^3 \nn \\
&=& g_R^1(p_0,p_\perp,p_z)\gamma^0 - g_R^2(p_0,p_\perp,p_z)(\gamma \cdot \hat{p}) + g_R^3(p_0,p_\perp,p_z)\gamma^3 \, , \label{r_rest}
\eea
where ${\hat p} ={\mathbf p}/|{\mathbf p}|$,  $p=|{\mathbf p}|$ and, $p_z$ and  $p_\perp$  are given, respectively,  in \eqref{p3} and \eqref{pperp}.
We also note that  though  $g_L^2=g_R^2;~g_L^3=g_R^3$, but they are treated  separately for the sake of notations that  we would be using, 
 for convenience, as $g_L^i$ and $g_R^i$. One can decompose the effective propagator into  six parts by separating 
 out the $\gamma$ matrices as
\bea
S^* &=& \mathcal{P}_-\gamma^0\mathcal{P}_+~\frac{g_L^1(p_0,p_\perp,p_z)}{L^2} 
- \mathcal{P}_-(\gamma \cdot \hat{p})\mathcal{P}_+~\frac{g_L^2(p_0,p_\perp,p_z)}{L^2} 
- \mathcal{P}_-\gamma^3\mathcal{P}_+~\frac{g_L^3(p_0,p_\perp,p_z)}{L^2} \nn\\
&+& \mathcal{P}_+\gamma^0\mathcal{P}_-~\frac{g_R^1(p_0,p_\perp,p_z)}{R^2} 
- \mathcal{P}_+(\gamma \cdot \hat{p})\mathcal{P}_-~\frac{g_R^2(p_0,p_\perp,p_z)}{R^2} 
+ \mathcal{P}_+\gamma^3\mathcal{P}_-~\frac{g_R^3(p_0,p_\perp,p_z)}{R^2}.
\label{prop_pre_spec}
\eea
In subsection~\ref{disp_rep} we have discussed that $L^2=0$ yields four poles, 
leading to four modes with both positive and negative  energy as  $\pm\omega_{L^{(+)}}(p_\perp,p_z)$ and $\pm\omega_{L^{(-)}}(p_\perp,p_z)$. 
Similarly, $R^2=0$  also yields four poles, namely $\pm\omega_{R^{(+)}}(p_\perp,p_z)$ and $\pm\omega_{R^{(-)}}(p_\perp,p_z)$. 

 With this information one can obtain the spectral representation~\cite{Bellac:2011kqa,Karsch:2000gi,Chakraborty:2001kx,braatendilepton} of  the effective propagator in  \eqref{prop_pre_spec}  as
\bea
\rho &=& \left(\mathcal{P}_-\gamma^0\mathcal{P}_+\right)~\rho_L^1 
- \left( \mathcal{P}_-(\gamma \cdot \hat{p})\mathcal{P}_+\right)~\rho_L^2 
- \left( \mathcal{P}_-\gamma^3\mathcal{P}_+\right)~\rho_L^3 \nn\\
&& + \left( \mathcal{P}_+\gamma^0\mathcal{P}_-\right)~\rho_R^1
 - \left(\mathcal{P}_+(\gamma \cdot \hat{p})\mathcal{P}_-\right)~\rho_R^2 
 + \left( \mathcal{P}_+\gamma^3\mathcal{P}_-\right)~\rho_R^3 \, .
\label{prop_spec}
\eea
where  the spectral function corresponding to each of the term can be written as 
 \bea
 \rho_L^i &=& \frac{1}{\pi} ~\mathrm{Im}\left(\frac{g_L^i}{L^2}\right)\nn\\
 &=& Z_{L^{(+)}}^{i+}(p_\perp,p_z)\delta(p_0-\omega_{L^{(+)}}(p_\perp,p_z))+Z_{L^{(+)}}^{i-}(p_\perp,p_z)\delta(p_0+\omega_{L^{(+)}}(p_\perp,p_z))\nn\\
 &&+Z_{L^{(-)}}^{i+}(p_\perp,p_z)\delta(p_0-\omega_{L^{(-)}}(p_\perp,p_z))+Z_{L^{(-)}}^{i-}(p_\perp,p_z)\delta(p_0+\omega_{L^{(-)}}(p_\perp,p_z))+\beta_L^i
 \, ,\label{spec_li}\\
 \rho_R^i &=& \frac{1}{\pi} ~\mathrm{Im}\left(\frac{g_R^i}{R^2}\right)\nn\\
 &=& Z_{R^{(+)}}^{i+}(p_\perp,p_z)\delta(p_0-\omega_{R^{(+)}}(p_\perp,p_z))+Z_{R^{(+)}}^{i-}(p_\perp,p_z)\delta(p_0+\omega_{R^{(+)}}(p_\perp,p_z))\nn\\
 &&+Z_{R^{(-)}}^{i+}(p_\perp,p_z)\delta(p_0-\omega_{R^{(-)}}(p_\perp,p_z))+Z_{R^{(-)}}^{i-}(p_\perp,p_z)\delta(p_0+\omega_{R^{(-)}}(p_\perp,p_z))+\beta_R^i\, ,
 \label{spec_residue}
 \eea
 where $i=1,2,3$. We note that the delta-functions are associated with pole parts originating from the time like domain ($p_0^2 >p^2)$ 
 whereas the cut parts  $\beta^i_{L(R)}$ are associated with the Landau damping arises from the space-like domain, $p_0^ 2 < p^2$, 
 of the propagator. The residues $Z^i_{L(R)}$ are  determined at the various poles as
 \bea
 Z_{L(R)}^{i \ {\mbox{sgn of pole }}}(p_\perp,p_z) = g_{L(R)}^i(p_0,p) \Bigg| \frac{\partial L^2(R^2)}{\partial p_0} \Bigg|^{-1}_{p_0=\mbox{ pole}} \, \label{residue}
 \eea
 As a demonstration,  we present analytical expressions of three residues  corresponding to the pole   $p_0=+\omega_{L^{(+)}}$ as 
 \bea
 Z_{L^{(+)}}^{1+} &=& \frac{p\left(p^2-\omega^2_{L^{(+)}}\right)\left[p^2\left(m_{th}^2\log\left(\frac{\omega_{L^{(+)}}+p}
 {\omega_{L^{(+)}}-p}\right)-2p\omega_{L^{(+)}}\right)+M'^2p_z(2p-\omega_{L^{(+)}})
\log\left(\frac{\omega_{L^{(+)}}+p}
 {\omega_{L^{(+)}}-p}\right)\right]}{m_{th}^2\left[8p^4\left(\omega_{L^{(+)}}+M'^2/m_{th}^2~p_z\right)+\log\left(\frac{\omega_{L^{(+)}}+p}
 {\omega_{L^{(+)}}-p}\right)
 X \right]}\, , \nn
 \eea
 \bea
 Z_{L^{(+)}}^{2+} &=&  \frac{p^2\left(p^2-\omega^2_{L^{(+)}}\right)\left[2p\left(m_{th}^2+p^2\right)-m_{th}^2\omega_{L^{(+)}}\log\left(\frac{\omega_{L^{(+)}}+p}
 {\omega_{L^{(+)}}-p}\right)\right]}{m_{th}^2\left[8p^4\left(\omega_{L^{(+)}}+M'^2/m_{th}^2~p_z\right)+\log\left(\frac{\omega_{L^{(+)}}+p}
 {\omega_{L^{(+)}}-p}\right) X
\right]} \, ,\nn\\
 Z_{L^{(+)}}^{3+} &=&  \frac{-M'^2p^5\left(p^2-\omega^2_{L^{(+)}}\right)\log\left(\frac{\omega_{L^{(+)}}+p}
 {\omega_{L^{(+)}}-p}\right)}{m_{th}^4\left[8p^4\left(\omega_{L^{(+)}}+M'^2/m_{th}^2~p_z\right)+\log\left(\frac{\omega_{L^{(+)}}+p}
 {\omega_{L^{(+)}}-p}\right)X \right]} \, , \nn
 \eea
 where   $X= 2p^3(M'^2-2m_{th}^2)+2M'^2pp_z^2+M'^2\omega_{L^{(+)}}p_\perp^2\log\left(\frac{\omega_{L^{(+)}}+p}
 {\omega_{L^{(+)}}-p}\right)$. The other poles of $L^2=0$  can trivially be found out by replacing $\omega_{L^{(+)}}$ in the above expressions. 
 The expressions for the residues for  $R$ parts can  similarly be expressed as the $L$ parts, but we do not show them. 

  Below in Fig.~\ref{residue_ll} we present the residues corresponding to the first Landau level where all the terms are present. 
  We take the value of the magnetic field as $m_\pi^2/2$ and temperature to be $200$ MeV.

\begin{center}
\begin{figure}[H]
\begin{center}
\hspace*{-1.0cm}  
 \includegraphics[scale=0.5]{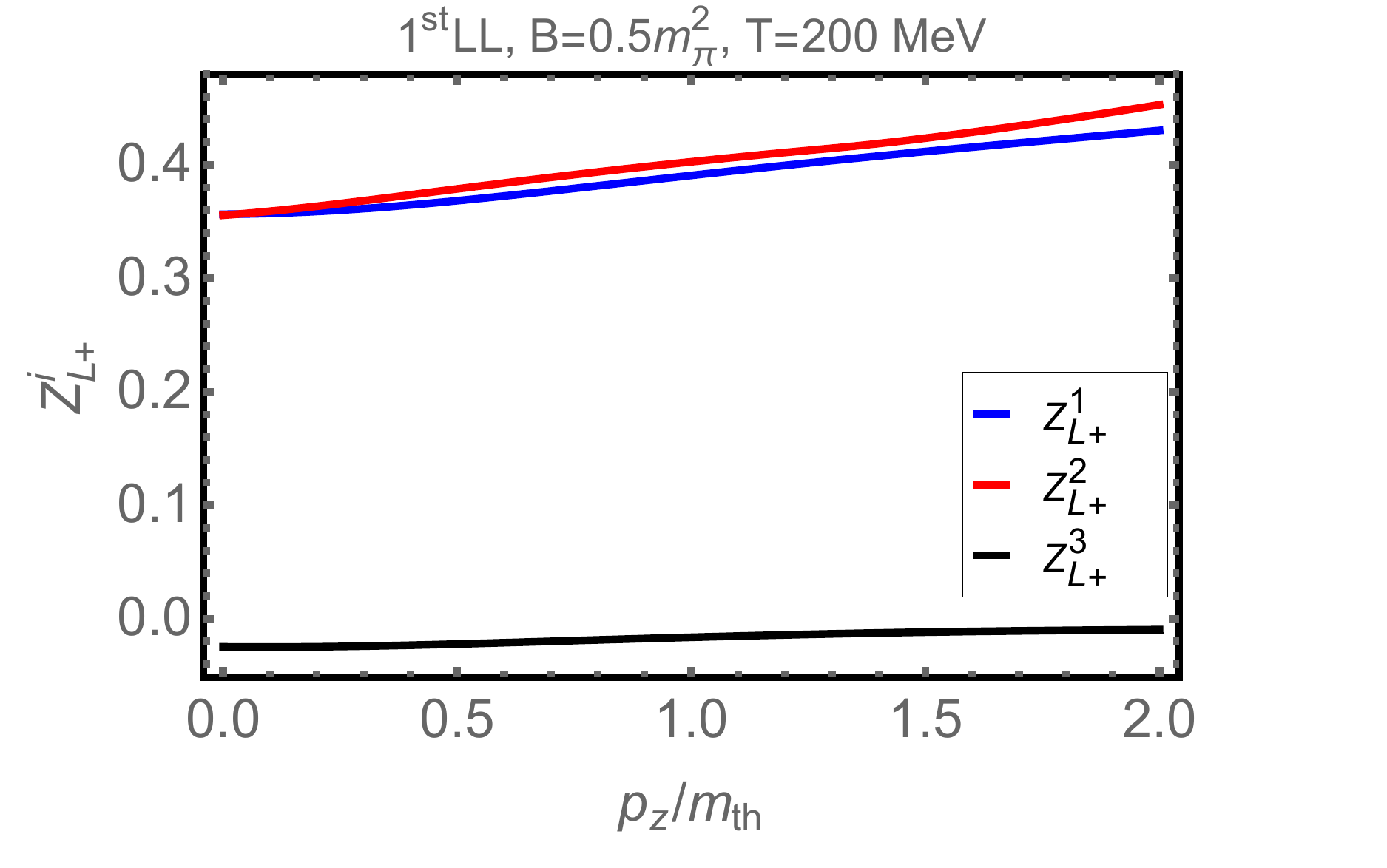}\hspace*{-1.0cm}\includegraphics[scale=0.5]{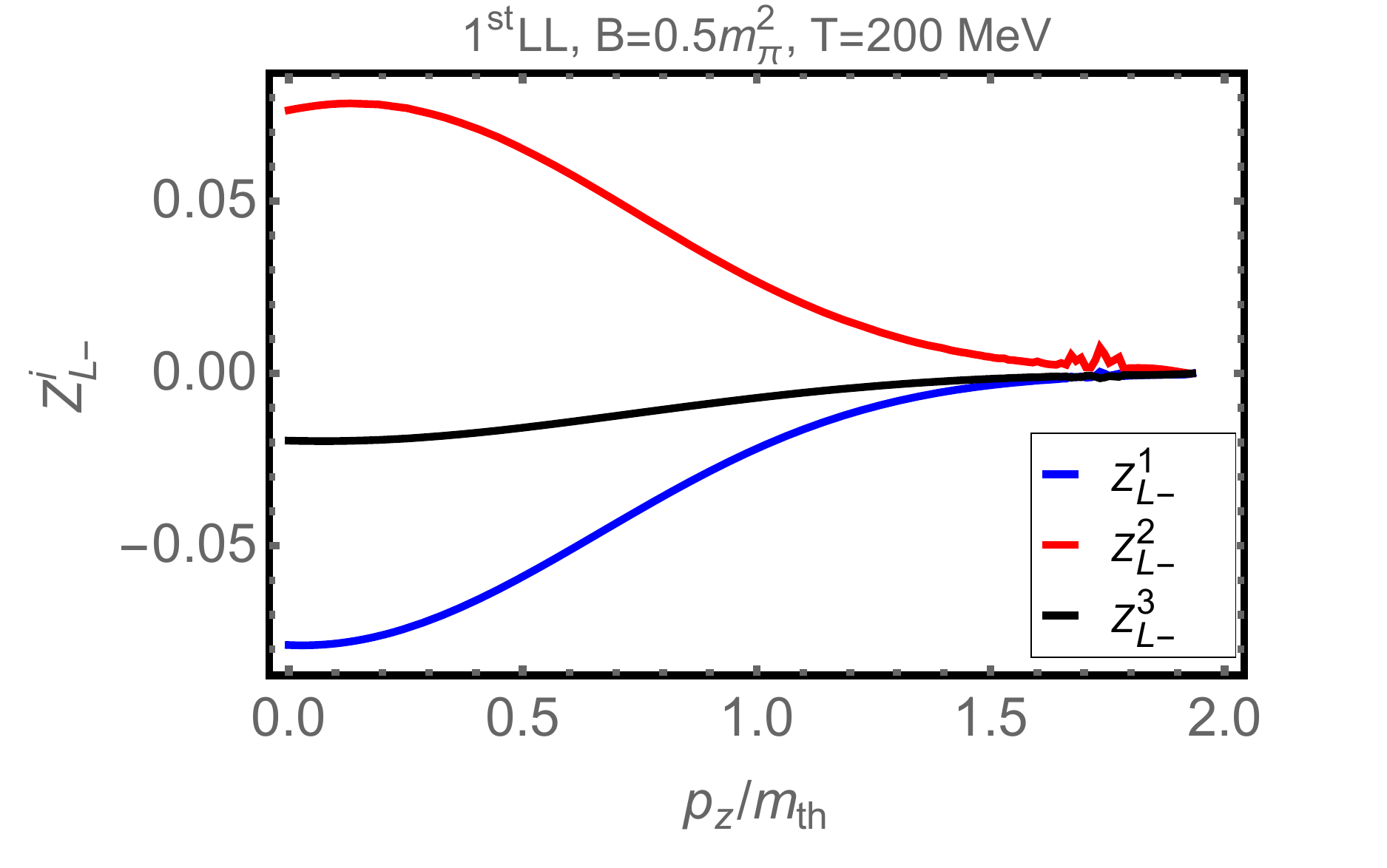}
\hspace*{-1.0cm}
 \includegraphics[scale=0.5]{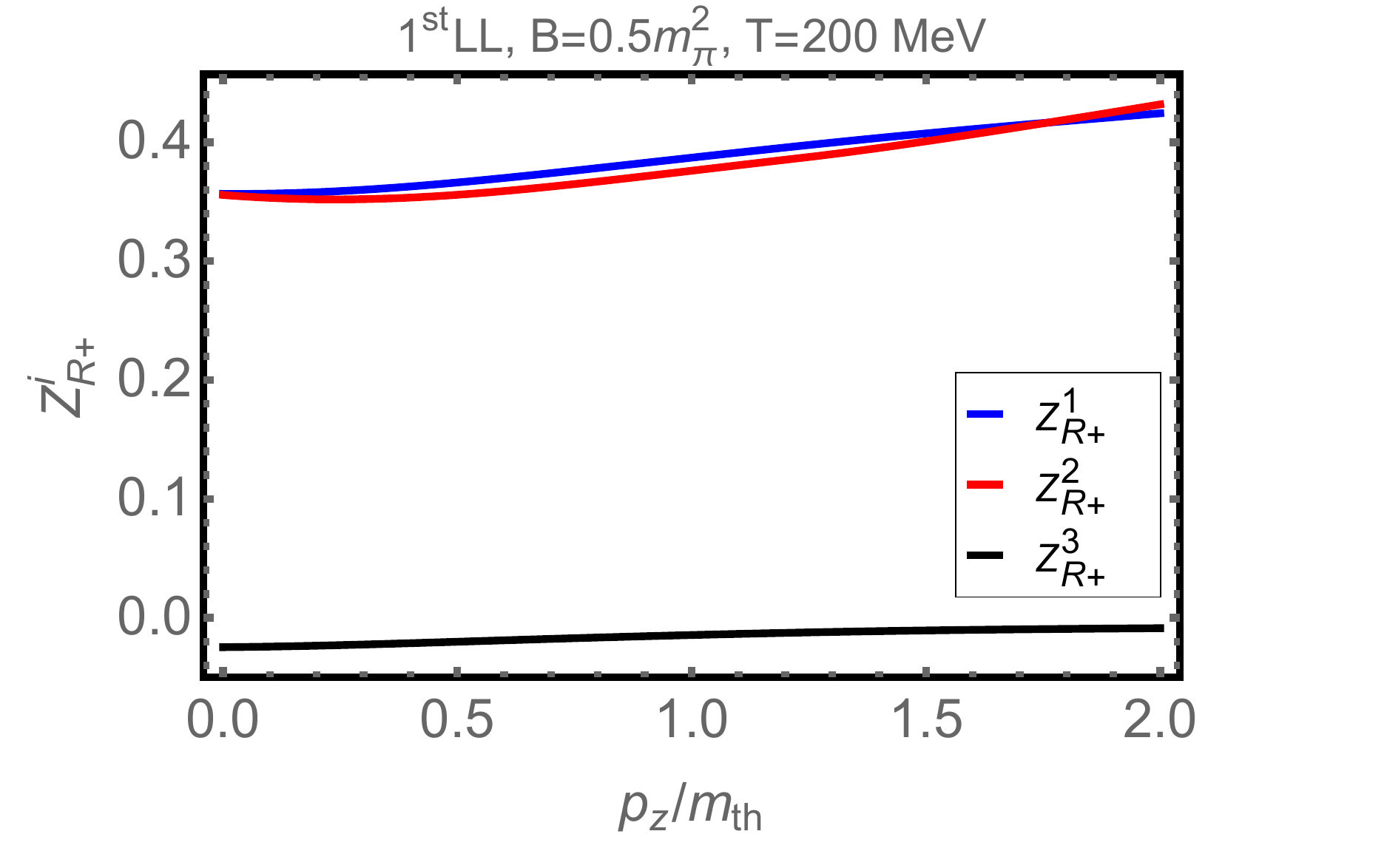}\hspace*{-1.0cm}\includegraphics[scale=0.5]{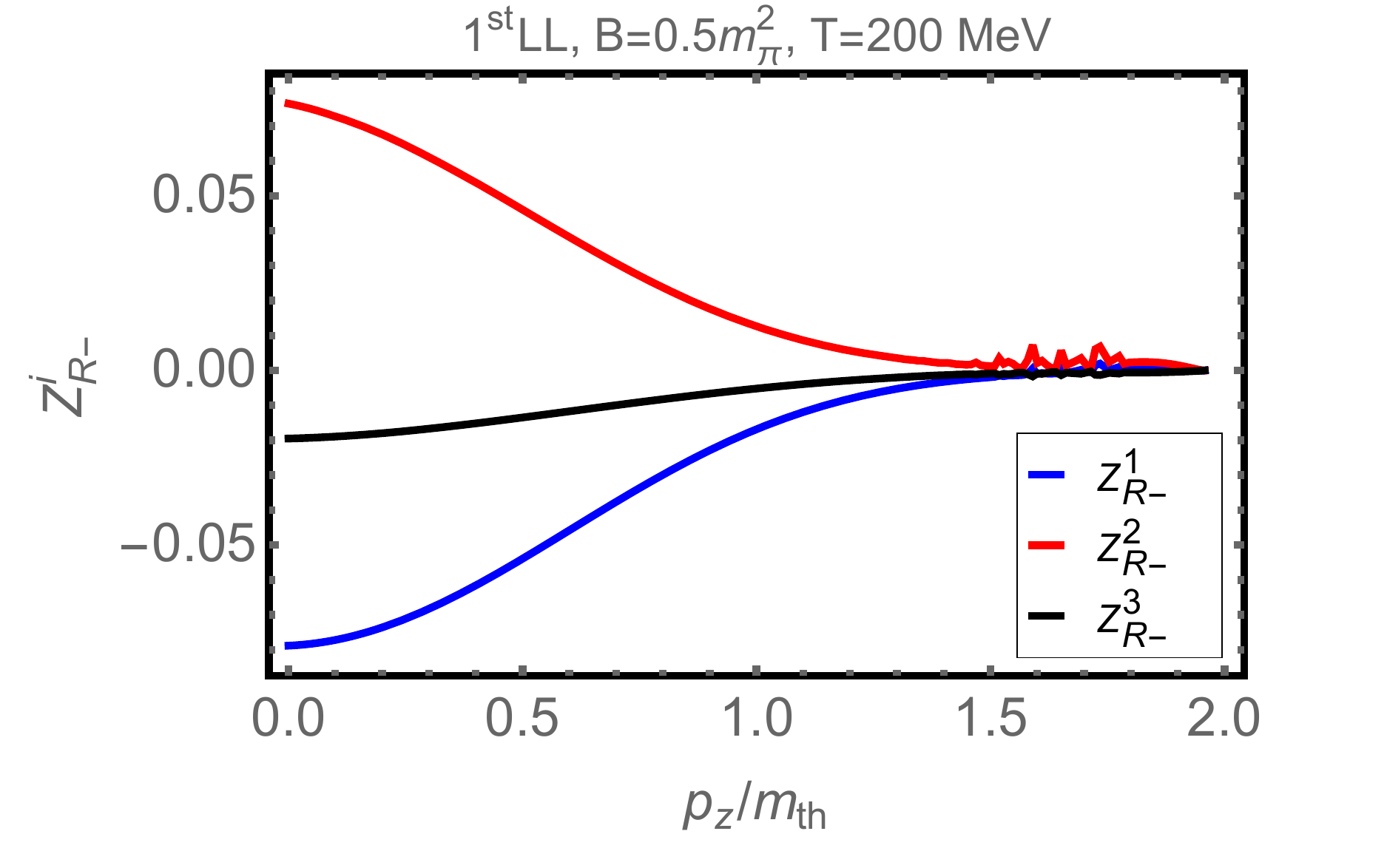}
 \caption{Different Residues for the first LL ($l=1$) are plotted with  scaled momentum along the magnetic field direction.}
  \label{residue_ll}
\end{center}
\end{figure}
\end{center}

Now, the expressions for the cut parts $\beta_{L(R)}^i$ are given below :
 
 \bea
 \beta_L^i &=& \frac{1}{\pi}\Theta(p^2-p_0^2)\frac{\mathrm{Im}(g_L^i)~\mathrm{Re}(L^2) - \mathrm{Im}(L^2)\mathrm{Re}(g_L^i)}{(\mathrm{Re}(L^2))^2+(\mathrm{Im}(L^2))^2}, \\
 \beta_R^i &=& \frac{1}{\pi}\Theta(p^2-p_0^2)\frac{\mathrm{Im}(g_R^i)~\mathrm{Re}(R^2) - \mathrm{Im}(R^2)\mathrm{Re}(g_R^i)}{(\mathrm{Re}(R^2))^2+(\mathrm{Im}(R^2))^2},
 \label{cut_LR_i}
 \eea
 where
 \bea
 \mathrm{Re}(g_L^1) &=& p_0-M'^2\frac{p_z}{p^2}-\frac{m_{th}^2}{p}\left(1-\frac{M'^2}{m_{th}^2}\frac{p_zp_0}{p^2}\right)Q_0\left(\left|\frac{p_0}{p}\right|\right),\\
 \mathrm{Im}(g_L^1) &=& \frac{\pi}{2}\frac{m_{th}^2}{p}\left(1-\frac{M'^2}{m_{th}^2}\frac{p_zp_0}{p^2}\right),\\
 \mathrm{Re}(g_L^2) &=&  \mathrm{Re}(g_R^2) = p+\frac{m^{2}_{th}}{p}\left[1-\frac{p_{0}}{p}Q_0\left(\left|\frac{p_0}{p}\right|\right)\right] ,\\
 \mathrm{Im}(g_L^2) &=& \mathrm{Im}(g_R^2) = \pi\,m^{2}_{th}\,\frac{p_{0}}{2\,p^{2}} ,\\
\mathrm{Re}(g_L^3) &=& \mathrm{Re}(g_R^3) = \frac{M'^2}{p}Q_0\left(\left|\frac{p_0}{p}\right|\right) , \\ 
\mathrm{Im}(g_L^3) &=& \mathrm{Im}(g_R^3) = - \frac{\pi M'^2}{2p} ,   \label{gl_re}
 \eea
 and
 \bea
 \mathrm{Re}(g_R^1) &=& p_0+M'^2\frac{p_z}{p^2}-\frac{m_{th}^2}{p}\left(1+\frac{M'^2}{m_{th}^2}\frac{p_zp_0}{p^2}\right)Q_0\left(\left|\frac{p_0}{p}\right|\right),\\
 \mathrm{Im}(g_R^1) &=& \frac{\pi}{2}\frac{m_{th}^2}{p}\left(1+\frac{M'^2}{m_{th}^2}\frac{p_zp_0}{p^2}\right) \, . \label{gl_im}
 \eea
 
 Also we obtain
 \bea
 \mathrm{Re}(L^2) &=& A_L+B_LQ_0\left(\left|\frac{p_0}{p}\right|\right)+C\left(Q_0^2\left(\left|\frac{p_0}{p}\right|\right)-\frac{\pi^2}{4}\right), \\
 \mathrm{Im}(L^2) &=& -\frac{\pi B_L}{2} - \pi Q_0\left(\left|\frac{p_0}{p}\right|\right)C, \\
 \mathrm{Re}(R^2) &=& A_R+B_RQ_0\left(\left|\frac{p_0}{p}\right|\right)+C\left(Q_0^2\left(\left|\frac{p_0}{p}\right|\right)-\frac{\pi^2}{4}\right),\\
 \mathrm{Im}(R^2) &=& -\frac{\pi B_R}{2} - \pi Q_0\left(\left|\frac{p_0}{p}\right|\right)C , 
 \eea
 where
 \bea
 A_L &=& p_0^2-p^2-2m_{th}^2-\frac{m_{th}^4}{p^2}-\frac{2M'^2p_0p_z}{p^2}+\frac{M'^4p_z^2}{p^4}, \\
 A_R &=& p_0^2-p^2-2m_{th}^2-\frac{m_{th}^4}{p^2}+\frac{2M'^2p_0p_z}{p^2}+\frac{M'^4p_z^2}{p^4} ,
 \eea
 \bea
 B_L &=& \frac{2m_{th}^4p_0}{p^3} - \frac{2M'^2p_z}{p} + \frac{2M'^2p_0^2p_z}{p^3} - \frac{2M'^4p_0p_z^2}{p^5}, \\
 B_R &=& \frac{2m_{th}^4p_0}{p^3} + \frac{2M'^2p_z}{p} - \frac{2M'^2p_0^2p_z}{p^3} - \frac{2M'^4p_0p_z^2}{p^5} ,\\
 C &=& \frac{m_{th}^4-M'^4}{p^2}-\frac{p_0^2m_{th}^4}{p^4} + \frac{M'^4p_0^2p_z^2}{p^6}.
 \eea
  
\subsection{LLL Case}
\label{spec_rep_lll}
 
For LLL, as $p_\perp=0$, so $g_{L(R)}^2$ and $g_{L(R)}^3$ in \eqref{l_rest} and \eqref{r_rest} can now  be merged as
 \bea
 g_L^{2+3} &=& \left[(1+a(p_0,p))p_z+c'(p_0,p)\right]\gamma^3\, , \nn\\
 g_R^{2+3} &=& \left[(1+a(p_0,p))p_z-c'(p_0,p)\right]\gamma^3 \, . \nn
 \eea
 \begin{center}
\begin{figure}[H]
\begin{center}
\hspace*{-1cm}
\includegraphics[scale=0.5]{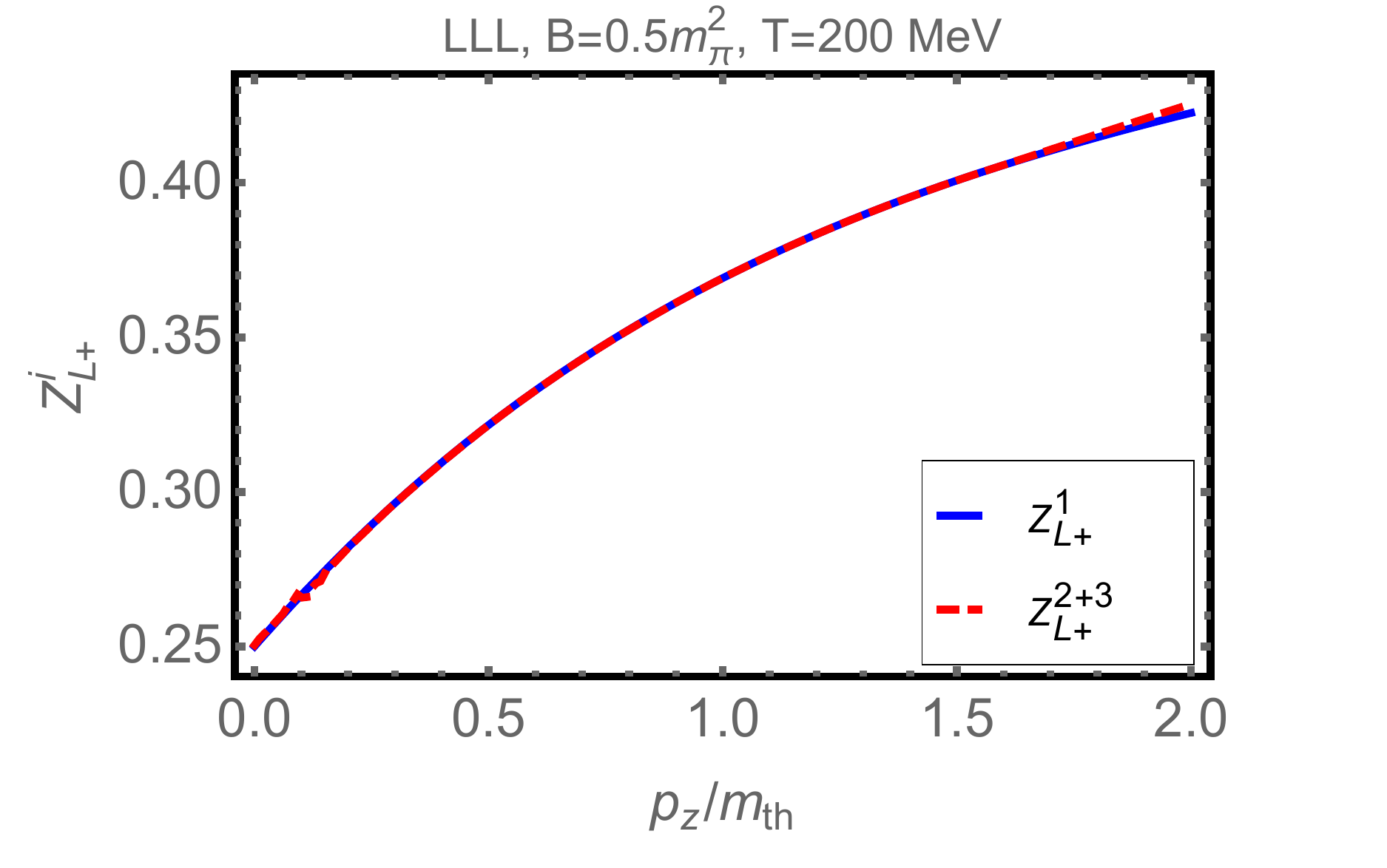}\hspace*{-1.0cm}\includegraphics[scale=0.5]{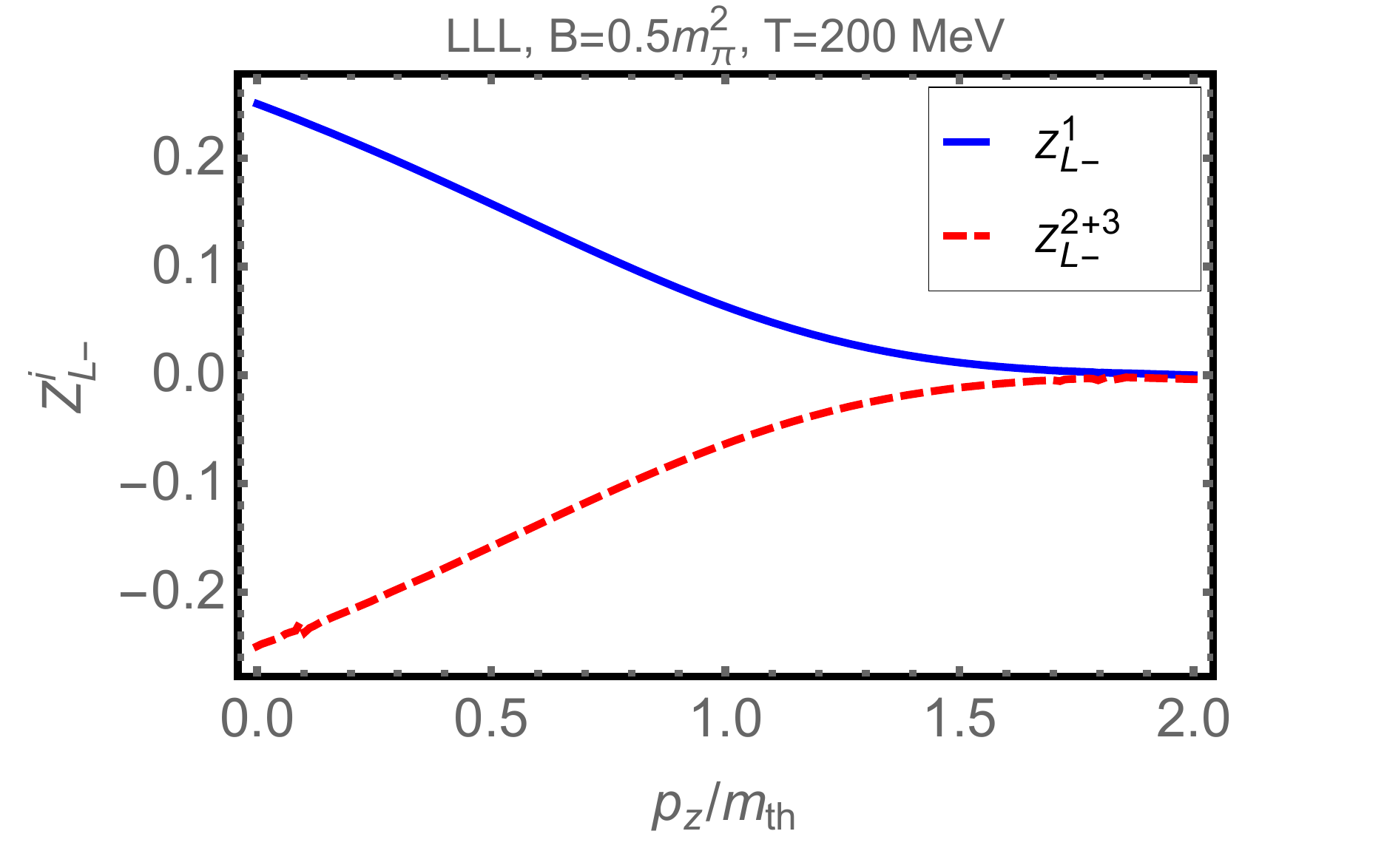}
    \hspace*{-1cm}
 \includegraphics[scale=0.5]{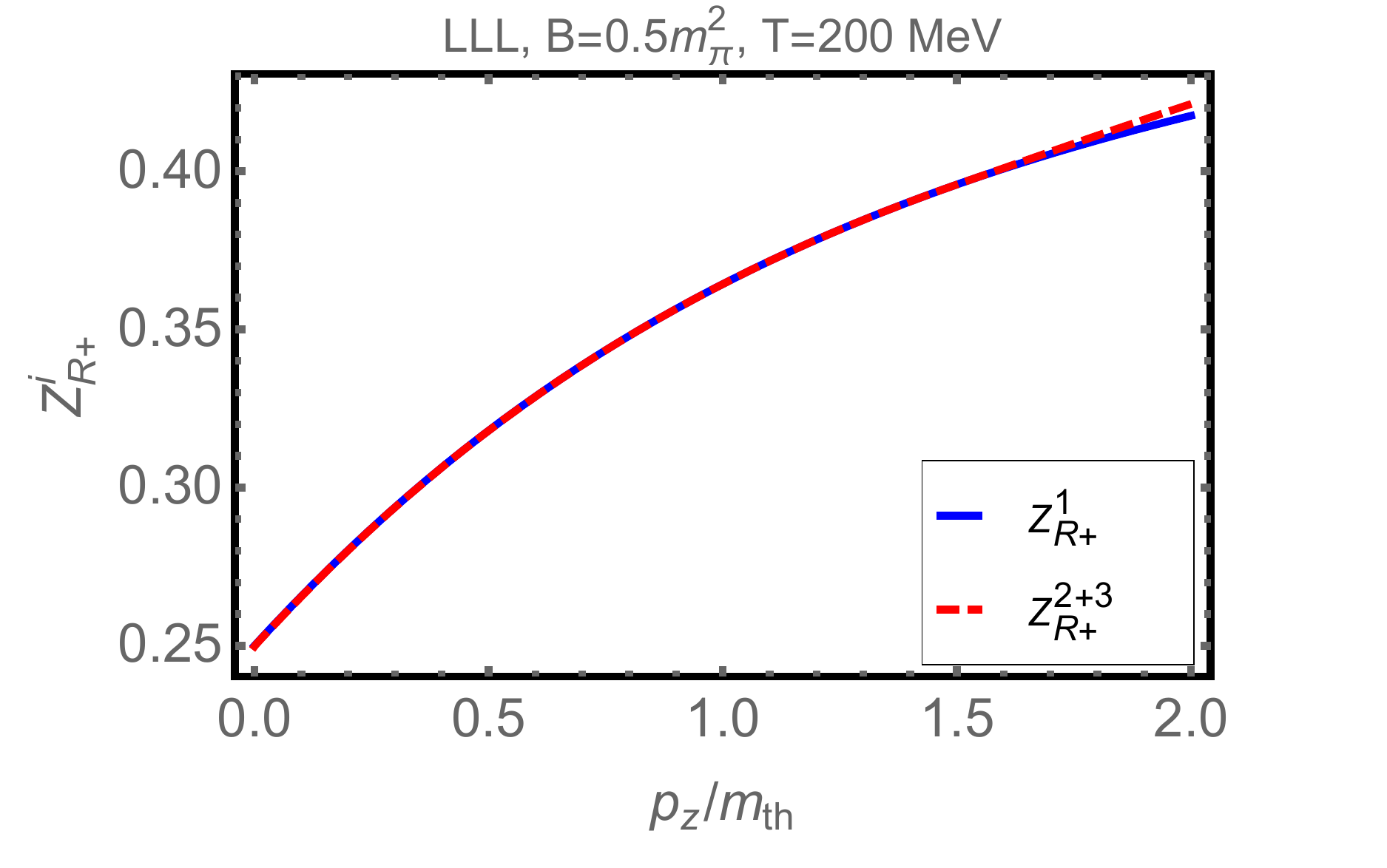}\hspace*{-1.0cm}\includegraphics[scale=0.5]{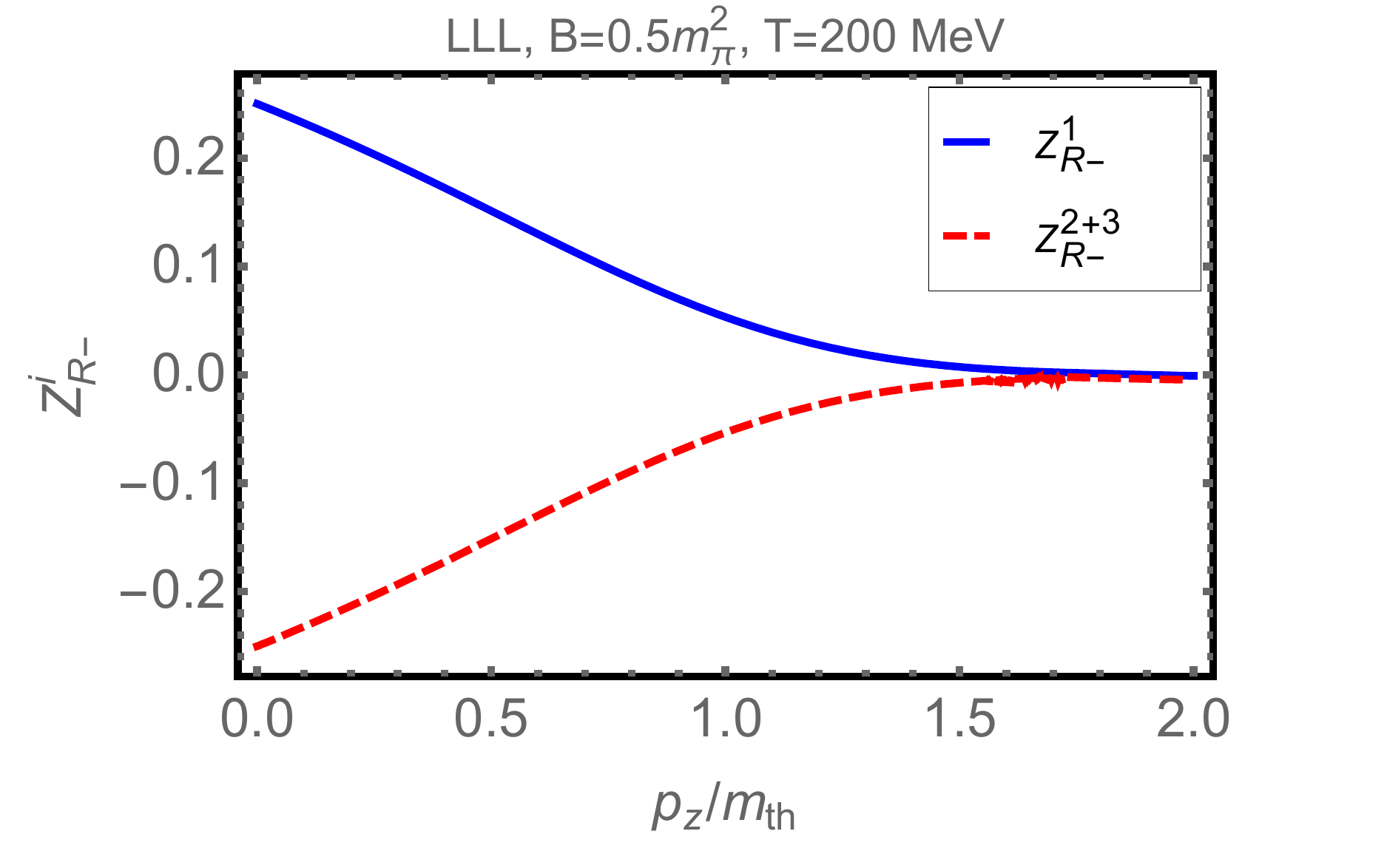}
 \caption{Different Residues for the LLL ($l=0$)  are plotted with  scaled momentum along the magnetic field direction.}
 \label{residue_lll}
    \end{center}
   \end{figure}
  \end{center}
  
The  spectral function corresponding to LLL reads as
  \bea
\rho_{LLL} &=&\left( \mathcal{P}_-\gamma^0\mathcal{P}_+\right)~\rho_L^1 - \left(\mathcal{P}_-\gamma^3\mathcal{P}_+\right)~\rho_L^{2+3} \nn\\
&+&\left( \mathcal{P}_+\gamma^0\mathcal{P}_-\right)~\rho_R^1 -\left( \mathcal{P}_+\gamma^3\mathcal{P}_-\right)~\rho_R^{2+3}.
\label{prop_spec_LLL}
\eea
where one needs to determine 
\bea
\rho_{L(R)}^{2+3} &=& \frac{1}{\pi} ~\mathrm{Im}\left(\frac{g_{L(R)}^{2+3}}{L^2(R^2)}\right)\nn,
\eea
which can again be represented in terms of different residues corresponding to different 
poles of $L^2(R^2)=0$ as in Eq.(\ref{spec_residue}). In Fig.~\ref{residue_lll}, the variation of 
the residues for the lowest Landau level are shown.   

In appendix~\ref{spec_htl} we have demonstrated how one gets back the HTL spectral functions when magnetic field is withdrawn 
from the thermal medium.
\section{Conclusions}
\label{remarks}
In this article the general structure of fermionic self-energy for a chirally invariant theory has been 
formulated  for a hot and magnetised medium. Using this we have obtained  a closed form of the
general structure of the effective fermion propagator.  The collective excitations in such a nontrivial
background has been obtained for  a time-like momenta in the weak field and HTL approximation
in the domain $m^2_{th}(\sim g^2T^2 < |q_fB| < T^2$. We found that 
the left and right handed modes get separated and become asymmetric in presence of magnetic field which 
were degenerate and symmetric otherwise.
The transformation of the effective propagator in a hot magnetised medium under some of the discrete symmetries
have been studied and its consequences  are also reflected in the collective fermion modes in the Landau levels. 
We have also obtained the Dirac spinors of the various collective modes by solving the Dirac equation
with the effective two-point function. Further,  we checked the general structure of the two-point function by obtaining the
three-point function using the Ward-Takahashi identity, which agrees with the direct calculation of one-loop order 
in weak field approximation.  We also  found  that only the longitudinal component
of the vertex would be relevant when there is only  background magnetic field. 
The spectral function corresponding to the effective propagator  
 is explicitly obtained for a hot magnetised medium which will be extremely useful
for studying the spectral properties, \textit{e.g.},  photon/dilepton production, damping rate, transport coefficients
for a hot magnetised medium. 
 This has pole contribution due to the various collective modes originating
 from the time-like domain and a Landau cut contribution appearing from the space-like domain.  
It has explicitly been shown that the spectral function reduces to that obtained for thermal
medium in absence of the magnetic field. Our formulation is in general applicable to both 
QED and QCD with nontrivial background like hot magnetised medium

\section{Acknowledgement}
The authors would like to acknowledge useful discussions with  Palash B Pal, Najmul Haque, 
Chowdhury Aminul Islam, 
Arghya Mukherjee and Bithika Karmakar.
AB and MGM were funded by the Department of Atomic Energy (DAE), India via the 
project TPAES whereas AD and PKR were funded by   the project DAE/ALICE/SINP. AB was also partially supported by the National Post Doctoral Program CAPES (PNPD/CAPES), Govt. of Brazil. 

\appendix
\section{Computations of structure functions in one-loop in a weak field approximation for hot magnetised QCD medium:}
\label{append_A}	 
\begin{figure}[h!]
\centering
\includegraphics[scale=1]{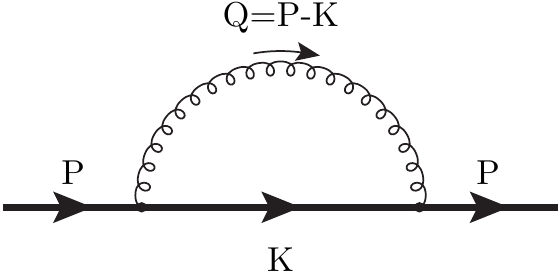}
\caption{One loop fermion self-energy in a hot magnetized medium.}
\label{fig:self_energy}
 \end{figure}

Here, we  present the computations of the various structure functions in (\ref{sta}) to (\ref{stcp})  
in 1-loop order (Fig.\ref{fig:self_energy})  in a weak field  and 
HTL approximations following the imaginary time formalism.
In  Fig.\ref{fig:self_energy}  the modified  quark propagator (bold line)  due to background magnetic  field is given in (\ref{wprop}).
Since glouns are chargeless, their  propagators do not change in presence of magnetic field. 
The gluon propagator in Feynman gauge, is given as\cite{chyiweak}
\begin{align}
D_{ab}^{\mu\nu}(Q)=-i\delta_{ab}\frac{g^{\mu\nu}}{Q^{2}} . \label{gluonprop}
\end{align}
We note that we would  like to explore the  fermion  spectrum  in a hot magnetised background in the limit $m_f^2 <  q_fB < T^2$.   
In this domain  the fermion propagator is obtained by expanding the sum over all Landau levels in powers 
of $q_{f}B$ in \eqref{mirprop}  and keeping  upto $\mathcal{O}([q_{f}B)^2]$, it reads as
\begin{small}
\begin{align}
S(K) &= i\frac{\slashed{K}+m_f}{K^{2}-m^{2}}+i\gamma_{1}\gamma_{2}\frac{\slashed{K}_{\shortparallel}+m_f}{(K^{2}-m_f^{2})^{2}}
q_{f}B+2\left[ \frac{\slashed{K}_{\shortparallel}+m_f}{(K^2-m_f^2)^{4}}\,K^{2}_{\perp}-\frac{K^{2}_{\shortparallel}-m_f^2}{(K^2-m_f^2)^{4}} 
\slashed{K}_{\perp}\right](q_{f}B)^{2} \nonumber \\
 &= i\frac{\slashed{K}+m_f}{K^{2}-m_f^{2}}+i\gamma_{1}\gamma_{2}\frac{\slashed{K}_{\shortparallel}+m_f}{(K^{2}-m_f^{2})^{2}}
q_{f}B + {\mathcal O}[(q_fB)^2], \label{wprop0}
\end{align}
\end{small}
where the first term is the free propagator and  the second one is  ${\mathcal O}[q_fB]$ correction to it. 
Now combining \eqref{wprop0} and \eqref{acom}  the fermion propagator in 
background magnetic field  reads as
 \begin{small}
\begin{align}
S(K) &= i\frac{\slashed{K}}{K^{2}-m_f^{2}}-\frac{\gamma_{5}\left[\left(K.n\right)\slashed{u}-\left(K.u\right)\slashed{n}\right]}{(K^{2}-m_f^{2})^{2}}
(q_{f}B)+ {\mathcal O}[(q_fB)^2] \nn \\
&=  S^{B=0}_1(K) +S^{B\ne 0}_2(K) +\mathcal{O}\left [(q_{f}B)^{2}  \right ],\label{wprop}
\end{align}
\end{small}
where the fermion mass in the numerator has been neglected in the weak field domain,  $m_f^2< (q_fB) < T^2$.

The one-loop quark self-energy  upto $\mathcal{O}(|q_{f}B|)$  can be written as
\begin{align}
\Sigma(P) &= g^{2}\,C_{F}\,T\,\sumint_{\{K\}} \gamma\indices{_\mu}\,\left(\frac{\slashed{K}}{K^{2}-m_f^{2}}-
\frac{\gamma_{5}\left[\left(K.n\right)\slashed{u}-\left(K.u\right)\slashed{n}\right]}{(K^{2}-m_f^{2})^{2}}\,q_{f}B\right)\,
\gamma\indices{^\mu}\,\frac{1}{(P-K)^{2}}  \nn \\
&\simeq \Sigma^{B=0}(P,T)+\Sigma^{B\neq 0}(P,T) \equiv \Sigma^0+\Sigma^B .
 \label{ferpropexp} 
\end{align}
 where $g$ is the QCD coupling constant, $C_{F}=4/3$ is the Casimir invariant of $SU(3)$ group, $T$ is the temperature of the system. 
 The first term is the thermal bath contribution in absence of magnetic field ($B=0)$ whereas the  second one is from the magnetised thermal bath.
 
Using \eqref{ferpropexp} in (\ref{sta}) and (\ref{stb}), the structure functions $a$ and $b$, respectively,  become 
\begin{subequations} 
 \begin{align}
 a(p_0,|\vec p|) &= \frac{1}{4}\,\,\frac{\Tr\left(\Sigma^0\slashed{P}\right)-(P.u)\,\Tr\left(\Sigma^0\slashed{u}\right)}{(P.u)^{2}-P^{2}}  , \label{ax}\\ 
b(p_0,|\vec p|)&= \frac{1}{4}\,\,\frac{-(P.u)\,\Tr\left(\Sigma^0\slashed{P}\right)+P^{2}\,\Tr\left(\Sigma^0\slashed{u}\right)}{(P.u)^{2}-P^{2}} , \label{bx} 
\end{align}
\end{subequations} 
where the contributions coming from $\Sigma^B$ vanish due to the trace of odd number of $\gamma$-matrices. 
\begin{subequations} 
Following the well known results in Ref.~\cite{weldonfermion}, one can write
 \begin{align}
 a(p_0,|\vec p|) & \, = \, -\frac{m^{2}_{th}}{|\vec{p}|^{2}}Q_{1}\left(\frac{p_{0}}{|\vec{p}|}\right), \label{a}\\ 
b(p_0,|\vec p|)&
\, = \, \frac{m^{2}_{th}}{|\vec{p}|}\left[\frac{p_{0}}{|\vec{p}|}Q_{1}\left(\frac{p_{0}}{|\vec{p}|}\right)-Q_{0}\left(\frac{p_{0}}{|\vec{p}|}\right)\right], \label{b} 
\end{align}
\end{subequations} 
where the Legendre functions of the second kind read as
\begin{subequations}
\begin{align}
Q_{0}(x) &= \frac{1}{2}\ln\left(\frac{x+1}{x-1}\right) , \label{Q0} \\
Q_{1}(x) &= x\,Q_{0}(x) -1 = \frac{x}{2}\ln\left(\frac{x+1}{x-1}\right) - 1 ,  \label{Q1}
\end{align}
\end{subequations}
and  the thermal mass~\cite{weldonfermion,Bellac:2011kqa} of the quark  is given as
\be
 m^{2}_{th}=C_{F}\frac{g^{2}T^{2}}{8}. \label{mth_q}
\ee
The thermal part of the self-energy in \eqref{ferpropexp}  becomes
\begin{align}
\Sigma^{B=0}(P,T)\equiv \Sigma^{0}(P,T) &= g^{2}\,C_{F}\,T\sumint_{K}\gamma\indices{_\mu}
\frac{\slashed{K}}{K^{2}-m^{2}}\gamma\indices{^\mu}\frac{1}{(P-K)^{2}} \nn \\
&=-a(p_0,|{\vec p}|) \slashed{P}  -b(p_0,|{\vec p}| ) \slashed{u} . \label{th_se}
\end{align}

Again using \eqref{ferpropexp} in (\ref{stbp}) and (\ref{stcp}), the structure functions $b'$ and $c'$, respectively,  become
\begin{align}
b^{\prime} &= -\frac{1}{4}\, \Tr (\slashed{u}\gamma_{5}\Sigma^B),  \label{bp}\\
c^{\prime} &= \frac{1}{4}\,\Tr(\slashed{n}\gamma_{5}\Sigma^B) , \label{cp}
\end{align} 
where the contributions coming from $\Sigma^0$ vanish due to the trace of odd number of $\gamma$-matrices.
For computing the above thermo-magnetic structure functions,  one needs to use the following two traces:
\begin{align}
\Tr \left[\slashed{u}\gamma_{5}\gamma\indices{_\mu}\gamma_{5}\left[(K.n)\slashed{u}-(K.u)\slashed{n}\right]\gamma\indices{^\mu}\right]&=  8\, (K.n)\, ,  \label{tbp} \\
\Tr \left[\slashed{n}\gamma_{5}\gamma\indices{_\mu}\gamma_{5}\left[(K.n)\slashed{u}-(K.u)\slashed{n}\right]\gamma\indices{^\mu}\right]&= 8\, (K.u) \, . \label{tcp} 
\end{align}

With this one can obtain
\begin{align}
b^{\prime} &= 2\,g^{2}\,C_{F}\,T\,q_{f}\,B\sumint_{\{K\}} (K.n)\,\Delta^{2}_{F}(K)\,\Delta_{B}(P-K) , \label{atbp}\\
c^{\prime} &= -2\,g^{2}\,C_{F}\,T\,q_{f}\,B\sumint_{\{K\}} (K.u)\,\Delta^{2}_{F}(K)\,\Delta_{B}(P-K), \label{atcp}
\end{align}
where the boson propagator in Saclay representation is given by
\begin{align*}
\Delta_{B}(K)=-\int_{0}^{\beta}d\tau e^{k_0\tau}\widetilde{\Delta}_{B}(\tau,k)
\end{align*}
and \begin{align*}
\widetilde{\Delta}_{B}(\tau,k) & = \sum_{k_0}e^{-k_0\tau}\Delta_{B}(K) \\
& = \frac{1}{2\omega_{k}}\left\lbrace \left[ 1+n_{B}(\omega_k)\right]e^{-\omega_{k}\tau}+n_{B}(\omega_{k})e^{\omega_{k}\tau}\right\rbrace
\end{align*}
where the sum is over $k_0=2\pi i nT$ and $\omega^{2}_{k}=k^2+m_f^2$.
Also the fermion propagator in Saclay representation reads
\begin{align*}
\Delta_{F}(K)=-\int_{0}^{\beta}d\tau e^{k_0\tau}\widetilde{\Delta}_{F}(\tau,k)
\end{align*}
and
\begin{align*}
\widetilde{\Delta}_{F}(\tau,k) & = \sum_{k_0}e^{-k_0\tau}\Delta_{F}(K) \\
& = \frac{1}{2\omega_{k}}\left\lbrace \left[ 1-n_{F}(\omega_k)\right]e^{-\omega_{k}\tau}-n_{F}(\omega_{k})e^{\omega_{k}\tau}\right\rbrace
\end{align*}
where the sum above is over $k_0=(2n+1)\pi i T$. 
Now following HTL approximation in presence of magnetic field~\cite{ayalafermionself,aritraweakpressure}  the \eqref{atbp} and \eqref{atcp} are simplified as
\begin{align*}
b^{\prime} 
&= -4\,g^{2}\,C_{F}\,M^{2}(T,m_f,q_{f}B)\,\int\frac{d\Omega}{4\pi}\frac{\hat{K}\cdot n}{P\cdot\hat{K}} \, ,  \\
c^{\prime} &= 4\,g^{2}\,C_{F}\,M^{2}(T,m_f,q_{f}B)\,\int\frac{d\Omega}{4\pi}\frac{\hat{K}\cdot u}{P\cdot\hat{K}} \, . 
\end{align*}
Using the results of  the HTL angular integrations~\cite{Haque:2017nxq}
\begin{align}
\int\frac{d\Omega}{4\pi}\,\frac{\hat{K}\cdot u}{P\cdot\hat{K}} &= \frac{1}{|\vec{p}|}\,Q_{0}\left(\frac{p\indices{^0}}{|\vec{p}|}\right),  \label{ang_u}\\
\int\frac{d\Omega}{4\pi}\,\frac{\hat{K}\cdot n}{P\cdot\hat{K}} &= -\frac{p\indices{^3}}{|\vec{p}|^{2}}Q_{1}\left(\frac{p\indices{^0}}{|\vec{p}|}\right) , \label{ang_n}
\end{align}
the thermo-magnetic structures functions become
\begin{align}
b^{\prime} &= 4g^{2}\,C_{F}\,M^{2}(T,m_f,q_{f}B)\,\frac{p\indices{^3}}{|\vec{p}|^{2}}Q_{1}\left(\frac{p\indices{^0}}{|\vec{p}|}\right), \label{fbp} \\
c^{\prime} &= 4g^{2}\,C_{F}\,M^{2}(T,m_f,q_{f}B)\,\frac{1}{|\vec{p}|}\,Q_{0}\left(\frac{p\indices{^0}}{|\vec{p}|}\right) \, , \label{fcp}
\end{align}
with the magnetic mass is obtained as
\begin{align}
M^{2}(T,m_f,q_{f}B) &= \frac{q_{f}B}{16\pi^{2}}\left[\ln(2)-\frac{T}{m_f}\frac{\pi}{2}\right] \, .\label{magneticmass}
\end{align}
We note here that for $m_f\rightarrow 0$, the magnetic mass diverges but it can be regulated by the the thermal mass $m_{th}$ in \eqref{mth_q} 
as is done in Refs.~\cite{Haque:2017nxq,ayalafermionself}. 
Then the domain of applicability becomes $m_{th}^2 (\sim g^2T^2)  < q_fB < T^2$ instead of $m_f^2 < q_fB < T^2$.

The  thermo-magnetic  part of the self-energy in \eqref{ferpropexp}  becomes
\begin{align}
\Sigma^{B\ne0}(P,T) \equiv \Sigma^B(P,T)= & -g^{2}\,C_{F}\,T\,q_{f}\,B\sumint \gamma\indices{_\mu}\frac{\gamma_{5}\left[(K.n)
\slashed{u}-(K.u)\slashed{n}\right]}{(K^{2}-m_f^{2})^{2}}\gamma\indices{^\mu}\frac{1}{(P-K)^{2}}  \nn \\
&= -b'(p_{0},|\vec{p}|)\gamma_5\slashed{u}-c'(p_{0},|\vec{p}|)\gamma_5 \slashed{n}.   \label{mag_se} 
\end{align}
 
Now combining \eqref{th_se}, \eqref{mag_se} and \eqref{ferpropexp}, the general structure of quark self-energy
in hot magnetised QCD  becomes		
\begin{align}
\Sigma(p_0,|\vec p|) &= -a(p_{0},|\vec{p}|)\slashed{P}-b(p_{0},|\vec{p}|)\slashed{u}-\gamma_{5}b^{\prime}(p_{0},|{\vec p}|)
\slashed{u}-\gamma_{5}c^{\prime}(p_{0},|\vec{p}|)\slashed{n}\, . \label{sigmap}
\end{align}
which agrees quite well with the general structure as discussed in \eqref{genstructselfenergy} and also with results directly
calculated in Refs.~\cite{Haque:2017nxq,ayalafermionself,aritraweakpressure}.

\section{Solution of the Modified Dirac equation at Lowest Landau Level (LLL)}
\label{app:LLLsolDirac}
\label{lll_app}
At LLL,  $l\rightarrow 0 \, \Rightarrow \, p_\perp=0$  and the effective Dirac equation becomes
\begin{align}
&\left(\mathcal{P}_{+}\slashed{L}+\mathcal{P}_{-}\slashed{R}\right)U = 0 \nonumber \\
& \begin{pmatrix}
0 && R\indices{_0}-\sigma^{3}R\indices{_z} \\
L\indices{_0}+\sigma^{3}L\indices{_z} && 0  
\end{pmatrix}U= 0 , \label{ll_sp0}
\end{align}
where  $U = \begin{pmatrix} \psi_{L} \\ \psi_{R} \end{pmatrix}$ with $\psi_{L(R)}$ are $2 \times 1$ blocks.
Now,  the condition for the non-trivial solution to exist is given as 
\begin{align}
\mbox{det}\begin{pmatrix}
0 && R\indices{_0}-\sigma^{3}R\indices{_z} \\
L\indices{_0}+\sigma^{3}L\indices{_z} && 0  
\end{pmatrix} &= 0 \nonumber \\
\left[(R\indices{_0})^{2}-(R\indices{_z})^{2}\right ]\left [(L\indices{_0})^{2}-(L\indices{_z})^{2}\right ] &= 0 \nonumber \\
\mbox{or},\,\,R\indices{_0}=\pm R\indices{_z},  \, \, \,  L\indices{_0}=\pm L\indices{_z}, \label{ll_sp1}
\end{align}

\begin{itemize}
\item {Case-I:  \,  \,  For $R\indices{_0} = R\indices{_z}$}  one can write \eqref{ll_sp0} as  
\begin{align} 
\begin{pmatrix}
0 && 0 && 0 && 0 \\
0 && 0 && 0 && 2R\indices{_z} \\
L\indices{_0}+L\indices{_z} && 0 && 0 && 0 \\
0 && L\indices{_0}-L\indices{_z} && 0 && 0
\end{pmatrix}.\begin{pmatrix}
\psi^{(1)}_{L} \\
\psi^{(2)}_{L} \\
\psi^{(1)}_{R} \\
\psi^{(2)}_{R}
\end{pmatrix} = 0 , \label{ll_sp2}
\end{align}
which leads to the following conditions:
\begin{align}
2R\indices{_z}\,\psi^{(2)}_{R} &= 0 , \nonumber \\
(L\indices{_0}+L\indices{_z})\,\psi^{(1)}_{L} &= 0 , \nonumber \\
(L\indices{_0}-L\indices{_z})\,\psi^{(2)}_{L} &= 0 , \nn \\
\psi^{(1)}_{R} &= \mbox{Arbitrary} . 
\end{align}
For normalisation, we choose  only non-zero component, \(\psi^{(1)}_{R} = 1\) which leads to
\begin{align} 
U^{(+)}_{R} = \begin{pmatrix}
0 \\
0 \\
1 \\
0
\end{pmatrix} . \label{ll_sp3}
\end{align}
 Now, for $R\indices{_0} = -R\indices{_z}$ , similarly one can obtain as 
\begin{align}
U^{(-)}_{R} = \begin{pmatrix}
0 \\
0 \\
0 \\
1
\end{pmatrix} . \label{ll_sp4}
\end{align}
\item Case-II: \, \, For $L\indices{_0} = L\indices{_z}$ , one gets 
\begin{align}
U^{(+)}_{L} = \begin{pmatrix}
0 \\
1 \\
0 \\
0
\end{pmatrix} , \label{ll_sp5}
\end{align}

whereas for $L\indices{_0} = -L\indices{_z}$ , one finds
\begin{align}
U^{(-)}_{L} = \begin{pmatrix}
1 \\
0 \\
0 \\
0
\end{pmatrix} . \label{ll_sp6}
\end{align}
\end{itemize}
\section{Verification of the Three Point Function from Direct Calculation}
\label{vert_direct}
In this appendix we would verify the general structure of the temporal  3-point function as obtained in 
sec.~\ref{vert_func} using the general
structure of the self-energy.

We begin with the  one-loop level  $3$-point function in a hot magnetised medium in~\cite{Haque:2017nxq}  
within HTL approximation~\cite{brateennucl337,Frenkel:1989br} as
\bea
\Gamma^\mu(P,K;Q) &=& \gamma^\mu +  \delta \Gamma^\mu_{\tiny \mbox{HTL}}(P,K) +  \delta \Gamma^\mu_{\tiny \mbox{TM}}(P,K) , \label{gen_3pt}
\eea
where the external four-momentum $Q=P-K$.  The HTL correction part~\cite{braatendilepton,Chakraborty:2001kx,Frenkel:1989br} is given as
\bea
\delta \Gamma^\mu_{\tiny \mbox{HTL}}(P,K)  &=& m_{th}^2 G^{\mu\nu}\gamma_\nu  
=  m_{th}^2 \int\frac{d\Omega}{4\pi}\frac{{\hat Y}^\mu{\hat Y}^\nu}{(P\cdot \hat{Y})(K\cdot \hat{Y})} \gamma_\nu
= \delta \Gamma^\mu_{\tiny \mbox{HTL}}(-P,-K) ,
\label{gen_htl_corr}
\eea
 where  ${\hat Y}_\mu = (1,\hat{y})$ is a light like four vector and the  thermo-magnetic (TM) correction part~\cite{ayalafermionself,Haque:2017nxq} is given  
\bea
\delta \Gamma^\mu_{\tiny\mbox{TM}}(P,K) = 4 \gamma_5 g^2 C_F M^2 ~ \int\frac{d\Omega}{4\pi}\frac{1}{(P\cdot \hat{Y})(K\cdot \hat{Y})}
\left [ (\hat{Y}\cdot n)\slashed{u}-(\hat{Y}\cdot u)\slashed{n}\right]  \hat{Y}^\mu \, .  \label{gen_tm_3}
\label{vertex_najmul}
\eea

 Now, choosing the temporal component 
 of the thermo-magnetic correction part of the 3-point function  and external three momentum $\vec{q}=0$, we get 
\bea
\left .  \delta \Gamma^0_{\tiny \mbox{TM}}(P,K) \right |_{{\vec q} =0} &=& \gamma_5 M'^2~ \int\frac{d\Omega}{4\pi}\frac{1}{(P\cdot \hat{Y})(K\cdot \hat{Y})}
\, \left [ (\hat{Y}\cdot n)\slashed{u}-(\hat{Y}\cdot u)\slashed{n}\right ]\nn\\
&=& \gamma_5 M'^2 ~ \int\frac{d\Omega}{4\pi}\frac{1}{(P\cdot \hat{Y})(K\cdot \hat{Y})}\, \left[(\hat{Y}\cdot n)\gamma_0+(\hat{Y}\cdot u)\gamma^3\right]
\eea
 
 Along with this following identity:
 \bea
 \left(\frac{1}{K\cdot \hat{Y}}-\frac{1}{P\cdot \hat{Y}}\right) &=& \frac{Q\cdot \hat{Y}}{(P\cdot \hat{Y})(K\cdot \hat{Y})} = \frac{q_0}{(P\cdot \hat{Y})(K\cdot \hat{Y})} , \nn 
 \eea
and ,   \eqref{ang_u} and \eqref{ang_n}, we one finally obtain
\bea
 \left. \delta \Gamma^0_{\tiny \mbox{TM}} (P,K) \right |_{{\vec q}\rightarrow}
  &=& \frac{M'^2p_z}{p^2q_0}\delta Q_1\gamma_5\gamma^0 -\frac{M'^2}{pq_0}\delta Q_0\gamma_5\gamma^3 \nn \\
  &=&  -   \frac{{M'}^2}{pq_0} \left [ \delta Q_0 \, \gamma_5 + \, \frac{p_z}{p} \,  \delta Q_1 \, (i\gamma^1\gamma^2) \right ] \, \gamma^3 \,  ,
  \label{tm_g_0}
 \eea
 where $\delta Q_n = Q_n\left(\frac{p_0}{p}\right) - Q_n\left(\frac{k_0}{p}\right)$.  We note that this expression matches exactly with the expression 
 obtained  in \eqref{wi_3pt_mag_g0} from the general structure of fermion self-energy.

\section{Analytical Solution of the Dispersion Relations and the Effective Mass in  LLL}
\label{eff_lll_mass}
The dispersion relations at LLL can be written the equations \eqref{defLsquare_l0} and \eqref{defRsquare_r0} as
\begin{subequations}
 \begin{align}
L^2_{LLL} & = \left(\mathcal{A}p_{0}+\mathcal{B}_{+}\right)^{2}-\left(\mathcal{A}p_{z} +c^{\prime}\right)^{2} = L_0^2-L_z^2=0 \, ,  \label{disp_l0}\\
R^{2}_{LLL}  &= \left(\mathcal{A}p_{0}+\mathcal{B}_{-}\right)^{2}-\left(\mathcal{A}p_{z}-c^{\prime}\right)^{2} =R_0^2-R_z^2= 0 \, , \label{disp_r0}
\end{align}
\end{subequations} 
each of which leads to two modes, respectively, as
\begin{subequations}
 \begin{align}
L_0& =  \pm L_z \nn \\
\mathcal{A}p_{0}+\mathcal{B}_{+}&= \pm \left(\mathcal{A}p_{z} +c^{\prime}\right) \, ,  \label{disp_lpm}
\end{align}
\end{subequations} 
and 
\begin{subequations}
 \begin{align}
R_0 &= \pm R_z \nn \\
 \mathcal{A}p_{0}+\mathcal{B}_{-} &=\pm \left(\mathcal{A}p_{z}-c^{\prime}\right)\, . \label{disp_rpm}
\end{align}
\end{subequations} 
Below we  try to get approximate analytical solution of these equations at small and high $p_z$ limits.  \subsection{Low $p_z$ limit}

In the low $p_{z}$ region, one needs to expand $a(p_{0},|p_{z}|)$, $b(p_{0},|p_{z}|)$, $b^{\prime}(p_{0},0,p_{z})$ and 
$c^{\prime}(p_{0},|p_{z}|)$ defined in  \eqref{at}, \eqref{bt}, \eqref{bprime} and \eqref{cprime}, respectively, 
which depend on  Legendre function 
of second kind $Q_{0}(x)$ and $Q_{1}(x)$ as given  in  equations \eqref{Q0} and \eqref{Q1}, respectively.
The  Legendre function $Q_0$ and structure coefficients are expanded in powers of $\displaystyle \frac{|p_{z}|}{p_{0}}$ as
\begin{align}
Q_0\left (\frac{p_0}{|p_z|}\right )&= \frac{|p_z|}{p_0} +\frac{1}{3}  \frac{|p_z|^3}{p^3_0} +\frac{1}{5}  \frac{|p_z|^5}{p_0^5} + \cdots  \\
a(p_{0},|p_{z}|) &= -\frac{m^{2}_{th}}{p^{2}_{0}}\,\left(\frac{1}{3}+\frac{1}{5}\,\frac{|p_{z}|^{2}}{p^{2}_{0}}+\cdots\right) \, , \label{exa}\\
b(p_{0},|p_{z}|) &= -2\,\frac{m^{2}_{th}}{p_{0}}\,\left(\frac{1}{3}+\frac{1}{15}\,\frac{|p_{z}|^{2}}{p^{2}_{0}}+\cdots\right)\, ,  \label{exb}\\
b^{\p}(p_{0},0,p_{z}) &= 4\,g^{2}\,C_{\sss{F}}\,M^{2}(T,m,qB)\,p_{z}\,\left(\frac{1}{3\,p^{2}_{0}}+\frac{|p_{z}|^{2}}{5\,p^{4}_{0}}+\cdots\right)\, , \label{exbp}\\
c^{\p}(p_{0},|p_{z}|) &= 4\,g^{2}\,C_{\sss{F}}\,M^{2}(T,m,qB)\,\left(\frac{1}{p_{0}}+\frac{|p_{z}|^{2}}{p^{3}_{0}}+\cdots\right)\, . \label{excp}
\end{align}
Now retaining the terms that are \emph{upto the order of  $p_{z}$} in  \eqref{exa}, \eqref{exb}, \eqref{exbp}, \eqref{excp}, 
we obtain the following expressions for  the dispersion relation of various modes:

\begin{enumerate}
\item $L_0=L_z$ leads to a mode $L^{(+)}$ as
\begin{align}
\omega_{L^{(+)}}(p_{z})&= m^{*+}_{\scriptscriptstyle LLL}+\frac{1}{3}\,p_{z} \, . \label{lll_l+_a}
\end{align}
\item $L_0=-L_z$ leads to a mode $L^{(-)}$ as
\begin{align}
\omega_{L^{(-)}}(p_{z})&= m^{*-}_{\scriptscriptstyle LLL}-\frac{1}{3}\,p_{z} \, . \label{lll_l+_a}
\end{align}
\item $R_0=R_z$ leads to a mode $R^{(+)}$ as
\begin{align}
\omega_{R^{(+)}}(p_{z})&= m^{*-}_{\scriptscriptstyle LLL}+\frac{1}{3}\,p_{z} \, . \label{lll_l+_a}
\end{align}
\item $R_0=-R_z$ leads to a mode $R^{(-)}$ as
\begin{align}
\omega_{R^{(-)}}(p_{z})&= m^{*+}_{\scriptscriptstyle LLL}-\frac{1}{3}\,p_{z} \, . \label{lll_l+_a}
\end{align}
\end{enumerate}
where the effective masses of various modes are given as
\begin{align}
m^{*\pm}_{\scriptscriptstyle LLL} = \begin{cases} \sqrt{m^{2}_{th}+4g^{2}C_{\scriptstyle F}M^{2}(T,M,q_{f}B)},
\qquad\text{for}\qquad L^{(+)}\,\&\, R^{(-)} , \\ \\
 \sqrt{m^{2}_{th}-4g^{2}C_{\scriptstyle F}M^{2}(T,M,q_{f}B)},
 \qquad\text{for}\qquad R^{(+)}\,\&\,L^{(-)} . \end{cases} \, \label{mp}
\end{align}

\subsection{High $p_{z}$ limit}
We note that $p_{z}$ can be written as
\begin{align*}
p_{z} = \begin{cases*}
|p_{z}|,\hspace{1cm}\text{for}\hspace{2mm}p_{z}>0 \\
-|p_{z}|.\hspace{0.8cm}\text{for}\hspace{2mm}p_{z}<0
\end{cases*}
\end{align*}
In high $p_{z}$ limit, we obtain 
\begin{enumerate}[(i)]
\item \begin{align}
[1+a(p\indices{_0},|p_{z}|)]\,(p\indices{_0}-p_{z})+b(p\indices{_0},|p_{z}|) = \begin{cases}
p\indices{_0}-|p_{z}|-\frac{m^2_{th}}{|p_{z}|},\hspace{3.7cm}\text{for}\hspace{2mm}p_z>0\\
2\,|p_{z}|+\frac{m^{2}_{th}}{|p_{z}|}-\frac{m^{2}_{th}}{|p_{z}|}\,\ln\left(\frac{2\,|p_{z}|}{p\indices{_0}-
|p_{z}|}\right),\hspace{1cm}\text{for}\hspace{2mm}p_z<0
\end{cases}
\end{align}
\item \begin{align}
[1+a(p\indices{_0},|p_{z}|)]\,(p\indices{_0}+p_{z})+b(p\indices{_0},|p_{z}|)=\begin{cases}
2\,|p_{z}|+\frac{m^{2}_{th}}{|p_{z}|}-\frac{m^{2}_{th}}{|p_{z}|}\,\ln\left(\frac{2\,|p_{z}|}{p\indices{_0}-
|p_{z}|}\right),\hspace{1cm}\text{for}\hspace{2mm}p_z>0\\
p\indices{_0}-|p_{z}|-\frac{m^2_{th}}{|p_{z}|},\hspace{3.7cm}\text{for}\hspace{2mm}p_z<0
\end{cases}
\end{align}
\item \begin{align}
b^{\p}(p\indices{_0},0,p_{z})+c^{\p}(p\indices{_0},|p_{z}|) = \begin{cases}
\frac{4g^{2}C_{\sss{F}}M^{2}}{|p_{z}|}\,\ln\left(\frac{2|p_{z}|}{p\indices{_0}-|p_{z}|}\right)-
\frac{4g^{2}C_{\sss{F}}M^{2}}{|p_{z}|},\hspace{2cm}\text{for}\hspace{2mm}p_{z}>0\\
\frac{4g^{2}C_{\sss{F}}M^{2}}{|p_{z}|}\hspace{6cm}\text{for}\hspace{2mm}p_{z}<0
\end{cases}
\end{align}
\item
\begin{align}
b^{\p}(p\indices{_0},0,p_{z})-c^{\p}(p\indices{_0},|p_{z}|) = \begin{cases}
-\frac{4g^{2}C_{\sss{F}}M^{2}}{|p_{z}|}\hspace{6cm}\text{for}\hspace{2mm}p_{z}>0\\
-\frac{4g^{2}C_{\sss{F}}M^{2}}{|p_{z}|}\,\ln\left(\frac{2|p_{z}|}{p\indices{_0}-|p_{z}|}\right)+
\frac{4g^{2}C_{\sss{F}}M^{2}}{|p_{z}|}.\hspace{2cm}\text{for}\hspace{2mm}p_{z}<0
\end{cases}
\end{align}
\end{enumerate}

\begin{enumerate}
\item $L_0=L_z$ leads to a mode $L^{(+)}$:

For ${p_{z}>0}$,
\begin{align}
\omega_{L^{(+)}}(p_{z}) = |p_{z}|+\frac{(m^{*+}_{\scriptscriptstyle LLL})^2}{|p_{z}|} .
\end{align}
For ${p_{z}<0}$,
\begin{align}
\omega_{L^{(+)}}(p_{z}) =|p_{z}|+\frac{2\,|p_{z}|}{e}\,\exp\left(-\frac{2\,p^{2}_{z}}{(m^{*+}_{\scriptscriptstyle LLL})^2}\right)\,  .
\end{align}

\item $L_0=-L_z$ leads to a mode $L^{(-)}$:

For ${p_{z}>0}$,
\begin{align}
\omega_{L^{(-)}}(p_{z}) =|p_{z}|+\frac{2\,|p_{z}|}{e}\,\exp\left(-\frac{2\,p^{2}_{z}}{(m^{*-}_{\scriptscriptstyle LLL})^2}\right)\, .
\end{align}
For ${p_{z}<0}$,
\begin{align}
\omega_{L^{(-)}}(p_{z}) = |p_{z}|+\frac{(m^{*-}_{\scriptscriptstyle LLL})^2}{|p_{z}|}\, .
\end{align}

\item $R_0=R_z$ leads to a mode $R^{(+)}$:

For ${p_{z}>0}$,
\begin{align}
\omega_{R^{(+)}}(p_{z}) = |p_{z}|+\frac{(m^{*-}_{\scriptscriptstyle LLL})^2}{|p_{z}|} \, .
\end{align}
For ${p_{z}<0}$,
\begin{align}
\omega_{R^{(+)}}(p_{z}) =|p_{z}|+\frac{2\,|p_{z}|}{e}\,\exp\left(-\frac{2\,p^{2}_{z}}{(m^{*-}_{\scriptscriptstyle LLL})^2}\right)\, .
\end{align}
\item $R_0=-R_z$ leads to a mode $R^{(-)}$:

For ${p_{z}>0}$,
\begin{align}
\omega_{R^{(-)}}(p_{z}) =|p_{z}|+\frac{2\,|p_{z}|}{e}\,\exp\left(-\frac{2\,p^{2}_{z}}{(m^{*+}_{\scriptscriptstyle LLL})^2}\right)\, .
\end{align}
For ${p_{z}<0}$,
\begin{align}
\omega_{R^{(-)}}(p_{z}) = |p_{z}|+\frac{(m^{*+}_{\scriptscriptstyle LLL})^2}{|p_{z}|}\, .
\end{align}
\end{enumerate}
Note that In the high momentum limit the above dispersion relations are given in terms of absolute 
values of $p_{z}$, i.e. $|p_{z}|$. 

We further note that  the above dispersion relations in the absence of the magnetic field 
reduce to  HTL results, where left and right handed are
degenerate. 

\section{Recovering HTL Spectral Function}
\label{spec_htl}
  
 One can easily get back to the HTL thermal spectral function from \eqref{prop_spec} by turning off the magnetic field,  \textit{i.e.}, 
  $B=0 \, \Rightarrow \, b'=c'=0$ and one gets the following simplifications:
 \bea
 g_L^1\Big|_{B=0}&=&g_R^1\Big|_{B=0}=g^1 ;\, \, \,   g_L^2\Big|_{B=0}=g_R^2\Big|_{B=0}=g^2 ; \, \, \,  g_L^3\Big|_{B=0}=g_R^3\Big|_{B=0}=0 , \\
L^2\Big|_{B=0} &=& R^2\Big|_{B=0} = H^2; \, \, \,  \omega_{L^{(\pm)}} \Big|_{B=0}=  \omega_{R^{(\pm)}}\Big|_{B=0} =\omega_\pm ,  \\
\rho_L^1\Big|_{B=0} &=& \rho_R^1\Big|_{B=0} = \rho^1; \, \, \, \rho_L^2\Big|_{B=0} = \rho_R^2 \Big|_{B=0} =\rho^2 \, , \, \, \,  \rho^3 \Big|_{B=0}=0 \, .
 \label{reduced_quant}
 \eea
 These implies that the spectral function can be written as
 \bea
 \rho\Big|_{B=0} = \gamma^0\rho^1 - (\gamma \cdot \hat{p})\rho^2  \, .     \label{reduced_htl}
 \eea
 Now the HTL spectral function~\cite{braatendilepton,Karsch:2000gi}  is given by
 \bea
 \rho_{\mathrm{HTL}} &=& \frac{1}{2}(\gamma^0-\gamma\cdot\hat{p})\rho_+ +\frac{1}{2} (\gamma^0+\gamma\cdot\hat{p})\rho_-   \nn\\
 &=&\frac{1}{2} \gamma^0 (\rho_++\rho_-) - \frac{1}{2} (\gamma \cdot \hat{p}) (\rho_+-\rho_-)  \, , \label{htl_spec}
 \eea
 where $\rho_\pm$  are the HTL spectral function.
Since the spectral has both pole and cut part,  comparing  \eqref{reduced_htl} and \eqref{htl_spec} one gets for the pole parts as
 \bea 
 \rho^1 \Big|^{\text{B=0}}_{\text{pole}} &=&  \frac{1}{2}\left[\rho_+\Big|^{\text{HTL}}_{\text{pole}}+\rho_-\Big|^{\text{HTL}}_{\text{pole}} \right]\, \label{pole1}\\ 
 \rho^2 \Big|^{\text{B=0}}_{\text{pole}} &=&  \frac{1}{2}\left[\rho_+\Big|^{\text{HTL}}_{\text{pole}}-\rho_-\Big|^{\text{HTL}}_{\text{pole}}\right ]\,  . \label{pole2} 
 \eea 
 and for the cut parts as 
  \bea 
 \beta^1 \Big|^{\text{B=0}}_{\text{cut}} &=&  \frac{1}{2}\left[\beta_+\Big|^{\text{HTL}}_{\text{cut}}+\beta_-\Big|^{\text{HTL}}_{\text{cut}} \right]\, \label{cut1}\\ 
 \beta^2 \Big|^{\text{B=0}}_{\text{cut}} &=&  \frac{1}{2}\left[\beta_+\Big|^{\text{HTL}}_{\text{cut}}-\beta_-\Big|^{\text{HTL}}_{\text{cut}}\right]\,  . \label{cut2} 
 \eea 
Now one can obtain  either \eqref{spec_li} or \eqref{spec_residue} 
 \bea
 \rho^1 \Big|^{\text{B=0}}_{\text{pole}}  &=& \left[Z^{i+}_+\delta(p_0-\omega_+)+Z^{i-}_+\delta(p_0+\omega_+)\right]
 +\left[Z_-^{1+}\delta(p_0-\omega_-)+Z_-^{1-}\delta(p_0+\omega_-)\right] . \label{spec_rho1_b0}
 \eea
 When the magnetic field is turned off, the different residues can be read from 
 their analytical expressions as given in sec.~\ref{spec_rep} as
 \bea
 Z^{1+}_+ = Z^{1-}_+ &=& \frac{\omega^2_+-p^2}{4m_{th}^2} = \frac{1}{2}Z_+ \, , \\
 Z^{1+}_- = Z^{1+}_- &=& \frac{\omega^2_--p^2}{4m_{th}^2} =\frac{1}{2} Z_-  \, ,
 \eea
 where,  $Z_\pm= ({\omega_\pm}^2-p^2)/2m_{th}^2$, are the residues corresponding to the  modes $\omega^\pm$.  
 Using these one can write \eqref{spec_rho1_b0} as
 \bea
 \rho^1 \Big|^{\text{B=0}}_{\text{pole}}  &=&\frac{1}{2} \left[Z_+\delta(p_0-\omega_+)+Z_+\delta(p_0+\omega_+)\right]
 +\frac{1}{2} \left[Z_-\delta(p_0-\omega_-)+Z_-\delta(p_0+\omega_-)\right]  \nn \\
 &=&\frac{1}{2} \left[Z_+\delta(p_0-\omega_+)+Z_-\delta(p_0+\omega_-)\right]
 +\frac{1}{2} \left[Z_-\delta(p_0-\omega_-)+Z_+\delta(p_0+\omega_+)\right]  \nn \\
 &=& \frac{1}{2}\left[ \rho_+\Big|^{\text{HTL}}_{\text{pole}}+\rho_-\Big|^{\text{HTL}}_{\text{pole}}\right]\, , \label{spec_rho1}
 \eea 
 which agree with \eqref{pole1}.
 
 Now the other non-zero component of the spectral function can be reduced as
 \bea
 \rho^2 \Big|^{\text{B=0}}_{\text{pole}}  &=& \left[Z_+^{2+}\delta(p_0-\omega_+)+Z_+^{2-}\delta(p_0+\omega_+)\right]
+\left[Z_-^{2+}\delta(p_0-\omega_-)+Z_-^{2-}\delta(p_0+\omega_-)\right] \, ,
 \label{spec_rho2_b0}
 \eea
along with  the remaining non-zero residues  as
 \bea
 Z^{2+}_+ = - Z^{2-}_+&=& \frac{\omega^2_+-p^2}{4m_{th}^2} \times \frac{
\omega_+m_{th}^2\log\left(\frac{\omega_++p}{\omega_+-p}\right)-2p(m_{th}^2+p^2)}{2p^2\omega_+-p~m_{th}^2\log\left(\frac{\omega_++p}{\omega^+-p}\right)}\nn\\
 &=&  \frac{\omega^2_+-p^2}{4m_{th}^2} =\frac{1}{2} Z_+ ,  \label{z+}
 \eea
 
\noindent and
 \bea
 Z^{2+}_- = - Z^{2-}_-  &=& \frac{\omega^2_- -p^2}{4m_{th}^2} \times 
 \frac{\omega_-m_{th}^2\log\left(\frac{\omega_-+p}{\omega^--p}\right)-2p(m_{th}^2+p^2)}{2p^2\omega_--p~m_{th}^2\log\left(\frac{\omega_-+p}{\omega_--p}\right)}\nn\\
 &=&  \frac{\omega^2_- -p^2}{4m_{th}^2} \, = \frac{1}{2} Z_- \, . \label{z-}  
 \eea
 Note that we have used the respective dispersion relations coming from $H^2=0$,  in the last line of \eqref{z+} and \eqref{z-} for further simplifications.
 Now \eqref{spec_rho2_b0} can be rewritten as
 \bea
 \rho^2 \Big|^{\text{B=0}}_{\text{pole}} &=& \frac{1}{2} \left[Z_+\delta(p_0-\omega_+)-Z_+\delta(p_0+\omega_+)\right]- \frac{1}{2} 
 \left[Z_-\delta(p_0-\omega_-)-Z_- \delta(p_0+\omega_-)\right] \,  \nn \\ 
  &=& \frac{1}{2} \left[Z_+\delta(p_0-\omega_+)+Z_-\delta(p_0+\omega_-)\right]- \frac{1}{2} 
 \left[Z_-\delta(p_0-\omega_-)+Z_+\delta(p_0+\omega_+)\right] \,  \nn \\ 
  &=& \frac{1}{2}\left[\rho_+\Big|^{\text{HTL}}_{\text{pole}}-\rho_-\Big|^{\text{HTL}}_{\text{pole}}\right] \, , \label{spec_rho2}
 \eea
 which agree with \eqref{pole2}. 
 
In absence of magnetic field one write the cut parts from  the general expression of $\beta_L^i$ in \eqref{cut_LR_i} as
 \bea
 \beta^1 &=& \frac{1}{\pi}~\Theta(p^2-p_0^2)~ \frac{\mathrm{Im}(g^1)~\mathrm{Re}(H^2) - \mathrm{Im}(H^2)\mathrm{Re}(g^1)}{|H^2|^2}, \\
  \beta^2 &=& \frac{1}{\pi}~\Theta(p^2-p_0^2)~ \frac{\mathrm{Im}(g^2)~\mathrm{Re}(H^2) - \mathrm{Im}(H^2)\mathrm{Re}(g^2)}{|H^2|^2}, 
 \eea
where for zero magnetic field case 
\bea
H^2 = L^2\Big|_{B=0} = R^2\Big|_{B=0} = (g^1+g^2)(g^1-g^2) = H_-H_+ \, ,
\eea
where following the same convention as before $H_-=g^1+g^2$ and $H_+=g^1-g^2$.

The real and imaginary parts of $H^2$ can be written in terms of $H_-$ and $H_+$ as 
 \bea
 \mathrm{Re}(H^2)+i~\mathrm{Im}(H^2) &=& (\mathrm{Re}(H_-)+i~\mathrm{Im}(H_-))(\mathrm{Re}(H_+)+i~\mathrm{Im}(H_+))\nn\\
 &=& \left[\mathrm{Re}(H_-)\mathrm{Re}(H_+)-\mathrm{Im}(H_-)\mathrm{Im}(H_+)\right]\nn \\
 && +i~\left[\mathrm{Re}(H_-)\mathrm{Im}(H_+)+\mathrm{Re}(H_+)\mathrm{Im}(H_-)\right] \, . 
 \eea
 
Now we can write down
 \bea
 \beta^1+\beta^2 &=& \frac{1}{\pi}~\Theta(p^2-p_0^2)~ \frac{(\mathrm{Im}(g^1)+\mathrm{Im}(g^2))~\mathrm{Re}(H^2) - \mathrm{Im}(H^2)(\mathrm{Re}(g^1)+\mathrm{Re}(g^2))}{|H_-|^2|H_+|^2}\nn \\
 &=& \frac{1}{\pi}~\Theta(p^2-p_0^2)~ \frac{\mathrm{Im}(H_-)~\mathrm{Re}(H^2) - \mathrm{Im}(H^2)\mathrm{Re}(H_-)}{|H_1|^2|H_2|^2}\nn \\
 &=&-~\frac{1}{\pi}~\Theta(p^2-p_0^2)~\frac{\mathrm{Im}(H_+)}{|H_+|^2} \nn\\
 &=& -~\frac{\frac{1}{2p}m_{th}^2\left(1-\frac{p_0}{p}\right)}{\left[\left\{p_0-p+\frac{m_{th}^2}{2p}\left(\left(1-\frac{p_0}{p}\right)\log\Big|\frac{p_0+p}{p_0-p}\Big|+2\right)\right\}^2+\frac{\pi^2m_{th}^4}{4p^2}\left(1-\frac{p_0}{p}\right)^2\right]}\nn\\
 &=& \beta_+ \, . \label{cut+}
 \eea
 and similarly
 \bea
 \beta^1-\beta^2  &=&-~\frac{1}{\pi}~\Theta(p^2-p_0^2)~\frac{\mathrm{Im}(H_-)}{|H_-|^2},\nn\\
 &=& -~\frac{\frac{1}{2p}m_{th}^2\left(1+\frac{p_0}{p}\right)}{\left[\left\{p_0+p-\frac{m_{th}^2}{2p}\left(\left(1+\frac{p_0}{p}\right)\log\Big|\frac{p_0+p}{p_0-p}\Big|-2\right)\right\}^2+\frac{\pi^2m_{th}^4}{4p^2}\left(1+\frac{p_0}{p}\right)^2\right]},\nn\\
 &=& \beta_- \, , \label{cut-}
 \eea
where both  \eqref{cut+} and \eqref{cut-} agree with the HTL cut parts~\cite{braatendilepton,Karsch:2000gi}.

\bibliography{prd_submit.bib}

\begin{thebibliography}{75}%
\makeatletter
\providecommand \@ifxundefined [1]{%
 \@ifx{#1\undefined}
}%
\providecommand \@ifnum [1]{%
 \ifnum #1\expandafter \@firstoftwo
 \else \expandafter \@secondoftwo
 \fi
}%
\providecommand \@ifx [1]{%
 \ifx #1\expandafter \@firstoftwo
 \else \expandafter \@secondoftwo
 \fi
}%
\providecommand \natexlab [1]{#1}%
\providecommand \enquote  [1]{``#1''}%
\providecommand \bibnamefont  [1]{#1}%
\providecommand \bibfnamefont [1]{#1}%
\providecommand \citenamefont [1]{#1}%
\providecommand \href@noop [0]{\@secondoftwo}%
\providecommand \href [0]{\begingroup \@sanitize@url \@href}%
\providecommand \@href[1]{\@@startlink{#1}\@@href}%
\providecommand \@@href[1]{\endgroup#1\@@endlink}%
\providecommand \@sanitize@url [0]{\catcode `\\12\catcode `\$12\catcode
  `\&12\catcode `\#12\catcode `\^12\catcode `\_12\catcode `\%12\relax}%
\providecommand \@@startlink[1]{}%
\providecommand \@@endlink[0]{}%
\providecommand \url  [0]{\begingroup\@sanitize@url \@url }%
\providecommand \@url [1]{\endgroup\@href {#1}{\urlprefix }}%
\providecommand \urlprefix  [0]{URL }%
\providecommand \Eprint [0]{\href }%
\providecommand \doibase [0]{http://dx.doi.org/}%
\providecommand \selectlanguage [0]{\@gobble}%
\providecommand \bibinfo  [0]{\@secondoftwo}%
\providecommand \bibfield  [0]{\@secondoftwo}%
\providecommand \translation [1]{[#1]}%
\providecommand \BibitemOpen [0]{}%
\providecommand \bibitemStop [0]{}%
\providecommand \bibitemNoStop [0]{.\EOS\space}%
\providecommand \EOS [0]{\spacefactor3000\relax}%
\providecommand \BibitemShut  [1]{\csname bibitem#1\endcsname}%
\let\auto@bib@innerbib\@empty
\bibitem [{\citenamefont {Adare}\ \emph {et~al.}(2012)\citenamefont {Adare}
  \emph {et~al.}}]{phenixphoton}%
  \BibitemOpen
  \bibfield  {author} {\bibinfo {author} {\bibfnamefont {A.}~\bibnamefont
  {Adare}} \emph {et~al.} (\bibinfo {collaboration} {PHENIX}),\ }\href
  {\doibase 10.1103/PhysRevLett.109.122302} {\bibfield  {journal} {\bibinfo
  {journal} {Phys. Rev. Lett.}\ }\textbf {\bibinfo {volume} {109}},\ \bibinfo
  {pages} {122302} (\bibinfo {year} {2012})},\ \Eprint
  {http://arxiv.org/abs/1105.4126} {arXiv:1105.4126 [nucl-ex]} \BibitemShut
  {NoStop}%
\bibitem [{\citenamefont {Skokov}\ \emph {et~al.}(2009)\citenamefont {Skokov},
  \citenamefont {Illarionov},\ and\ \citenamefont {Toneev}}]{VSkokov2009}%
  \BibitemOpen
  \bibfield  {author} {\bibinfo {author} {\bibfnamefont {V.}~\bibnamefont
  {Skokov}}, \bibinfo {author} {\bibfnamefont {A.~{\relax Yu}.}\ \bibnamefont
  {Illarionov}}, \ and\ \bibinfo {author} {\bibfnamefont {V.}~\bibnamefont
  {Toneev}},\ }\href {\doibase 10.1142/S0217751X09047570} {\bibfield  {journal}
  {\bibinfo  {journal} {Int. J. Mod. Phys.}\ }\textbf {\bibinfo {volume}
  {A24}},\ \bibinfo {pages} {5925} (\bibinfo {year} {2009})},\ \Eprint
  {http://arxiv.org/abs/0907.1396} {arXiv:0907.1396 [nucl-th]} \BibitemShut
  {NoStop}%
\bibitem [{\citenamefont {de~la Incera}(2011)}]{VdelaIncera}%
  \BibitemOpen
  \bibfield  {author} {\bibinfo {author} {\bibfnamefont {V.}~\bibnamefont
  {de~la Incera}},\ }\bibfield  {booktitle} {\emph {\bibinfo {booktitle}
  {{Proceedings, 12th Mexican Workshop on Particles and fields (MWPF 2009):
  Mazatlan, Mexico, November 9-14, 2009}}},\ }\href {\doibase
  10.1063/1.3622687} {\bibfield  {journal} {\bibinfo  {journal} {AIP Conf.
  Proc.}\ }\textbf {\bibinfo {volume} {1361}},\ \bibinfo {pages} {74} (\bibinfo
  {year} {2011})},\ \Eprint {http://arxiv.org/abs/1004.4931} {arXiv:1004.4931
  [hep-ph]} \BibitemShut {NoStop}%
\bibitem [{\citenamefont {Bandyopadhyay}\ \emph {et~al.}(1997)\citenamefont
  {Bandyopadhyay}, \citenamefont {Chakrabarty},\ and\ \citenamefont
  {Pal}}]{Bandyopadhyay:1997kh}%
  \BibitemOpen
  \bibfield  {author} {\bibinfo {author} {\bibfnamefont {D.}~\bibnamefont
  {Bandyopadhyay}}, \bibinfo {author} {\bibfnamefont {S.}~\bibnamefont
  {Chakrabarty}}, \ and\ \bibinfo {author} {\bibfnamefont {S.}~\bibnamefont
  {Pal}},\ }\href {\doibase 10.1103/PhysRevLett.79.2176} {\bibfield  {journal}
  {\bibinfo  {journal} {Phys. Rev. Lett.}\ }\textbf {\bibinfo {volume} {79}},\
  \bibinfo {pages} {2176} (\bibinfo {year} {1997})},\ \Eprint
  {http://arxiv.org/abs/astro-ph/9703066} {arXiv:astro-ph/9703066 [astro-ph]}
  \BibitemShut {NoStop}%
\bibitem [{\citenamefont {Chakrabarty}\ \emph {et~al.}(1997)\citenamefont
  {Chakrabarty}, \citenamefont {Bandyopadhyay},\ and\ \citenamefont
  {Pal}}]{Chakrabarty:1997ef}%
  \BibitemOpen
  \bibfield  {author} {\bibinfo {author} {\bibfnamefont {S.}~\bibnamefont
  {Chakrabarty}}, \bibinfo {author} {\bibfnamefont {D.}~\bibnamefont
  {Bandyopadhyay}}, \ and\ \bibinfo {author} {\bibfnamefont {S.}~\bibnamefont
  {Pal}},\ }\href {\doibase 10.1103/PhysRevLett.78.2898} {\bibfield  {journal}
  {\bibinfo  {journal} {Phys. Rev. Lett.}\ }\textbf {\bibinfo {volume} {78}},\
  \bibinfo {pages} {2898} (\bibinfo {year} {1997})},\ \Eprint
  {http://arxiv.org/abs/astro-ph/9703034} {arXiv:astro-ph/9703034 [astro-ph]}
  \BibitemShut {NoStop}%
\bibitem [{\citenamefont {Andersen}\ \emph {et~al.}(2016)\citenamefont
  {Andersen}, \citenamefont {Naylor},\ and\ \citenamefont
  {Tranberg}}]{andersenphase}%
  \BibitemOpen
  \bibfield  {author} {\bibinfo {author} {\bibfnamefont {J.~O.}\ \bibnamefont
  {Andersen}}, \bibinfo {author} {\bibfnamefont {W.~R.}\ \bibnamefont
  {Naylor}}, \ and\ \bibinfo {author} {\bibfnamefont {A.}~\bibnamefont
  {Tranberg}},\ }\href {\doibase 10.1103/RevModPhys.88.025001} {\bibfield
  {journal} {\bibinfo  {journal} {Rev. Mod. Phys.}\ }\textbf {\bibinfo {volume}
  {88}},\ \bibinfo {pages} {025001} (\bibinfo {year} {2016})},\ \Eprint
  {http://arxiv.org/abs/1411.7176} {arXiv:1411.7176 [hep-ph]} \BibitemShut
  {NoStop}%
\bibitem [{\citenamefont {Basar}\ \emph {et~al.}(2012)\citenamefont {Basar},
  \citenamefont {Kharzeev}, \citenamefont {Kharzeev},\ and\ \citenamefont
  {Skokov}}]{Basar:2012bp}%
  \BibitemOpen
  \bibfield  {author} {\bibinfo {author} {\bibfnamefont {G.}~\bibnamefont
  {Basar}}, \bibinfo {author} {\bibfnamefont {D.}~\bibnamefont {Kharzeev}},
  \bibinfo {author} {\bibfnamefont {D.}~\bibnamefont {Kharzeev}}, \ and\
  \bibinfo {author} {\bibfnamefont {V.}~\bibnamefont {Skokov}},\ }\href
  {\doibase 10.1103/PhysRevLett.109.202303} {\bibfield  {journal} {\bibinfo
  {journal} {Phys. Rev. Lett.}\ }\textbf {\bibinfo {volume} {109}},\ \bibinfo
  {pages} {202303} (\bibinfo {year} {2012})},\ \Eprint
  {http://arxiv.org/abs/1206.1334} {arXiv:1206.1334 [hep-ph]} \BibitemShut
  {NoStop}%
\bibitem [{\citenamefont {Sadooghi}\ and\ \citenamefont
  {Taghinavaz}(2017)}]{sadooghidilepton}%
  \BibitemOpen
  \bibfield  {author} {\bibinfo {author} {\bibfnamefont {N.}~\bibnamefont
  {Sadooghi}}\ and\ \bibinfo {author} {\bibfnamefont {F.}~\bibnamefont
  {Taghinavaz}},\ }\href {\doibase 10.1016/j.aop.2016.11.008} {\bibfield
  {journal} {\bibinfo  {journal} {Annals Phys.}\ }\textbf {\bibinfo {volume}
  {376}},\ \bibinfo {pages} {218} (\bibinfo {year} {2017})},\ \Eprint
  {http://arxiv.org/abs/1601.04887} {arXiv:1601.04887 [hep-ph]} \BibitemShut
  {NoStop}%
\bibitem [{\citenamefont {Bandyopadhyay}\ \emph {et~al.}(2016)\citenamefont
  {Bandyopadhyay}, \citenamefont {Islam},\ and\ \citenamefont
  {Mustafa}}]{aritraemspectrum}%
  \BibitemOpen
  \bibfield  {author} {\bibinfo {author} {\bibfnamefont {A.}~\bibnamefont
  {Bandyopadhyay}}, \bibinfo {author} {\bibfnamefont {C.~A.}\ \bibnamefont
  {Islam}}, \ and\ \bibinfo {author} {\bibfnamefont {M.~G.}\ \bibnamefont
  {Mustafa}},\ }\href {\doibase 10.1103/PhysRevD.94.114034} {\bibfield
  {journal} {\bibinfo  {journal} {Phys. Rev.}\ }\textbf {\bibinfo {volume}
  {D94}},\ \bibinfo {pages} {114034} (\bibinfo {year} {2016})},\ \Eprint
  {http://arxiv.org/abs/1602.06769} {arXiv:1602.06769 [hep-ph]} \BibitemShut
  {NoStop}%
\bibitem [{\citenamefont {Bandyopadhyay}\ and\ \citenamefont
  {Mallik}(2017)}]{aritradilepton}%
  \BibitemOpen
  \bibfield  {author} {\bibinfo {author} {\bibfnamefont {A.}~\bibnamefont
  {Bandyopadhyay}}\ and\ \bibinfo {author} {\bibfnamefont {S.}~\bibnamefont
  {Mallik}},\ }\href {\doibase 10.1103/PhysRevD.95.074019} {\bibfield
  {journal} {\bibinfo  {journal} {Phys. Rev.}\ }\textbf {\bibinfo {volume}
  {D95}},\ \bibinfo {pages} {074019} (\bibinfo {year} {2017})},\ \Eprint
  {http://arxiv.org/abs/1704.01364} {arXiv:1704.01364 [hep-ph]} \BibitemShut
  {NoStop}%
\bibitem [{\citenamefont {Tuchin}(2013{\natexlab{a}})}]{Tuchin:2013apa}%
  \BibitemOpen
  \bibfield  {author} {\bibinfo {author} {\bibfnamefont {K.}~\bibnamefont
  {Tuchin}},\ }\href {\doibase 10.1103/PhysRevC.88.024911} {\bibfield
  {journal} {\bibinfo  {journal} {Phys. Rev.}\ }\textbf {\bibinfo {volume}
  {C88}},\ \bibinfo {pages} {024911} (\bibinfo {year} {2013}{\natexlab{a}})},\
  \Eprint {http://arxiv.org/abs/1305.5806} {arXiv:1305.5806 [hep-ph]}
  \BibitemShut {NoStop}%
\bibitem [{\citenamefont {Tuchin}(2013{\natexlab{b}})}]{Tuchin:2012mf}%
  \BibitemOpen
  \bibfield  {author} {\bibinfo {author} {\bibfnamefont {K.}~\bibnamefont
  {Tuchin}},\ }\href {\doibase 10.1103/PhysRevC.87.024912} {\bibfield
  {journal} {\bibinfo  {journal} {Phys. Rev.}\ }\textbf {\bibinfo {volume}
  {C87}},\ \bibinfo {pages} {024912} (\bibinfo {year} {2013}{\natexlab{b}})},\
  \Eprint {http://arxiv.org/abs/1206.0485} {arXiv:1206.0485 [hep-ph]}
  \BibitemShut {NoStop}%
\bibitem [{\citenamefont {Tuchin}(2013{\natexlab{c}})}]{Tuchin:2013bda}%
  \BibitemOpen
  \bibfield  {author} {\bibinfo {author} {\bibfnamefont {K.}~\bibnamefont
  {Tuchin}},\ }\href {\doibase 10.1103/PhysRevC.88.024910} {\bibfield
  {journal} {\bibinfo  {journal} {Phys. Rev.}\ }\textbf {\bibinfo {volume}
  {C88}},\ \bibinfo {pages} {024910} (\bibinfo {year} {2013}{\natexlab{c}})},\
  \Eprint {http://arxiv.org/abs/1305.0545} {arXiv:1305.0545 [nucl-th]}
  \BibitemShut {NoStop}%
\bibitem [{\citenamefont {Kharzeev}(2014)}]{dmitrikharzeevCME}%
  \BibitemOpen
  \bibfield  {author} {\bibinfo {author} {\bibfnamefont {D.~E.}\ \bibnamefont
  {Kharzeev}},\ }\href {\doibase 10.1016/j.ppnp.2014.01.002} {\bibfield
  {journal} {\bibinfo  {journal} {Prog. Part. Nucl. Phys.}\ }\textbf {\bibinfo
  {volume} {75}},\ \bibinfo {pages} {133} (\bibinfo {year} {2014})},\ \Eprint
  {http://arxiv.org/abs/1312.3348} {arXiv:1312.3348 [hep-ph]} \BibitemShut
  {NoStop}%
\bibitem [{\citenamefont {Fukushima}(2013)}]{FukushimaCME}%
  \BibitemOpen
  \bibfield  {author} {\bibinfo {author} {\bibfnamefont {K.}~\bibnamefont
  {Fukushima}},\ }\href {\doibase 10.1007/978-3-642-37305-3_9} {\bibfield
  {journal} {\bibinfo  {journal} {Lect. Notes Phys.}\ }\textbf {\bibinfo
  {volume} {871}},\ \bibinfo {pages} {241} (\bibinfo {year} {2013})},\ \Eprint
  {http://arxiv.org/abs/1209.5064} {arXiv:1209.5064 [hep-ph]} \BibitemShut
  {NoStop}%
\bibitem [{\citenamefont {Basar}\ and\ \citenamefont {Dunne}(2013)}]{CMEaxial}%
  \BibitemOpen
  \bibfield  {author} {\bibinfo {author} {\bibfnamefont {G.}~\bibnamefont
  {Basar}}\ and\ \bibinfo {author} {\bibfnamefont {G.~V.}\ \bibnamefont
  {Dunne}},\ }\href {\doibase 10.1007/978-3-642-37305-3_10} {\bibfield
  {journal} {\bibinfo  {journal} {Lect. Notes Phys.}\ }\textbf {\bibinfo
  {volume} {871}},\ \bibinfo {pages} {261} (\bibinfo {year} {2013})},\ \Eprint
  {http://arxiv.org/abs/1207.4199} {arXiv:1207.4199 [hep-th]} \BibitemShut
  {NoStop}%
\bibitem [{\citenamefont {Gusynin}\ \emph {et~al.}(1996)\citenamefont
  {Gusynin}, \citenamefont {Miransky},\ and\ \citenamefont
  {Shovkovy}}]{miranskydimred}%
  \BibitemOpen
  \bibfield  {author} {\bibinfo {author} {\bibfnamefont {V.~P.}\ \bibnamefont
  {Gusynin}}, \bibinfo {author} {\bibfnamefont {V.~A.}\ \bibnamefont
  {Miransky}}, \ and\ \bibinfo {author} {\bibfnamefont {I.~A.}\ \bibnamefont
  {Shovkovy}},\ }\href {\doibase 10.1016/0550-3213(96)00021-1} {\bibfield
  {journal} {\bibinfo  {journal} {Nucl. Phys.}\ }\textbf {\bibinfo {volume}
  {B462}},\ \bibinfo {pages} {249} (\bibinfo {year} {1996})},\ \Eprint
  {http://arxiv.org/abs/hep-ph/9509320} {arXiv:hep-ph/9509320 [hep-ph]}
  \BibitemShut {NoStop}%
\bibitem [{\citenamefont {Shovkovy}(2013)}]{igormagneticcatalysis}%
  \BibitemOpen
  \bibfield  {author} {\bibinfo {author} {\bibfnamefont {I.~A.}\ \bibnamefont
  {Shovkovy}},\ }\href {\doibase 10.1007/978-3-642-37305-3_2} {\bibfield
  {journal} {\bibinfo  {journal} {Lect. Notes Phys.}\ }\textbf {\bibinfo
  {volume} {871}},\ \bibinfo {pages} {13} (\bibinfo {year} {2013})},\ \Eprint
  {http://arxiv.org/abs/1207.5081} {arXiv:1207.5081 [hep-ph]} \BibitemShut
  {NoStop}%
\bibitem [{\citenamefont {Preis}\ \emph {et~al.}(2013)\citenamefont {Preis},
  \citenamefont {Rebhan},\ and\ \citenamefont
  {Schmitt}}]{inversemagneticcatalysis}%
  \BibitemOpen
  \bibfield  {author} {\bibinfo {author} {\bibfnamefont {F.}~\bibnamefont
  {Preis}}, \bibinfo {author} {\bibfnamefont {A.}~\bibnamefont {Rebhan}}, \
  and\ \bibinfo {author} {\bibfnamefont {A.}~\bibnamefont {Schmitt}},\ }\href
  {\doibase 10.1007/978-3-642-37305-3_3} {\bibfield  {journal} {\bibinfo
  {journal} {Lect. Notes Phys.}\ }\textbf {\bibinfo {volume} {871}},\ \bibinfo
  {pages} {51} (\bibinfo {year} {2013})},\ \Eprint
  {http://arxiv.org/abs/1208.0536} {arXiv:1208.0536 [hep-ph]} \BibitemShut
  {NoStop}%
\bibitem [{\citenamefont {Farias}\ \emph {et~al.}(2014)\citenamefont {Farias},
  \citenamefont {Gomes}, \citenamefont {Krein},\ and\ \citenamefont
  {Pinto}}]{Farias:2014eca}%
  \BibitemOpen
  \bibfield  {author} {\bibinfo {author} {\bibfnamefont {R.~L.~S.}\
  \bibnamefont {Farias}}, \bibinfo {author} {\bibfnamefont {K.~P.}\
  \bibnamefont {Gomes}}, \bibinfo {author} {\bibfnamefont {G.~I.}\ \bibnamefont
  {Krein}}, \ and\ \bibinfo {author} {\bibfnamefont {M.~B.}\ \bibnamefont
  {Pinto}},\ }\href {\doibase 10.1103/PhysRevC.90.025203} {\bibfield  {journal}
  {\bibinfo  {journal} {Phys. Rev.}\ }\textbf {\bibinfo {volume} {C90}},\
  \bibinfo {pages} {025203} (\bibinfo {year} {2014})},\ \Eprint
  {http://arxiv.org/abs/1404.3931} {arXiv:1404.3931 [hep-ph]} \BibitemShut
  {NoStop}%
\bibitem [{\citenamefont {Chernodub}(2013)}]{vacuumsuperconductivity}%
  \BibitemOpen
  \bibfield  {author} {\bibinfo {author} {\bibfnamefont {M.~N.}\ \bibnamefont
  {Chernodub}},\ }\href {\doibase 10.1007/978-3-642-37305-3_6} {\bibfield
  {journal} {\bibinfo  {journal} {Lect. Notes Phys.}\ }\textbf {\bibinfo
  {volume} {871}},\ \bibinfo {pages} {143} (\bibinfo {year} {2013})},\ \Eprint
  {http://arxiv.org/abs/1208.5025} {arXiv:1208.5025 [hep-ph]} \BibitemShut
  {NoStop}%
\bibitem [{\citenamefont {Fraga}(2013)}]{chiralconfinedeconfine}%
  \BibitemOpen
  \bibfield  {author} {\bibinfo {author} {\bibfnamefont {E.~S.}\ \bibnamefont
  {Fraga}},\ }\href {\doibase 10.1007/978-3-642-37305-3_5} {\bibfield
  {journal} {\bibinfo  {journal} {Lect. Notes Phys.}\ }\textbf {\bibinfo
  {volume} {871}},\ \bibinfo {pages} {121} (\bibinfo {year} {2013})},\ \Eprint
  {http://arxiv.org/abs/1208.0917} {arXiv:1208.0917 [hep-ph]} \BibitemShut
  {NoStop}%
\bibitem [{\citenamefont {Kharzeev}\ \emph {et~al.}(2008)\citenamefont
  {Kharzeev}, \citenamefont {McLerran},\ and\ \citenamefont
  {Warringa}}]{kharzeevtopo}%
  \BibitemOpen
  \bibfield  {author} {\bibinfo {author} {\bibfnamefont {D.~E.}\ \bibnamefont
  {Kharzeev}}, \bibinfo {author} {\bibfnamefont {L.~D.}\ \bibnamefont
  {McLerran}}, \ and\ \bibinfo {author} {\bibfnamefont {H.~J.}\ \bibnamefont
  {Warringa}},\ }\href {\doibase 10.1016/j.nuclphysa.2008.02.298} {\bibfield
  {journal} {\bibinfo  {journal} {Nucl. Phys.}\ }\textbf {\bibinfo {volume}
  {A803}},\ \bibinfo {pages} {227} (\bibinfo {year} {2008})},\ \Eprint
  {http://arxiv.org/abs/0711.0950} {arXiv:0711.0950 [hep-ph]} \BibitemShut
  {NoStop}%
\bibitem [{\citenamefont {Landsteiner}\ \emph {et~al.}(2013)\citenamefont
  {Landsteiner}, \citenamefont {Megias},\ and\ \citenamefont
  {Pena-Benitez}}]{anomaloustransport}%
  \BibitemOpen
  \bibfield  {author} {\bibinfo {author} {\bibfnamefont {K.}~\bibnamefont
  {Landsteiner}}, \bibinfo {author} {\bibfnamefont {E.}~\bibnamefont {Megias}},
  \ and\ \bibinfo {author} {\bibfnamefont {F.}~\bibnamefont {Pena-Benitez}},\
  }\href {\doibase 10.1007/978-3-642-37305-3_17} {\bibfield  {journal}
  {\bibinfo  {journal} {Lect. Notes Phys.}\ }\textbf {\bibinfo {volume}
  {871}},\ \bibinfo {pages} {433} (\bibinfo {year} {2013})},\ \Eprint
  {http://arxiv.org/abs/1207.5808} {arXiv:1207.5808 [hep-th]} \BibitemShut
  {NoStop}%
\bibitem [{\citenamefont {Hattori}\ and\ \citenamefont
  {Itakura}(2013{\natexlab{a}})}]{Hattori:2012je}%
  \BibitemOpen
  \bibfield  {author} {\bibinfo {author} {\bibfnamefont {K.}~\bibnamefont
  {Hattori}}\ and\ \bibinfo {author} {\bibfnamefont {K.}~\bibnamefont
  {Itakura}},\ }\href {\doibase 10.1016/j.aop.2012.11.010} {\bibfield
  {journal} {\bibinfo  {journal} {Annals Phys.}\ }\textbf {\bibinfo {volume}
  {330}},\ \bibinfo {pages} {23} (\bibinfo {year} {2013}{\natexlab{a}})},\
  \Eprint {http://arxiv.org/abs/1209.2663} {arXiv:1209.2663 [hep-ph]}
  \BibitemShut {NoStop}%
\bibitem [{\citenamefont {Hattori}\ and\ \citenamefont
  {Itakura}(2013{\natexlab{b}})}]{Hattori:2012ny}%
  \BibitemOpen
  \bibfield  {author} {\bibinfo {author} {\bibfnamefont {K.}~\bibnamefont
  {Hattori}}\ and\ \bibinfo {author} {\bibfnamefont {K.}~\bibnamefont
  {Itakura}},\ }\href {\doibase 10.1016/j.aop.2013.03.016} {\bibfield
  {journal} {\bibinfo  {journal} {Annals Phys.}\ }\textbf {\bibinfo {volume}
  {334}},\ \bibinfo {pages} {58} (\bibinfo {year} {2013}{\natexlab{b}})},\
  \Eprint {http://arxiv.org/abs/1212.1897} {arXiv:1212.1897 [hep-ph]}
  \BibitemShut {NoStop}%
\bibitem [{\citenamefont {Fayazbakhsh}\ \emph {et~al.}(2012)\citenamefont
  {Fayazbakhsh}, \citenamefont {Sadeghian},\ and\ \citenamefont
  {Sadooghi}}]{mesonrefractive}%
  \BibitemOpen
  \bibfield  {author} {\bibinfo {author} {\bibfnamefont {S.}~\bibnamefont
  {Fayazbakhsh}}, \bibinfo {author} {\bibfnamefont {S.}~\bibnamefont
  {Sadeghian}}, \ and\ \bibinfo {author} {\bibfnamefont {N.}~\bibnamefont
  {Sadooghi}},\ }\href {\doibase 10.1103/PhysRevD.86.085042} {\bibfield
  {journal} {\bibinfo  {journal} {Phys. Rev.}\ }\textbf {\bibinfo {volume}
  {D86}},\ \bibinfo {pages} {085042} (\bibinfo {year} {2012})},\ \Eprint
  {http://arxiv.org/abs/1206.6051} {arXiv:1206.6051 [hep-ph]} \BibitemShut
  {NoStop}%
\bibitem [{\citenamefont {Fayazbakhsh}\ and\ \citenamefont
  {Sadooghi}(2013)}]{mesondecay}%
  \BibitemOpen
  \bibfield  {author} {\bibinfo {author} {\bibfnamefont {S.}~\bibnamefont
  {Fayazbakhsh}}\ and\ \bibinfo {author} {\bibfnamefont {N.}~\bibnamefont
  {Sadooghi}},\ }\href {\doibase 10.1103/PhysRevD.88.065030} {\bibfield
  {journal} {\bibinfo  {journal} {Phys. Rev.}\ }\textbf {\bibinfo {volume}
  {D88}},\ \bibinfo {pages} {065030} (\bibinfo {year} {2013})},\ \Eprint
  {http://arxiv.org/abs/1306.2098} {arXiv:1306.2098 [hep-ph]} \BibitemShut
  {NoStop}%
\bibitem [{\citenamefont {Strickland}\ \emph {et~al.}(2012)\citenamefont
  {Strickland}, \citenamefont {Dexheimer},\ and\ \citenamefont
  {Menezes}}]{Strickland:2012vu}%
  \BibitemOpen
  \bibfield  {author} {\bibinfo {author} {\bibfnamefont {M.}~\bibnamefont
  {Strickland}}, \bibinfo {author} {\bibfnamefont {V.}~\bibnamefont
  {Dexheimer}}, \ and\ \bibinfo {author} {\bibfnamefont {D.~P.}\ \bibnamefont
  {Menezes}},\ }\href {\doibase 10.1103/PhysRevD.86.125032} {\bibfield
  {journal} {\bibinfo  {journal} {Phys. Rev.}\ }\textbf {\bibinfo {volume}
  {D86}},\ \bibinfo {pages} {125032} (\bibinfo {year} {2012})},\ \Eprint
  {http://arxiv.org/abs/1209.3276} {arXiv:1209.3276 [nucl-th]} \BibitemShut
  {NoStop}%
\bibitem [{\citenamefont {Andersen}(2012)}]{Andersen:2012zc}%
  \BibitemOpen
  \bibfield  {author} {\bibinfo {author} {\bibfnamefont {J.~O.}\ \bibnamefont
  {Andersen}},\ }\href {\doibase 10.1007/JHEP10(2012)005} {\bibfield  {journal}
  {\bibinfo  {journal} {JHEP}\ }\textbf {\bibinfo {volume} {10}},\ \bibinfo
  {pages} {005} (\bibinfo {year} {2012})},\ \Eprint
  {http://arxiv.org/abs/1205.6978} {arXiv:1205.6978 [hep-ph]} \BibitemShut
  {NoStop}%
\bibitem [{\citenamefont {Mamo}(2013)}]{Mamo:2013efa}%
  \BibitemOpen
  \bibfield  {author} {\bibinfo {author} {\bibfnamefont {K.~A.}\ \bibnamefont
  {Mamo}},\ }\href {\doibase 10.1007/JHEP08(2013)083} {\bibfield  {journal}
  {\bibinfo  {journal} {JHEP}\ }\textbf {\bibinfo {volume} {08}},\ \bibinfo
  {pages} {083} (\bibinfo {year} {2013})},\ \Eprint
  {http://arxiv.org/abs/1210.7428} {arXiv:1210.7428 [hep-th]} \BibitemShut
  {NoStop}%
\bibitem [{\citenamefont {Sadooghi}\ and\ \citenamefont
  {Taghinavaz}(2015)}]{Sadooghi:2015hha}%
  \BibitemOpen
  \bibfield  {author} {\bibinfo {author} {\bibfnamefont {N.}~\bibnamefont
  {Sadooghi}}\ and\ \bibinfo {author} {\bibfnamefont {F.}~\bibnamefont
  {Taghinavaz}},\ }\href {\doibase 10.1103/PhysRevD.92.025006} {\bibfield
  {journal} {\bibinfo  {journal} {Phys. Rev.}\ }\textbf {\bibinfo {volume}
  {D92}},\ \bibinfo {pages} {025006} (\bibinfo {year} {2015})},\ \Eprint
  {http://arxiv.org/abs/1504.04268} {arXiv:1504.04268 [hep-ph]} \BibitemShut
  {NoStop}%
\bibitem [{\citenamefont {Elmfors}\ \emph {et~al.}(1996)\citenamefont
  {Elmfors}, \citenamefont {Persson},\ and\ \citenamefont
  {Skagerstam}}]{elmforsfermion}%
  \BibitemOpen
  \bibfield  {author} {\bibinfo {author} {\bibfnamefont {P.}~\bibnamefont
  {Elmfors}}, \bibinfo {author} {\bibfnamefont {D.}~\bibnamefont {Persson}}, \
  and\ \bibinfo {author} {\bibfnamefont {B.-S.}\ \bibnamefont {Skagerstam}},\
  }\href {\doibase 10.1016/0550-3213(96)00042-9} {\bibfield  {journal}
  {\bibinfo  {journal} {Nucl. Phys.}\ }\textbf {\bibinfo {volume} {B464}},\
  \bibinfo {pages} {153} (\bibinfo {year} {1996})},\ \Eprint
  {http://arxiv.org/abs/hep-ph/9509418} {arXiv:hep-ph/9509418 [hep-ph]}
  \BibitemShut {NoStop}%
\bibitem [{\citenamefont {Bhattacharya}\ and\ \citenamefont
  {Pal}(2004{\natexlab{a}})}]{Bhattacharya:2002qf}%
  \BibitemOpen
  \bibfield  {author} {\bibinfo {author} {\bibfnamefont {K.}~\bibnamefont
  {Bhattacharya}}\ and\ \bibinfo {author} {\bibfnamefont {P.~B.}\ \bibnamefont
  {Pal}},\ }\href {\doibase 10.1007/BF02705251} {\bibfield  {journal} {\bibinfo
   {journal} {Pramana}\ }\textbf {\bibinfo {volume} {62}},\ \bibinfo {pages}
  {1041} (\bibinfo {year} {2004}{\natexlab{a}})},\ \Eprint
  {http://arxiv.org/abs/hep-ph/0209053} {arXiv:hep-ph/0209053 [hep-ph]}
  \BibitemShut {NoStop}%
\bibitem [{\citenamefont {Bhattacharya}\ and\ \citenamefont
  {Pal}(2004{\natexlab{b}})}]{Bhattacharya:2002aj}%
  \BibitemOpen
  \bibfield  {author} {\bibinfo {author} {\bibfnamefont {K.}~\bibnamefont
  {Bhattacharya}}\ and\ \bibinfo {author} {\bibfnamefont {P.~B.}\ \bibnamefont
  {Pal}},\ }\bibfield  {booktitle} {\emph {\bibinfo {booktitle} {{Proceedings,
  Neutrino 2001 Meeting: Chennai, India, February, 2001}}},\ }\href@noop {}
  {\bibfield  {journal} {\bibinfo  {journal} {Proc. Indian Natl. Sci. Acad.}\
  }\textbf {\bibinfo {volume} {A70}},\ \bibinfo {pages} {145} (\bibinfo {year}
  {2004}{\natexlab{b}})},\ \Eprint {http://arxiv.org/abs/hep-ph/0212118}
  {arXiv:hep-ph/0212118 [hep-ph]} \BibitemShut {NoStop}%
\bibitem [{\citenamefont {Ganguly}\ \emph {et~al.}(1999)\citenamefont
  {Ganguly}, \citenamefont {Konar},\ and\ \citenamefont
  {Pal}}]{Ganguly:1999ts}%
  \BibitemOpen
  \bibfield  {author} {\bibinfo {author} {\bibfnamefont {A.~K.}\ \bibnamefont
  {Ganguly}}, \bibinfo {author} {\bibfnamefont {S.}~\bibnamefont {Konar}}, \
  and\ \bibinfo {author} {\bibfnamefont {P.~B.}\ \bibnamefont {Pal}},\ }\href
  {\doibase 10.1103/PhysRevD.60.105014} {\bibfield  {journal} {\bibinfo
  {journal} {Phys. Rev.}\ }\textbf {\bibinfo {volume} {D60}},\ \bibinfo {pages}
  {105014} (\bibinfo {year} {1999})},\ \Eprint
  {http://arxiv.org/abs/hep-ph/9905206} {arXiv:hep-ph/9905206 [hep-ph]}
  \BibitemShut {NoStop}%
\bibitem [{\citenamefont {D'Olivo}\ \emph {et~al.}(2003)\citenamefont
  {D'Olivo}, \citenamefont {Nieves},\ and\ \citenamefont
  {Sahu}}]{DOlivo:2002omk}%
  \BibitemOpen
  \bibfield  {author} {\bibinfo {author} {\bibfnamefont {J.~C.}\ \bibnamefont
  {D'Olivo}}, \bibinfo {author} {\bibfnamefont {J.~F.}\ \bibnamefont {Nieves}},
  \ and\ \bibinfo {author} {\bibfnamefont {S.}~\bibnamefont {Sahu}},\ }\href
  {\doibase 10.1103/PhysRevD.67.025018} {\bibfield  {journal} {\bibinfo
  {journal} {Phys. Rev.}\ }\textbf {\bibinfo {volume} {D67}},\ \bibinfo {pages}
  {025018} (\bibinfo {year} {2003})},\ \Eprint
  {http://arxiv.org/abs/hep-ph/0208146} {arXiv:hep-ph/0208146 [hep-ph]}
  \BibitemShut {NoStop}%
\bibitem [{\citenamefont {Schwinger}(1951)}]{schwinger1951}%
  \BibitemOpen
  \bibfield  {author} {\bibinfo {author} {\bibfnamefont {J.}~\bibnamefont
  {Schwinger}},\ }\href {\doibase 10.1103/PhysRev.82.664} {\bibfield  {journal}
  {\bibinfo  {journal} {Phys. Rev.}\ }\textbf {\bibinfo {volume} {82}},\
  \bibinfo {pages} {664} (\bibinfo {year} {1951})}\BibitemShut {NoStop}%
\bibitem [{\citenamefont {Tsai}(1974{\natexlab{a}})}]{tsaifermion}%
  \BibitemOpen
  \bibfield  {author} {\bibinfo {author} {\bibfnamefont {W.-y.}\ \bibnamefont
  {Tsai}},\ }\href {\doibase 10.1103/PhysRevD.10.1342} {\bibfield  {journal}
  {\bibinfo  {journal} {Phys. Rev. D}\ }\textbf {\bibinfo {volume} {10}},\
  \bibinfo {pages} {1342} (\bibinfo {year} {1974}{\natexlab{a}})}\BibitemShut
  {NoStop}%
\bibitem [{\citenamefont {Tsai}(1974{\natexlab{b}})}]{tsaivacuumpol}%
  \BibitemOpen
  \bibfield  {author} {\bibinfo {author} {\bibfnamefont {W.-y.}\ \bibnamefont
  {Tsai}},\ }\href {\doibase 10.1103/PhysRevD.10.2699} {\bibfield  {journal}
  {\bibinfo  {journal} {Phys. Rev. D}\ }\textbf {\bibinfo {volume} {10}},\
  \bibinfo {pages} {2699} (\bibinfo {year} {1974}{\natexlab{b}})}\BibitemShut
  {NoStop}%
\bibitem [{\citenamefont {Chyi}\ \emph {et~al.}(2000)\citenamefont {Chyi},
  \citenamefont {Hwang}, \citenamefont {Kao}, \citenamefont {Lin},
  \citenamefont {Ng},\ and\ \citenamefont {Tseng}}]{chyiweak}%
  \BibitemOpen
  \bibfield  {author} {\bibinfo {author} {\bibfnamefont {T.-K.}\ \bibnamefont
  {Chyi}}, \bibinfo {author} {\bibfnamefont {C.-W.}\ \bibnamefont {Hwang}},
  \bibinfo {author} {\bibfnamefont {W.~F.}\ \bibnamefont {Kao}}, \bibinfo
  {author} {\bibfnamefont {G.-L.}\ \bibnamefont {Lin}}, \bibinfo {author}
  {\bibfnamefont {K.-W.}\ \bibnamefont {Ng}}, \ and\ \bibinfo {author}
  {\bibfnamefont {J.-J.}\ \bibnamefont {Tseng}},\ }\href {\doibase
  10.1103/PhysRevD.62.105014} {\bibfield  {journal} {\bibinfo  {journal} {Phys.
  Rev.}\ }\textbf {\bibinfo {volume} {D62}},\ \bibinfo {pages} {105014}
  (\bibinfo {year} {2000})},\ \Eprint {http://arxiv.org/abs/hep-th/9912134}
  {arXiv:hep-th/9912134 [hep-th]} \BibitemShut {NoStop}%
\bibitem [{\citenamefont {Adhya}\ \emph {et~al.}(2016)\citenamefont {Adhya},
  \citenamefont {Mandal}, \citenamefont {Biswas},\ and\ \citenamefont
  {Roy}}]{pradiproypionself}%
  \BibitemOpen
  \bibfield  {author} {\bibinfo {author} {\bibfnamefont {S.~P.}\ \bibnamefont
  {Adhya}}, \bibinfo {author} {\bibfnamefont {M.}~\bibnamefont {Mandal}},
  \bibinfo {author} {\bibfnamefont {S.}~\bibnamefont {Biswas}}, \ and\ \bibinfo
  {author} {\bibfnamefont {P.~K.}\ \bibnamefont {Roy}},\ }\href {\doibase
  10.1103/PhysRevD.93.074033} {\bibfield  {journal} {\bibinfo  {journal} {Phys.
  Rev.}\ }\textbf {\bibinfo {volume} {D93}},\ \bibinfo {pages} {074033}
  (\bibinfo {year} {2016})},\ \Eprint {http://arxiv.org/abs/1601.04578}
  {arXiv:1601.04578 [nucl-th]} \BibitemShut {NoStop}%
\bibitem [{\citenamefont {Mukherjee}\ \emph {et~al.}(2017)\citenamefont
  {Mukherjee}, \citenamefont {Ghosh}, \citenamefont {Mandal}, \citenamefont
  {Roy},\ and\ \citenamefont {Sarkar}}]{Mukherjee:2017dls}%
  \BibitemOpen
  \bibfield  {author} {\bibinfo {author} {\bibfnamefont {A.}~\bibnamefont
  {Mukherjee}}, \bibinfo {author} {\bibfnamefont {S.}~\bibnamefont {Ghosh}},
  \bibinfo {author} {\bibfnamefont {M.}~\bibnamefont {Mandal}}, \bibinfo
  {author} {\bibfnamefont {P.}~\bibnamefont {Roy}}, \ and\ \bibinfo {author}
  {\bibfnamefont {S.}~\bibnamefont {Sarkar}},\ }\href {\doibase
  10.1103/PhysRevD.96.016024} {\bibfield  {journal} {\bibinfo  {journal} {Phys.
  Rev.}\ }\textbf {\bibinfo {volume} {D96}},\ \bibinfo {pages} {016024}
  (\bibinfo {year} {2017})},\ \Eprint {http://arxiv.org/abs/1708.02385}
  {arXiv:1708.02385 [hep-ph]} \BibitemShut {NoStop}%
\bibitem [{\citenamefont {Ghosh}\ \emph {et~al.}(2016)\citenamefont {Ghosh},
  \citenamefont {Mukherjee}, \citenamefont {Mandal}, \citenamefont {Sarkar},\
  and\ \citenamefont {Roy}}]{arghyarho}%
  \BibitemOpen
  \bibfield  {author} {\bibinfo {author} {\bibfnamefont {S.}~\bibnamefont
  {Ghosh}}, \bibinfo {author} {\bibfnamefont {A.}~\bibnamefont {Mukherjee}},
  \bibinfo {author} {\bibfnamefont {M.}~\bibnamefont {Mandal}}, \bibinfo
  {author} {\bibfnamefont {S.}~\bibnamefont {Sarkar}}, \ and\ \bibinfo {author}
  {\bibfnamefont {P.}~\bibnamefont {Roy}},\ }\href {\doibase
  10.1103/PhysRevD.94.094043} {\bibfield  {journal} {\bibinfo  {journal} {Phys.
  Rev.}\ }\textbf {\bibinfo {volume} {D94}},\ \bibinfo {pages} {094043}
  (\bibinfo {year} {2016})},\ \Eprint {http://arxiv.org/abs/1612.02966}
  {arXiv:1612.02966 [nucl-th]} \BibitemShut {NoStop}%
\bibitem [{\citenamefont {Bandyopadhyay}\ and\ \citenamefont
  {Mallik}(2016)}]{Bandyopadhyay:2016cpf}%
  \BibitemOpen
  \bibfield  {author} {\bibinfo {author} {\bibfnamefont {A.}~\bibnamefont
  {Bandyopadhyay}}\ and\ \bibinfo {author} {\bibfnamefont {S.}~\bibnamefont
  {Mallik}},\ }\href@noop {} {\  (\bibinfo {year} {2016})},\ \Eprint
  {http://arxiv.org/abs/1610.07887} {arXiv:1610.07887 [hep-ph]} \BibitemShut
  {NoStop}%
\bibitem [{\citenamefont {Ghosh}\ \emph {et~al.}(2017)\citenamefont {Ghosh},
  \citenamefont {Mukherjee}, \citenamefont {Mandal}, \citenamefont {Sarkar},\
  and\ \citenamefont {Roy}}]{arghyarhothermal}%
  \BibitemOpen
  \bibfield  {author} {\bibinfo {author} {\bibfnamefont {S.}~\bibnamefont
  {Ghosh}}, \bibinfo {author} {\bibfnamefont {A.}~\bibnamefont {Mukherjee}},
  \bibinfo {author} {\bibfnamefont {M.}~\bibnamefont {Mandal}}, \bibinfo
  {author} {\bibfnamefont {S.}~\bibnamefont {Sarkar}}, \ and\ \bibinfo {author}
  {\bibfnamefont {P.}~\bibnamefont {Roy}},\ }\href@noop {} {\  (\bibinfo {year}
  {2017})},\ \Eprint {http://arxiv.org/abs/1704.05319} {arXiv:1704.05319
  [hep-ph]} \BibitemShut {NoStop}%
\bibitem [{\citenamefont {Andersen}\ \emph {et~al.}(1999)\citenamefont
  {Andersen}, \citenamefont {Braaten},\ and\ \citenamefont
  {Strickland}}]{Andersen:1999fw}%
  \BibitemOpen
  \bibfield  {author} {\bibinfo {author} {\bibfnamefont {J.~O.}\ \bibnamefont
  {Andersen}}, \bibinfo {author} {\bibfnamefont {E.}~\bibnamefont {Braaten}}, \
  and\ \bibinfo {author} {\bibfnamefont {M.}~\bibnamefont {Strickland}},\
  }\href {\doibase 10.1103/PhysRevLett.83.2139} {\bibfield  {journal} {\bibinfo
   {journal} {Phys. Rev. Lett.}\ }\textbf {\bibinfo {volume} {83}},\ \bibinfo
  {pages} {2139} (\bibinfo {year} {1999})},\ \Eprint
  {http://arxiv.org/abs/hep-ph/9902327} {arXiv:hep-ph/9902327 [hep-ph]}
  \BibitemShut {NoStop}%
\bibitem [{\citenamefont {Braaten}\ and\ \citenamefont
  {Pisarski}(1990{\natexlab{a}})}]{brateennucl337}%
  \BibitemOpen
  \bibfield  {author} {\bibinfo {author} {\bibfnamefont {E.}~\bibnamefont
  {Braaten}}\ and\ \bibinfo {author} {\bibfnamefont {R.~D.}\ \bibnamefont
  {Pisarski}},\ }\href {\doibase
  http://dx.doi.org/10.1016/0550-3213(90)90508-B} {\bibfield  {journal}
  {\bibinfo  {journal} {Nuclear Physics B}\ }\textbf {\bibinfo {volume}
  {337}},\ \bibinfo {pages} {569 } (\bibinfo {year}
  {1990}{\natexlab{a}})}\BibitemShut {NoStop}%
\bibitem [{\citenamefont {Braaten}\ \emph {et~al.}(1990)\citenamefont
  {Braaten}, \citenamefont {Pisarski},\ and\ \citenamefont
  {Yuan}}]{braatendilepton}%
  \BibitemOpen
  \bibfield  {author} {\bibinfo {author} {\bibfnamefont {E.}~\bibnamefont
  {Braaten}}, \bibinfo {author} {\bibfnamefont {R.~D.}\ \bibnamefont
  {Pisarski}}, \ and\ \bibinfo {author} {\bibfnamefont {T.~C.}\ \bibnamefont
  {Yuan}},\ }\href {\doibase 10.1103/PhysRevLett.64.2242} {\bibfield  {journal}
  {\bibinfo  {journal} {Phys. Rev. Lett.}\ }\textbf {\bibinfo {volume} {64}},\
  \bibinfo {pages} {2242} (\bibinfo {year} {1990})}\BibitemShut {NoStop}%
\bibitem [{\citenamefont {Braaten}\ and\ \citenamefont
  {Pisarski}(1990{\natexlab{b}})}]{HTLgluondamping}%
  \BibitemOpen
  \bibfield  {author} {\bibinfo {author} {\bibfnamefont {E.}~\bibnamefont
  {Braaten}}\ and\ \bibinfo {author} {\bibfnamefont {R.~D.}\ \bibnamefont
  {Pisarski}},\ }\href {\doibase 10.1103/PhysRevD.42.2156} {\bibfield
  {journal} {\bibinfo  {journal} {Phys. Rev. D}\ }\textbf {\bibinfo {volume}
  {42}},\ \bibinfo {pages} {2156} (\bibinfo {year}
  {1990}{\natexlab{b}})}\BibitemShut {NoStop}%
\bibitem [{\citenamefont {Haque}\ \emph {et~al.}(2011)\citenamefont {Haque},
  \citenamefont {Mustafa},\ and\ \citenamefont {Thoma}}]{Haque:2011iz}%
  \BibitemOpen
  \bibfield  {author} {\bibinfo {author} {\bibfnamefont {N.}~\bibnamefont
  {Haque}}, \bibinfo {author} {\bibfnamefont {M.~G.}\ \bibnamefont {Mustafa}},
  \ and\ \bibinfo {author} {\bibfnamefont {M.~H.}\ \bibnamefont {Thoma}},\
  }\href {\doibase 10.1103/PhysRevD.84.054009} {\bibfield  {journal} {\bibinfo
  {journal} {Phys. Rev.}\ }\textbf {\bibinfo {volume} {D84}},\ \bibinfo {pages}
  {054009} (\bibinfo {year} {2011})},\ \Eprint {http://arxiv.org/abs/1103.3394}
  {arXiv:1103.3394 [hep-ph]} \BibitemShut {NoStop}%
\bibitem [{\citenamefont {Haque}\ and\ \citenamefont
  {Mustafa}(2011)}]{Haque:2011vt}%
  \BibitemOpen
  \bibfield  {author} {\bibinfo {author} {\bibfnamefont {N.}~\bibnamefont
  {Haque}}\ and\ \bibinfo {author} {\bibfnamefont {M.~G.}\ \bibnamefont
  {Mustafa}},\ }\bibfield  {booktitle} {\emph {\bibinfo {booktitle}
  {{Proceedings, 6th International Conference on Physics and Astrophysics of
  Quark Gluon Plasma (ICPAQGP 2010): Goa, India, December 6-10, 2010}}},\
  }\href {\doibase 10.1016/j.nuclphysa.2011.05.070} {\bibfield  {journal}
  {\bibinfo  {journal} {Nucl. Phys.}\ }\textbf {\bibinfo {volume} {A862-863}},\
  \bibinfo {pages} {271} (\bibinfo {year} {2011})},\ \Eprint
  {http://arxiv.org/abs/1109.0799} {arXiv:1109.0799 [hep-ph]} \BibitemShut
  {NoStop}%
\bibitem [{\citenamefont {Haque}\ and\ \citenamefont
  {Mustafa}(2010)}]{Haque:2010rb}%
  \BibitemOpen
  \bibfield  {author} {\bibinfo {author} {\bibfnamefont {N.}~\bibnamefont
  {Haque}}\ and\ \bibinfo {author} {\bibfnamefont {M.~G.}\ \bibnamefont
  {Mustafa}},\ }\href@noop {} {\  (\bibinfo {year} {2010})},\ \Eprint
  {http://arxiv.org/abs/1007.2076} {arXiv:1007.2076 [hep-ph]} \BibitemShut
  {NoStop}%
\bibitem [{\citenamefont {Andersen}\ \emph {et~al.}(2002)\citenamefont
  {Andersen}, \citenamefont {Braaten}, \citenamefont {Petitgirard},\ and\
  \citenamefont {Strickland}}]{Andersen:2002ey}%
  \BibitemOpen
  \bibfield  {author} {\bibinfo {author} {\bibfnamefont {J.~O.}\ \bibnamefont
  {Andersen}}, \bibinfo {author} {\bibfnamefont {E.}~\bibnamefont {Braaten}},
  \bibinfo {author} {\bibfnamefont {E.}~\bibnamefont {Petitgirard}}, \ and\
  \bibinfo {author} {\bibfnamefont {M.}~\bibnamefont {Strickland}},\ }\href
  {\doibase 10.1103/PhysRevD.66.085016} {\bibfield  {journal} {\bibinfo
  {journal} {Phys. Rev.}\ }\textbf {\bibinfo {volume} {D66}},\ \bibinfo {pages}
  {085016} (\bibinfo {year} {2002})},\ \Eprint
  {http://arxiv.org/abs/hep-ph/0205085} {arXiv:hep-ph/0205085 [hep-ph]}
  \BibitemShut {NoStop}%
\bibitem [{\citenamefont {Andersen}\ \emph {et~al.}(2004)\citenamefont
  {Andersen}, \citenamefont {Petitgirard},\ and\ \citenamefont
  {Strickland}}]{Andersen:2003zk}%
  \BibitemOpen
  \bibfield  {author} {\bibinfo {author} {\bibfnamefont {J.~O.}\ \bibnamefont
  {Andersen}}, \bibinfo {author} {\bibfnamefont {E.}~\bibnamefont
  {Petitgirard}}, \ and\ \bibinfo {author} {\bibfnamefont {M.}~\bibnamefont
  {Strickland}},\ }\href {\doibase 10.1103/PhysRevD.70.045001} {\bibfield
  {journal} {\bibinfo  {journal} {Phys. Rev.}\ }\textbf {\bibinfo {volume}
  {D70}},\ \bibinfo {pages} {045001} (\bibinfo {year} {2004})},\ \Eprint
  {http://arxiv.org/abs/hep-ph/0302069} {arXiv:hep-ph/0302069 [hep-ph]}
  \BibitemShut {NoStop}%
\bibitem [{\citenamefont {Haque}\ \emph
  {et~al.}(2013{\natexlab{a}})\citenamefont {Haque}, \citenamefont {Mustafa},\
  and\ \citenamefont {Strickland}}]{Haque:2012my}%
  \BibitemOpen
  \bibfield  {author} {\bibinfo {author} {\bibfnamefont {N.}~\bibnamefont
  {Haque}}, \bibinfo {author} {\bibfnamefont {M.~G.}\ \bibnamefont {Mustafa}},
  \ and\ \bibinfo {author} {\bibfnamefont {M.}~\bibnamefont {Strickland}},\
  }\href {\doibase 10.1103/PhysRevD.87.105007} {\bibfield  {journal} {\bibinfo
  {journal} {Phys. Rev.}\ }\textbf {\bibinfo {volume} {D87}},\ \bibinfo {pages}
  {105007} (\bibinfo {year} {2013}{\natexlab{a}})},\ \Eprint
  {http://arxiv.org/abs/1212.1797} {arXiv:1212.1797 [hep-ph]} \BibitemShut
  {NoStop}%
\bibitem [{\citenamefont {Haque}\ \emph
  {et~al.}(2013{\natexlab{b}})\citenamefont {Haque}, \citenamefont {Mustafa},\
  and\ \citenamefont {Strickland}}]{Haque:2013qta}%
  \BibitemOpen
  \bibfield  {author} {\bibinfo {author} {\bibfnamefont {N.}~\bibnamefont
  {Haque}}, \bibinfo {author} {\bibfnamefont {M.~G.}\ \bibnamefont {Mustafa}},
  \ and\ \bibinfo {author} {\bibfnamefont {M.}~\bibnamefont {Strickland}},\
  }\href {\doibase 10.1007/JHEP07(2013)184} {\bibfield  {journal} {\bibinfo
  {journal} {JHEP}\ }\textbf {\bibinfo {volume} {07}},\ \bibinfo {pages} {184}
  (\bibinfo {year} {2013}{\natexlab{b}})},\ \Eprint
  {http://arxiv.org/abs/1302.3228} {arXiv:1302.3228 [hep-ph]} \BibitemShut
  {NoStop}%
\bibitem [{\citenamefont {Andersen}\ \emph
  {et~al.}(2010{\natexlab{a}})\citenamefont {Andersen}, \citenamefont
  {Strickland},\ and\ \citenamefont {Su}}]{Andersen:2009tc}%
  \BibitemOpen
  \bibfield  {author} {\bibinfo {author} {\bibfnamefont {J.~O.}\ \bibnamefont
  {Andersen}}, \bibinfo {author} {\bibfnamefont {M.}~\bibnamefont
  {Strickland}}, \ and\ \bibinfo {author} {\bibfnamefont {N.}~\bibnamefont
  {Su}},\ }\href {\doibase 10.1103/PhysRevLett.104.122003} {\bibfield
  {journal} {\bibinfo  {journal} {Phys. Rev. Lett.}\ }\textbf {\bibinfo
  {volume} {104}},\ \bibinfo {pages} {122003} (\bibinfo {year}
  {2010}{\natexlab{a}})},\ \Eprint {http://arxiv.org/abs/0911.0676}
  {arXiv:0911.0676 [hep-ph]} \BibitemShut {NoStop}%
\bibitem [{\citenamefont {Andersen}\ \emph
  {et~al.}(2010{\natexlab{b}})\citenamefont {Andersen}, \citenamefont
  {Strickland},\ and\ \citenamefont {Su}}]{Andersen:2010ct}%
  \BibitemOpen
  \bibfield  {author} {\bibinfo {author} {\bibfnamefont {J.~O.}\ \bibnamefont
  {Andersen}}, \bibinfo {author} {\bibfnamefont {M.}~\bibnamefont
  {Strickland}}, \ and\ \bibinfo {author} {\bibfnamefont {N.}~\bibnamefont
  {Su}},\ }\href {\doibase 10.1007/JHEP08(2010)113} {\bibfield  {journal}
  {\bibinfo  {journal} {JHEP}\ }\textbf {\bibinfo {volume} {08}},\ \bibinfo
  {pages} {113} (\bibinfo {year} {2010}{\natexlab{b}})},\ \Eprint
  {http://arxiv.org/abs/1005.1603} {arXiv:1005.1603 [hep-ph]} \BibitemShut
  {NoStop}%
\bibitem [{\citenamefont {Andersen}\ \emph
  {et~al.}(2011{\natexlab{a}})\citenamefont {Andersen}, \citenamefont
  {Leganger}, \citenamefont {Strickland},\ and\ \citenamefont
  {Su}}]{Andersen:2010wu}%
  \BibitemOpen
  \bibfield  {author} {\bibinfo {author} {\bibfnamefont {J.~O.}\ \bibnamefont
  {Andersen}}, \bibinfo {author} {\bibfnamefont {L.~E.}\ \bibnamefont
  {Leganger}}, \bibinfo {author} {\bibfnamefont {M.}~\bibnamefont
  {Strickland}}, \ and\ \bibinfo {author} {\bibfnamefont {N.}~\bibnamefont
  {Su}},\ }\href {\doibase 10.1016/j.physletb.2010.12.070} {\bibfield
  {journal} {\bibinfo  {journal} {Phys. Lett.}\ }\textbf {\bibinfo {volume}
  {B696}},\ \bibinfo {pages} {468} (\bibinfo {year} {2011}{\natexlab{a}})},\
  \Eprint {http://arxiv.org/abs/1009.4644} {arXiv:1009.4644 [hep-ph]}
  \BibitemShut {NoStop}%
\bibitem [{\citenamefont {Andersen}\ \emph
  {et~al.}(2011{\natexlab{b}})\citenamefont {Andersen}, \citenamefont
  {Leganger}, \citenamefont {Strickland},\ and\ \citenamefont
  {Su}}]{Andersen:2011sf}%
  \BibitemOpen
  \bibfield  {author} {\bibinfo {author} {\bibfnamefont {J.~O.}\ \bibnamefont
  {Andersen}}, \bibinfo {author} {\bibfnamefont {L.~E.}\ \bibnamefont
  {Leganger}}, \bibinfo {author} {\bibfnamefont {M.}~\bibnamefont
  {Strickland}}, \ and\ \bibinfo {author} {\bibfnamefont {N.}~\bibnamefont
  {Su}},\ }\href {\doibase 10.1007/JHEP08(2011)053} {\bibfield  {journal}
  {\bibinfo  {journal} {JHEP}\ }\textbf {\bibinfo {volume} {08}},\ \bibinfo
  {pages} {053} (\bibinfo {year} {2011}{\natexlab{b}})},\ \Eprint
  {http://arxiv.org/abs/1103.2528} {arXiv:1103.2528 [hep-ph]} \BibitemShut
  {NoStop}%
\bibitem [{\citenamefont {Andersen}\ \emph
  {et~al.}(2011{\natexlab{c}})\citenamefont {Andersen}, \citenamefont
  {Leganger}, \citenamefont {Strickland},\ and\ \citenamefont
  {Su}}]{Andersen:2011ug}%
  \BibitemOpen
  \bibfield  {author} {\bibinfo {author} {\bibfnamefont {J.~O.}\ \bibnamefont
  {Andersen}}, \bibinfo {author} {\bibfnamefont {L.~E.}\ \bibnamefont
  {Leganger}}, \bibinfo {author} {\bibfnamefont {M.}~\bibnamefont
  {Strickland}}, \ and\ \bibinfo {author} {\bibfnamefont {N.}~\bibnamefont
  {Su}},\ }\href {\doibase 10.1103/PhysRevD.84.087703} {\bibfield  {journal}
  {\bibinfo  {journal} {Phys. Rev.}\ }\textbf {\bibinfo {volume} {D84}},\
  \bibinfo {pages} {087703} (\bibinfo {year} {2011}{\natexlab{c}})},\ \Eprint
  {http://arxiv.org/abs/1106.0514} {arXiv:1106.0514 [hep-ph]} \BibitemShut
  {NoStop}%
\bibitem [{\citenamefont {Haque}\ \emph
  {et~al.}(2014{\natexlab{a}})\citenamefont {Haque}, \citenamefont {Andersen},
  \citenamefont {Mustafa}, \citenamefont {Strickland},\ and\ \citenamefont
  {Su}}]{Haque:2013sja}%
  \BibitemOpen
  \bibfield  {author} {\bibinfo {author} {\bibfnamefont {N.}~\bibnamefont
  {Haque}}, \bibinfo {author} {\bibfnamefont {J.~O.}\ \bibnamefont {Andersen}},
  \bibinfo {author} {\bibfnamefont {M.~G.}\ \bibnamefont {Mustafa}}, \bibinfo
  {author} {\bibfnamefont {M.}~\bibnamefont {Strickland}}, \ and\ \bibinfo
  {author} {\bibfnamefont {N.}~\bibnamefont {Su}},\ }\href {\doibase
  10.1103/PhysRevD.89.061701} {\bibfield  {journal} {\bibinfo  {journal} {Phys.
  Rev.}\ }\textbf {\bibinfo {volume} {D89}},\ \bibinfo {pages} {061701}
  (\bibinfo {year} {2014}{\natexlab{a}})},\ \Eprint
  {http://arxiv.org/abs/1309.3968} {arXiv:1309.3968 [hep-ph]} \BibitemShut
  {NoStop}%
\bibitem [{\citenamefont {Haque}\ \emph
  {et~al.}(2014{\natexlab{b}})\citenamefont {Haque}, \citenamefont
  {Bandyopadhyay}, \citenamefont {Andersen}, \citenamefont {Mustafa},
  \citenamefont {Strickland},\ and\ \citenamefont {Su}}]{Haque:2014rua}%
  \BibitemOpen
  \bibfield  {author} {\bibinfo {author} {\bibfnamefont {N.}~\bibnamefont
  {Haque}}, \bibinfo {author} {\bibfnamefont {A.}~\bibnamefont
  {Bandyopadhyay}}, \bibinfo {author} {\bibfnamefont {J.~O.}\ \bibnamefont
  {Andersen}}, \bibinfo {author} {\bibfnamefont {M.~G.}\ \bibnamefont
  {Mustafa}}, \bibinfo {author} {\bibfnamefont {M.}~\bibnamefont {Strickland}},
  \ and\ \bibinfo {author} {\bibfnamefont {N.}~\bibnamefont {Su}},\ }\href
  {\doibase 10.1007/JHEP05(2014)027} {\bibfield  {journal} {\bibinfo  {journal}
  {JHEP}\ }\textbf {\bibinfo {volume} {05}},\ \bibinfo {pages} {027} (\bibinfo
  {year} {2014}{\natexlab{b}})},\ \Eprint {http://arxiv.org/abs/1402.6907}
  {arXiv:1402.6907 [hep-ph]} \BibitemShut {NoStop}%
\bibitem [{\citenamefont {Mustafa}\ \emph {et~al.}(2005)\citenamefont
  {Mustafa}, \citenamefont {Thoma},\ and\ \citenamefont
  {Chakraborty}}]{Mustafa:2004hf}%
  \BibitemOpen
  \bibfield  {author} {\bibinfo {author} {\bibfnamefont {M.~G.}\ \bibnamefont
  {Mustafa}}, \bibinfo {author} {\bibfnamefont {M.~H.}\ \bibnamefont {Thoma}},
  \ and\ \bibinfo {author} {\bibfnamefont {P.}~\bibnamefont {Chakraborty}},\
  }\href {\doibase 10.1103/PhysRevC.71.017901} {\bibfield  {journal} {\bibinfo
  {journal} {Phys. Rev.}\ }\textbf {\bibinfo {volume} {C71}},\ \bibinfo {pages}
  {017901} (\bibinfo {year} {2005})},\ \Eprint
  {http://arxiv.org/abs/hep-ph/0403279} {arXiv:hep-ph/0403279 [hep-ph]}
  \BibitemShut {NoStop}%
\bibitem [{\citenamefont {Ayala}\ \emph {et~al.}(2015)\citenamefont {Ayala},
  \citenamefont {Cobos-Martínez}, \citenamefont {Loewe}, \citenamefont
  {Tejeda-Yeomans},\ and\ \citenamefont {Zamora}}]{ayalafermionself}%
  \BibitemOpen
  \bibfield  {author} {\bibinfo {author} {\bibfnamefont {A.}~\bibnamefont
  {Ayala}}, \bibinfo {author} {\bibfnamefont {J.~J.}\ \bibnamefont
  {Cobos-Martínez}}, \bibinfo {author} {\bibfnamefont {M.}~\bibnamefont
  {Loewe}}, \bibinfo {author} {\bibfnamefont {M.~E.}\ \bibnamefont
  {Tejeda-Yeomans}}, \ and\ \bibinfo {author} {\bibfnamefont {R.}~\bibnamefont
  {Zamora}},\ }\href {\doibase 10.1103/PhysRevD.91.016007} {\bibfield
  {journal} {\bibinfo  {journal} {Phys. Rev.}\ }\textbf {\bibinfo {volume}
  {D91}},\ \bibinfo {pages} {016007} (\bibinfo {year} {2015})},\ \Eprint
  {http://arxiv.org/abs/1410.6388} {arXiv:1410.6388 [hep-ph]} \BibitemShut
  {NoStop}%
\bibitem [{\citenamefont {Haque}(2017)}]{Haque:2017nxq}%
  \BibitemOpen
  \bibfield  {author} {\bibinfo {author} {\bibfnamefont {N.}~\bibnamefont
  {Haque}},\ }\href {\doibase 10.1103/PhysRevD.96.014019} {\bibfield  {journal}
  {\bibinfo  {journal} {Phys. Rev.}\ }\textbf {\bibinfo {volume} {D96}},\
  \bibinfo {pages} {014019} (\bibinfo {year} {2017})},\ \Eprint
  {http://arxiv.org/abs/1704.05833} {arXiv:1704.05833 [hep-ph]} \BibitemShut
  {NoStop}%
\bibitem [{\citenamefont {Bandyopadhyay}\ \emph {et~al.}(2017)\citenamefont
  {Bandyopadhyay}, \citenamefont {Haque},\ and\ \citenamefont
  {Mustafa}}]{aritraweakpressure}%
  \BibitemOpen
  \bibfield  {author} {\bibinfo {author} {\bibfnamefont {A.}~\bibnamefont
  {Bandyopadhyay}}, \bibinfo {author} {\bibfnamefont {N.}~\bibnamefont
  {Haque}}, \ and\ \bibinfo {author} {\bibfnamefont {M.~G.}\ \bibnamefont
  {Mustafa}},\ }\href@noop {} {\  (\bibinfo {year} {2017})},\ \Eprint
  {http://arxiv.org/abs/1702.02875} {arXiv:1702.02875 [hep-ph]} \BibitemShut
  {NoStop}%
\bibitem [{\citenamefont {Weldon}(1982{\natexlab{a}})}]{weldonfermion}%
  \BibitemOpen
  \bibfield  {author} {\bibinfo {author} {\bibfnamefont {H.~A.}\ \bibnamefont
  {Weldon}},\ }\href {\doibase 10.1103/PhysRevD.26.2789} {\bibfield  {journal}
  {\bibinfo  {journal} {Phys. Rev. D}\ }\textbf {\bibinfo {volume} {26}},\
  \bibinfo {pages} {2789} (\bibinfo {year} {1982}{\natexlab{a}})}\BibitemShut
  {NoStop}%
\bibitem [{\citenamefont {Weldon}(1982{\natexlab{b}})}]{Weldon:1982aq}%
  \BibitemOpen
  \bibfield  {author} {\bibinfo {author} {\bibfnamefont {H.~A.}\ \bibnamefont
  {Weldon}},\ }\href {\doibase 10.1103/PhysRevD.26.1394} {\bibfield  {journal}
  {\bibinfo  {journal} {Phys. Rev.}\ }\textbf {\bibinfo {volume} {D26}},\
  \bibinfo {pages} {1394} (\bibinfo {year} {1982}{\natexlab{b}})}\BibitemShut
  {NoStop}%
\bibitem [{\citenamefont {Weldon}(2000)}]{Weldon:1999th}%
  \BibitemOpen
  \bibfield  {author} {\bibinfo {author} {\bibfnamefont {H.~A.}\ \bibnamefont
  {Weldon}},\ }\href {\doibase 10.1103/PhysRevD.61.036003} {\bibfield
  {journal} {\bibinfo  {journal} {Phys. Rev.}\ }\textbf {\bibinfo {volume}
  {D61}},\ \bibinfo {pages} {036003} (\bibinfo {year} {2000})},\ \Eprint
  {http://arxiv.org/abs/hep-ph/9908204} {arXiv:hep-ph/9908204 [hep-ph]}
  \BibitemShut {NoStop}%
\bibitem [{\citenamefont {Bellac}(2011)}]{Bellac:2011kqa}%
  \BibitemOpen
  \bibfield  {author} {\bibinfo {author} {\bibfnamefont {M.~L.}\ \bibnamefont
  {Bellac}},\ }\href
  {http://www.cambridge.org/mw/academic/subjects/physics/theoretical-physics-and-mathematical-physics/thermal-field-theory?format=AR}
  {\emph {\bibinfo {title} {{Thermal Field Theory}}}}\ (\bibinfo  {publisher}
  {Cambridge University Press},\ \bibinfo {year} {2011})\BibitemShut {NoStop}%
\bibitem [{\citenamefont {Karsch}\ \emph {et~al.}(2001)\citenamefont {Karsch},
  \citenamefont {Mustafa},\ and\ \citenamefont {Thoma}}]{Karsch:2000gi}%
  \BibitemOpen
  \bibfield  {author} {\bibinfo {author} {\bibfnamefont {F.}~\bibnamefont
  {Karsch}}, \bibinfo {author} {\bibfnamefont {M.~G.}\ \bibnamefont {Mustafa}},
  \ and\ \bibinfo {author} {\bibfnamefont {M.~H.}\ \bibnamefont {Thoma}},\
  }\href {\doibase 10.1016/S0370-2693(00)01322-8} {\bibfield  {journal}
  {\bibinfo  {journal} {Phys. Lett.}\ }\textbf {\bibinfo {volume} {B497}},\
  \bibinfo {pages} {249} (\bibinfo {year} {2001})},\ \Eprint
  {http://arxiv.org/abs/hep-ph/0007093} {arXiv:hep-ph/0007093 [hep-ph]}
  \BibitemShut {NoStop}%
\bibitem [{\citenamefont {Chakraborty}\ \emph {et~al.}(2002)\citenamefont
  {Chakraborty}, \citenamefont {Mustafa},\ and\ \citenamefont
  {Thoma}}]{Chakraborty:2001kx}%
  \BibitemOpen
  \bibfield  {author} {\bibinfo {author} {\bibfnamefont {P.}~\bibnamefont
  {Chakraborty}}, \bibinfo {author} {\bibfnamefont {M.~G.}\ \bibnamefont
  {Mustafa}}, \ and\ \bibinfo {author} {\bibfnamefont {M.~H.}\ \bibnamefont
  {Thoma}},\ }\href {\doibase 10.1007/s100520200899} {\bibfield  {journal}
  {\bibinfo  {journal} {Eur. Phys. J.}\ }\textbf {\bibinfo {volume} {C23}},\
  \bibinfo {pages} {591} (\bibinfo {year} {2002})},\ \Eprint
  {http://arxiv.org/abs/hep-ph/0111022} {arXiv:hep-ph/0111022 [hep-ph]}
  \BibitemShut {NoStop}%
\bibitem [{\citenamefont {Frenkel}\ and\ \citenamefont
  {Taylor}(1990)}]{Frenkel:1989br}%
  \BibitemOpen
  \bibfield  {author} {\bibinfo {author} {\bibfnamefont {J.}~\bibnamefont
  {Frenkel}}\ and\ \bibinfo {author} {\bibfnamefont {J.~C.}\ \bibnamefont
  {Taylor}},\ }\href {\doibase 10.1016/0550-3213(90)90661-V} {\bibfield
  {journal} {\bibinfo  {journal} {Nucl. Phys.}\ }\textbf {\bibinfo {volume}
  {B334}},\ \bibinfo {pages} {199} (\bibinfo {year} {1990})}\BibitemShut
  {NoStop}%
\end{thebibliography}%
\bibliographystyle{apsrev4-1}

\end{document}